\documentclass[12pt,preprint]{aastex}

\shorttitle{A Population of Faint Line Emitters}
\shortauthors{Rauch, Haehnelt, Bunker, Becker et al}

\begin{document}


\title{A Population of Faint Extended Line Emitters and the Host Galaxies of
Optically Thick QSO Absorption Systems\altaffilmark{1}}


\author{Michael Rauch}
\affil{Observatories of the Carnegie Institution of Washington, 813 Santa Barbara Street, Pasadena, CA 91101}
\email{mr@ociw.edu}
\author{Martin Haehnelt}
\affil{Institute of Astronomy, Madingley Road, Cambridge CB3 0HA, UK}
\email{haehnelt@ast.cam.ac.uk}
\author{Andrew Bunker}
\affil{School of Physics, Stocker Road, Exeter EX4 4QL, UK}
\email{bunker@astro.ex.ac.uk}
\author{George Becker}
\affil{Observatories of the Carnegie Institution of Washington, 813 Santa Barbara Street, Pasadena, CA 91101, USA}
\email{gdb@ociw.edu}
\author{Francine Marleau }
\affil{Spitzer Science Center, Caltech, Mail Stop 220-6, 1200 East California Blvd, Pasadena, CA 91125, USA}
\email{marleau@ipac.caltech.edu}
\author{James Graham }
\affil{601 Campbell Hall, University of California, Berkeley CA 94720-3411 ,USA}
\email{jrg@graham.berkeley.edu}
\author{Stefano Cristiani }
\affil{ Osservatorio Astronomico di Trieste, INAF, Via Tiepolo 11 34143 Trieste, Italy}
\email{cristiani@oats.inaf.it}
\author{Matt J.  Jarvis}
\affil{Centre for Astrophysics, Science \& Technology Research Institute, University of Hertfordshire, Hatfield AL10 9AB, UK}
\email{M.J.Jarvis@herts.ac.uk}
\author{Cedric Lacey }
\affil{Institute for Computational Cosmology, Department of Physics, Durham University, South Road, Durham DH1 3LE, UK}
\email{cedric.lacey@durham.ac.uk}
\author{Simon Morris }
\affil{Department of Physics, Durham University, South Road, Durham DH1 3LE, UK}
\email{simon.morris@durham.ac.uk}
\author{Celine Peroux }
\affil{Observatoire Astronomique de Marseille-Provence, 2, Place Le Verrier,13248 Marseille Cedex 04, France}
\email{celine.peroux@oamp.fr}

\author{Huub R\"ottgering}
\affil{Leiden Observatory, NL-2300 RA Leiden, The Netherlands}
\email{rottgeri@strw.leidenuniv.nl}

\author{Tom Theuns }
\affil{Department of Physics, Durham University, South Road, Durham DH1 3LE, UK}
\email{tom.theuns@durham.ac.uk}

\author{ }
\affil{}


\altaffiltext{1}{Based partly on observations made with ESO Telescopes at the Paranal Observatories under Program ID LP173.A-0440,
and partly on observations obtained at the Gemini Observatory, which is operated by the
Association of Universities for Research in Astronomy, Inc., under a cooperative agreement
with the NSF on behalf of the Gemini partnership: the National Science Foundation (United
States), the Science and Technology Facilities Council (United Kingdom), the
National Research Council (Canada), CONICYT (Chile), the Australian Research Council
(Australia), CNPq (Brazil) and CONICET (Argentina).}
\pagebreak
\begin{abstract}
We have conducted a long slit search for low surface brightness Lyman $\alpha$ emitters at redshift $2.67 < z < 3.75$. A 92 hour long exposure with the ESO VLT FORS2 instrument 
down to a  $1\sigma$ surface brightness detection limit of  $8\times 10^{-20}$ erg cm$^{-2}$ s$^{-1}$ $\sq\arcsec^{-1}$
per $\sq\arcsec$ aperture,
yielded a sample of 27 single 
line emitters with fluxes of a few $\times 10^{-18}$ erg s$^{-1}$cm$^{-2}$. We present arguments that most objects
are indeed Ly$\alpha$. 
The large comoving number density $3\times10^{-2}$ h$_{70}^3$ Mpc$^{-3}$, the large covering factor $dN/dz \sim 0.2- 1$, and the often extended Ly$\alpha$ emission suggest that the emitters
be identified with the elusive host population of damped Ly$\alpha$ systems (DLAS) and high column density
Lyman limit systems (LLS). A small inferred star formation rate, perhaps supplanted by cooling radiation,
appears to energetically dominate the Ly$\alpha$ emission, and is consistent with
the low metallicity, low dust content, and theoretically inferred low masses of DLAS,  and with the relative lack of success of earlier searches for their optical counterparts. Some of the line profiles show evidence for radiative transfer in galactic outflows. Stacking surface brightness profiles  we find emission out to at least 4\arcsec . 
The centrally concentrated emission of most objects appears to  light up the outskirts of the emitters (where LLS arise) down to a column density where the conversion from UV to Ly$\alpha$ photon becomes inefficient. DLAS, high column density LLS, and the emitter population
discovered in this survey  appear to be different observational manifestations  of the same low-mass, protogalactic building blocks of present day $L*$ galaxies. 

\end{abstract}





\section{Introduction}

Perhaps surprisingly, modern astronomy has found ways to study the material
structures of the high redshift universe over their entire 
vast range of densities and sizes, from the underdense large voids, via bright starforming
galaxies to the most luminous QSOs.

The dark stages of this sequence, with densities from the universal mean to virial
density, too low to produce 
detectable radiation, have been studied mainly with QSO absorption lines.
This approach has taught us where to find  most of the baryons (in the ionized  Lyman $\alpha$ forest gas; e.g., Rauch et al 1997a) and most of the neutral gas (in damped Lyman $\alpha$ systems; e.g., Wolfe et al 1995) at high redshift.
From the bright end of the cosmic matter distribution, serious inroads into the population of high redshift galaxies have been made by color-selecting stellar continuum emitters,
exploiting the Lyman limit continuum decrement (e.g., Steidel \& Hamilton 1993), or by
searching for Ly$\alpha$ line emission induced by star-formation (e.g., Cowie \& Hu 1998). 
The advent of space-based, broad-band galaxy surveys, in particular the Hubble Ultra Deep Field (Bunker et al 2004; Beckwith et al 2006; Bouwens et al 2007) has 
begun filling in the gap between the essentially dark, lower mass
range of sub-galactic or barely star-forming proto-galactic objects probed by absorption lines, and bright high redshift galaxies with large star-formation rates of tens of $M_{\odot}$yr$^{-1}$ (e.g., Erb et al 2006). 

The boundary between dark and bright is delineated by an important astrophysical transition, the one from ionized gas to neutral, self-shielded gas, on a galactic mass scale.  
The underlying objects in which this transition has happened,
known from absorption studies as Lyman limit systems (LLS), or, at higher column densities, damped Ly$\alpha$ systems (DLAS),  have been rather elusive.  We know quite a bit about their chemistry and ionization state
but little in terms of stellar contents, size,  kinematics  or  mass. 

Theoretical studies over the past two decades have suggested that
a survey of LLS and DLAS for HI Ly$\alpha$ line emission of sufficient depth may uncover several distinct
astrophysical sources of line emission that have the potential to shed
considerable light (literally) on the distribution of neutral hydrogen, and thus the
bedrock of galaxy formation. 
If we acknowledge that high redshift DLAS must have something to do with stars
(they contain the main reservoir of neutral gas, and have 
a median metallicity of [Z/H] = -1.5, more than an order of magnitude higher than the coeval abundances in the intergalactic medium) then we may expect 
the potentially strongest signal to be star formation induced Ly$\alpha$ and/or a stellar
continuum.
Attempts at identifying individual  LLS or DLAS with galaxy counterparts have often been frustrated by the difficulty of detecting an extremely faint object (the DLAS host) next to an extremely bright object (a QSO). 
At low redshift ($z<1$) where we can hope to learn most about the underlying galaxy population, observations have shown that DLA host galaxies represent
a range of galaxy types (e.g., LeBrun et al 1997, Chen, Kennicutt \& Rauch 2005) dominated by faint objects (e.g., Chun et al 2006) with  low star formation rates 
(Wild, Hewett \& Pettini 2007).  
Searches at high redshift (e.g., Warren et al 2001,  Fynbo et al 2003; Kulkarni et al 2000, 2001;  Christensen
et al 2007) have so far produced only a handful of confirmed detections of the underlying galaxies (Weatherley et al 2005).
Such efforts indicated that DLAS hosts at high redshift are generally drawn from the very faint end (Fynbo, M\o ller, \& Warren 1999; Bunker et al 1999)  of the general galaxy population at high redshift, intersected by the QSO line of sight at  small impact parameters (M\o ller et al
2002). Wolfe \& Chen (2007) have recently performed a search for
spatially extended continuum emission down to very faint levels using the Hubble Ultra Deep Field, and were able to place stringent upper limits on extended star formation in DLAS. 

These results agree well with theoretical CDM-based models of galaxy formation 
(Kauffmann 1996) that envisage DLAS hosts as numerous small, faint, low mass, merging protogalactic clumps (Haehnelt, Steinmetz
\& Rauch 1998, 2000; Ledoux et al 1998;  Johansson \& Efstathiou 2006; Nagamine et al 2007), rather than the large hypothetical disks (Prochaska \& Wolfe 1997) once popular.

\medskip

Cooling radiation from collapsing galaxies (e.g., Haiman \& Rees 2001; Haiman, Spaans \& Quataert 2000; Fardal et al 2001, Furlanetto et al 2005) is a second  process
able to produce line radiation at fluxes competitive with those from star formation,
at least in the halos of relatively massive galaxies (Dijkstra et al 2006a,b).
The peculiar spatial distribution of the radiation and certain asymmetries
of the spectral line profile, together with the local absence of a stellar continuum,
could conceivably help to distinguish this source of radiation from star formation.
The few  candidates for such galactic cooling flows reported so far (Keel et al 1999; Steidel
et al 2000; Francis et al 2001; Bower et al 2004; Dey et al 2005; Matsuda et al 2006; Nilsson et al 2006;
Smith \& Jarvis 2007), have been atypical for the galaxy population
as a whole.

Perhaps the most curious of these emission processes is  Ly$\alpha$ fluorescence,
where HI ionizing photons impinging on optical thick gas are absorbed and
converted with high efficiency to Ly$\alpha$ line radiation, albeit at a very low surface brightness level $ < 10^{-19}$erg cm$^{-2}$s$^{-1}\sq\arcsec^{-1}$; Hogan \& Weymann 1987; Binette et al 1993; Gould \& Weinberg 1996; Cantalupo, Lily \& Porciani 2005).  Lyman limit patches should light up the whole gaseous cosmic
web to yield a relatively uniform, very faint "glow",  perhaps
locally enhanced by the proximity of a QSO (e.g., Cantalupo et al 2005, 2007).
A number of increasingly deep searches for this effect at high redshift have been performed 
(Lowenthal et al 1990; Bunker et al 1998, 1999), in pursuit of a  repeatedly shrinking  theoretically
predicted signal. A detection, which at the same time 
would be a measurement of the general UV background, has so far eluded us. Most recently, a number of detections of the enhanced Ly$\alpha$ fluorescence in the proximity of QSOs have been reported (Fynbo, M\o ller, \& Warren 1999; Bunker et al 2003;  Weidinger et al 2004, 2005; Adelberger et al 2006,
Cantalupo et al 2007, Hennawi 2007) but the understanding of the effect has been complicated by the large
number of degrees of freedom in the properties and behavior of the QSO (e.g., orientation, opening angle of the beam,
life time).

Thus there is good reason to conduct a search for extremely low light level Ly$\alpha$ line emission
at high redshift. While most previous Ly$\alpha$ surveys have been performed as narrow band imaging
searches it is clear that the largest contrast with the sky background and the largest
redshift (=spatial) depth is obtained by 
(long) slit spectroscopy. Because of the small volume covered 
a blind search is not a viable way to find  objects as rare as Lyman break galaxies, which
are hard to hit with a single randomly positioned slit. 
In contrast, the rate of incidence of  LLS with neutral hydrogen column densities exceeding (N(HI) $> 10^{19}$cm$^{-2}$) per unit redshift is approximately unity at redshift $\sim 3$ (e.g., 
Peroux et al 2003), i.e., the objects  essentially cover the sky, and there should be numerous hits in a single setting for a typical long slit spectrograph.  We anticipate  seeing about 30 patches of emission on a 2\arcsec\ wide and
7\arcmin\ long slit at redshift $\sim 3$ (e.g., Bunker et al 1998).
Searching over a  redshift range near $z\sim 3$ achieves a reasonable comprise between avoiding
the ravages of $(1+z)^4$ dimming and succumbing to the relatively poor blue performance of 
most low resolution spectrographs worldwide. However, in principle it
is desirable to go as far to the blue as possible. The UV background
as the source of the Ly$\alpha$ fluorescence
is not expected to drop dramatically down to at least redshift 2,  whereas the 
atmospheric background as the dominant source of noise decreases rapidly toward
the blue.

Prior to the current ESO VLT project the most sensitive search for Ly$\alpha$ fluorescence off optically thick gas, in regions dominated
by the general UV background, had been performed by the longslit experiment of Bunker et al (1998, 1999). 
The first attempt at measuring this effect with the ESO VLT originated in a (shelved)  science verification
project for FORS, later submitted unsuccessfully during periods 64-66
as a large project (PI Rauch). The amount of exposure time required even with an 8m class telescope
considerably exceeds typical observing time awards so  the strategy was to exploit the newly available service mode observing
and insert the observations
during bad seeing periods, as the targets were expected to be extended, with radii of several arcseconds.   
The current incarnation, a combination of the two projects, re-observing  the original target field of Bunker et al (1999),
was awarded 120 hours as  ESO Large Project (LP173.A-0440: PI Haehnelt). 
A parallel attempt to use the Gemini
GMOS instruments on the same field chosen for the ESO project
led to time awards with both Gemini telescopes (GN 2004-A-Q-91, GN-2005-B-Q-52, GS-2004-B-Q-61, GS-2005-B-Q-36; PI Bunker) and (GN-2004-B-Q-35,GN-2004-B-Q-35, GS-2004A-Q-78, GS-2004-B-Q-8: PI Rauch).
The Gemini effort resulted in  $46\times 3000s$ exposures, of which
30 where taken with Gemini-North, 16 with Gemini-South.
The Gemini data were taken at lower resolution (to take advantage of
the most commonly used spectroscopic
setup) but cover a larger redshift path than the ESO data.
The current paper deals with the ESO dataset and we 
postpone a full analysis of all data to a future paper. 
However, the Gemini data in a preliminary reduction have  been consulted here for checking the reality
of the emitters found in the ESO dataset (see below).

The ESO FORS2 project  resulted in (after overheads) 92 hours of on-source exposure, finally  reaching a $1\sigma$ surface brightness detection threshold of $8\times 10^{-20}$ erg cm$^{-2}$ s$^{-1}$ $\sq\arcsec^{-1}$ (measured in a one $\sq\arcsec$ aperture). This is remarkably close to our expected sensitivity ($6.6\times 10^{-20}$ erg cm$^{-2}$ s$^{-1}$ $\sq\arcsec^{-1}$ in 120 h).

The experiment is described below. Section \ref{obssec} details the
observational setup, and the sensitivity and spectral resolution limits reached.
Section \ref{obsressec} spells out the individual properties of the emitters found, and gives estimates of the sizes, number densities, flux distributions,
and derived rates of incidence.
 Section \ref{whataretheysec} discusses the possible identification with low redshift galaxies,
and Section \ref{lyasec} ponders the consequences if the emitters are predominantly Ly$\alpha$.
Alternative origins of the Ly$\alpha$ photons (stellar, QSO-induced, cooling radiation) are discussed 
in section \ref{origsec}, including a detailed discussion of some unusual line profiles found.
Section \ref{dlassec} establishes a connection between the emitters and  
QSO DLAS, followed by a brief discussion of Ly$\alpha$ fluorescence in section
\ref{fluores} and the conclusions in section \ref{concsec}. An Appendix
discusses the importance of slit losses.


\section{Observations}
\label{obssec}
The QSO DMS B 2139-0405 (z=3.32, V =20.805; Hall et al 1996) was observed during 2004 - 2006
with the VLT FORS2 low resolution spectrograph and the volume-phased holographic grism 1400V.
A $2\arcsec$ wide and about $453\arcsec$ long slit gave a spectral resolution of $\lambda/\Delta\lambda_{\rm FWHM} = 1050$. The CCDs were readout in
$2\times2$ binned mode, giving pixels with an extent of $0.252\arcsec$ along the slit and about 0.64 \AA\ in the
dispersion direction. Thus, the spectral resolution FHWM is sampled by about 8 pixels. The spectrum on the detector ranges from 4457 to 5776 \AA , with a midpoint at 5099 \AA .

The slit was centered on the QSO and rotated by a position angle $-5.95\deg $, to repeat the orientation  of the earlier Keck LRIS observation
by Bunker et al (1999), where the particular position angle was chosen
to minimize intersecting bright foreground galaxies. 

A total of 110 exposures were taken between May 2004 and August 2006,  giving a total integration time of 92 hours.  The exposures were  divided roughly evenly into three dither positions separated on the  sky by 10\arcsec .  The data were reduced using a custom set of IDL  routines.  Individual exposures were bias-subtracted and flat-fielded.  The sky was then subtracted from each exposure using an  optimal sky subtraction technique based on Kelson (2003), whereby the  sky counts are modeled as a function of wavelength and slit position  without rectifying the original data.  Object traces that were  visible in a single 3000 second exposure were masked when fitting the  sky.  The reduced two-dimensional spectra from all exposures and both  FORS2 CCDs was then combined into single array with spectral  dispersion closely matching the original exposures.  Nearest-neighbor  sampling was used when combining the frames to avoid correlating  adjacent pixels.  In order to avoid false detections, hot and cold  pixels, bad rows, charge traps, and other defects were identified in  the flat fields and reduced frames and aggressively excluded when  producing the combined spectrum.  Pixels with significant dark  current, identified by combining large numbers of reduced exposures,  were also rejected.  In total, roughly 2.5\% of the illuminated area  of each CCD was masked when producing the combined spectrum.   Combinations of various sub-samples of the data were also made to  check that no spurious features remained.

\begin{figure}[p]
%
\includegraphics*[scale=1.1,angle=0.]{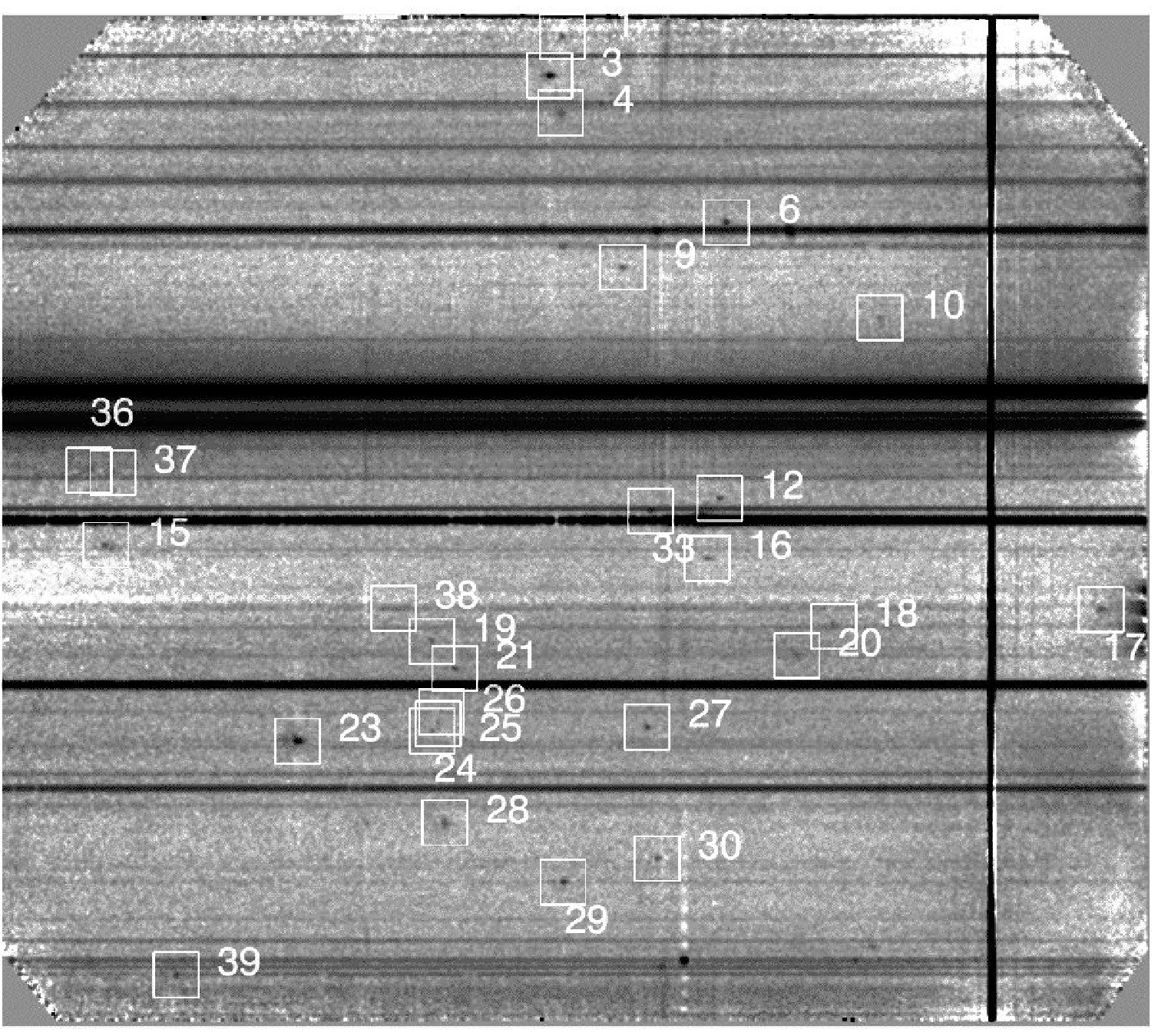}
\caption{\small Two-dimensional spectrum obtained in 92 hours of exposure time., showing the line emitter candidates for HI Ly$\alpha$ (boxes).  The dispersion direction is horizontal, with
blue to the left and red to the right; the spatial direction along
the slit is vertical. The QSO spectrum (multiple
absorption lines) is visible close to the center of the image.
The numbers refer to the column entry
'ID' in table \ref{bigtable}.\label{2dspec}}
\end{figure}

The resulting 2-dimensional spectrum is shown in fig \ref{2dspec}. A flux calibration using several spectrophotometric standard stars reproduced the published flux of the QSO to within about 20\% . 
In the center of the final, sky-subtracted spectrum, a flux density of $1.6 \times 10^{-20}$ erg cm$^{-2}$ s$^{-1}$ $\AA^{-1}$ produced 1 ADU per $0.252\arcsec\times2\arcsec\times0.64$ \AA\ wide pixel. The observed standard deviation near the center of the spectrum is  2.0 ADU. We can reconstruct a surface brightness profile of a line emitter  by integrating over
the line in the spectral direction (over one FWHM, $\sim 286$ kms$^{-1}$).  This gives a $1-\sigma$ surface brightness detection limit of    $8.1\times 10^{-20}$ erg cm$^{-2}$ s$^{-1}$ $\sq\arcsec^{-1}$, if
measured in a one $\sq\arcsec$ aperture.

The "seeing" profile as measured off the QSO near the center of the combined final spectrum is 1.07\arcsec\ FWHM. The seeing conditions
generally where not as bad as anticipated, with 89\% of the seeing
better than 1.5\arcsec .

The large number of continuum sources on the slit, the presence of
a few brighter stars and galaxies with PSF wings visible at
large distances,  together with
charge transfer and other cosmetic problems, exascerbated by a finite
number of dithering positions, made a selection of objects with
an automatic method impractical. Thus, emission line objects were selected by eye. The selected depth clearly varies with
the character of the object (extended, or point sources), but we estimate
that for an extended object (angular extent $> 1 \sq\arcsec$) we get a 
visual detection at a surface brightness $2\times 10^{-19}$ erg cm$^{-2}$ s$^{-1}$ $\sq\arcsec^{-1}$, reflected in the dearth of objects with a 
central surface brightness below that value in fig. \ref{2dspec}. Quantitatively, this corresponds to about three standard deviations in
the more precise surface brightness detection threshold based on pixel
noise given above.

We have tested the reality of the emitters found by
splitting the sample into halves. We were able to cross-identify all but two objects (numbers 18 and 25; see below) between the halves. One more (relatively
 bright) object (ID 2)
was entirely absent in one half but had already been excluded from the original sample because
of its suspicious shape. Perusing the combined spectrum from the Gemini
telescopes we are able to cross-identify 9 of the 22 ESO objects of our emitter
sample with common coverage by both the ESO and GMOS slits. However, most
objects are barely detected in the  shallower Gemini data. 

The list in table \ref{bigtable} contains 4 objects with a
significance (based on the standard deviation of the flux) of less
than $4\sigma$.  Of these, object 18 may or may not be real. 19 and 26 certainly
are present in both halves of the dataset. Object 25 
would not have been considered a detection everywhere in the two-dimensional spectrum
but struck the eye because it seemed to constitute a group with the nearby objects 24, 26, and possibly 19 and 21.
Thus we estimate that about two objects may be spurious detections. On the
other hand,
there are also a couple of candidates which  could have been included in the sample but were not.

\section{Observational Results}
\label{obsressec}

\begin{figure}[p]
%
\vspace{-1.cm}
\includegraphics*[scale=0.7,angle=90.]{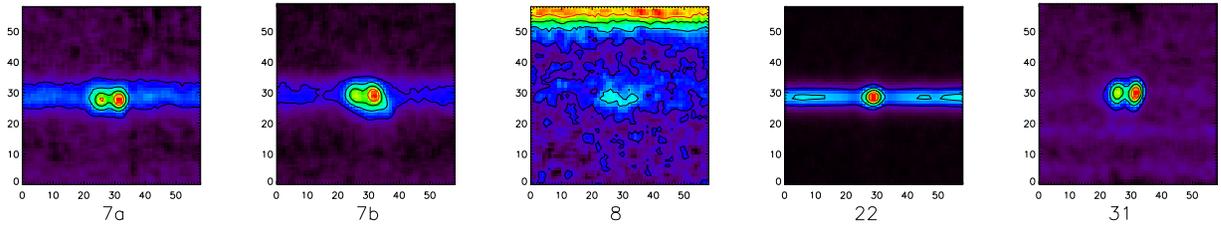}

\caption{\small Spectra of the emission line regions of five foreground line emitting galaxies identified
from their [OII] doublet (objects 7a, 7b, 31) or Balmer emission ($H_{\beta}$ for object 8, $H_{\gamma}$ for object 22) features. The coordinates are in pixel units ($0.252\arcsec \times 0.67 \AA $).
The sections of the spectra shown here are 15.12"  wide in the spatial direction and about
2266 kms$^{-1}$ long in the spectral direction (i.e., horizontally). 
The numbers refer to the column entry
'ID' in table \ref{oiitable}. \label{oiigals}}
\end{figure}

Spurious detection is a less serious concern than misidentification and 
the resulting contamination of the emitter sample by foreground low redshift galaxies, as lower redshift [OII] 3727\AA\ (a doublet marginally
resolved at our resolution), [OIII] 5007 \AA , or the HI  Balmer series in emission could
be mistaken for HI Ly$\alpha$. We checked for the presence of these features and for others that could give away a low z object, like 
the absence of a Lyman $\alpha$ forest decrement (when a continuum
was present), and  the existence of extended  emission 
in the broad band image. 
Among the emission line objects, five line emitters could be identified with foreground galaxies. They are listed in table \ref{oiitable} and shown in fig. \ref{oiigals}.
The table gives the ID numbers, redshifts, the sky-subtracted flux $F$ in units of $10^{-18}$erg cm$^{-2}$ s$^{-1}$ measured in a   $2\arcsec \times 2 \arcsec \times 755$ km s$^{-1}$ aperture, the maximum surface brightness along the slit in units of $10^{-18}$erg cm$^{-2}$ s$^{-1}$ arcsec$^{-2}$, and the source of the identification as a foreground galaxy.
 As we were interested only in line emitters we ignored the continuum-only objects for the time being.
Three of the emission line objects accidentally on the slit
were identified with foreground galaxies based on the presence of [OII]
(two spatially almost coincident objects at z=0.39278 and z=0.4336, with IDs 7a and 7b; a galaxy at z=0.4019, ID 31).
A fourth object appears to show  an  [OII] doublet on the very edge of the detector, and is clearly identified by several HI Balmer emission lines as a z=0.198 galaxy (ID 22). A fifth
object shows H$\beta$ and the Mg$\ b$ triplet at z=0.0458 (ID 8).  The contamination of the remaining sample of emitters by [OII] and [OIII]
is still a concern because we do not have information about the continuum for most objects.
We will treat the sample of emitters formally as Ly$\alpha$ in most of the paper
but will return to a discussion of contamination below.
The remaining 27 line emitters are listed in table \ref{bigtable}. 

\begin{figure}[p]
\includegraphics*[scale=0.7,angle=90.]{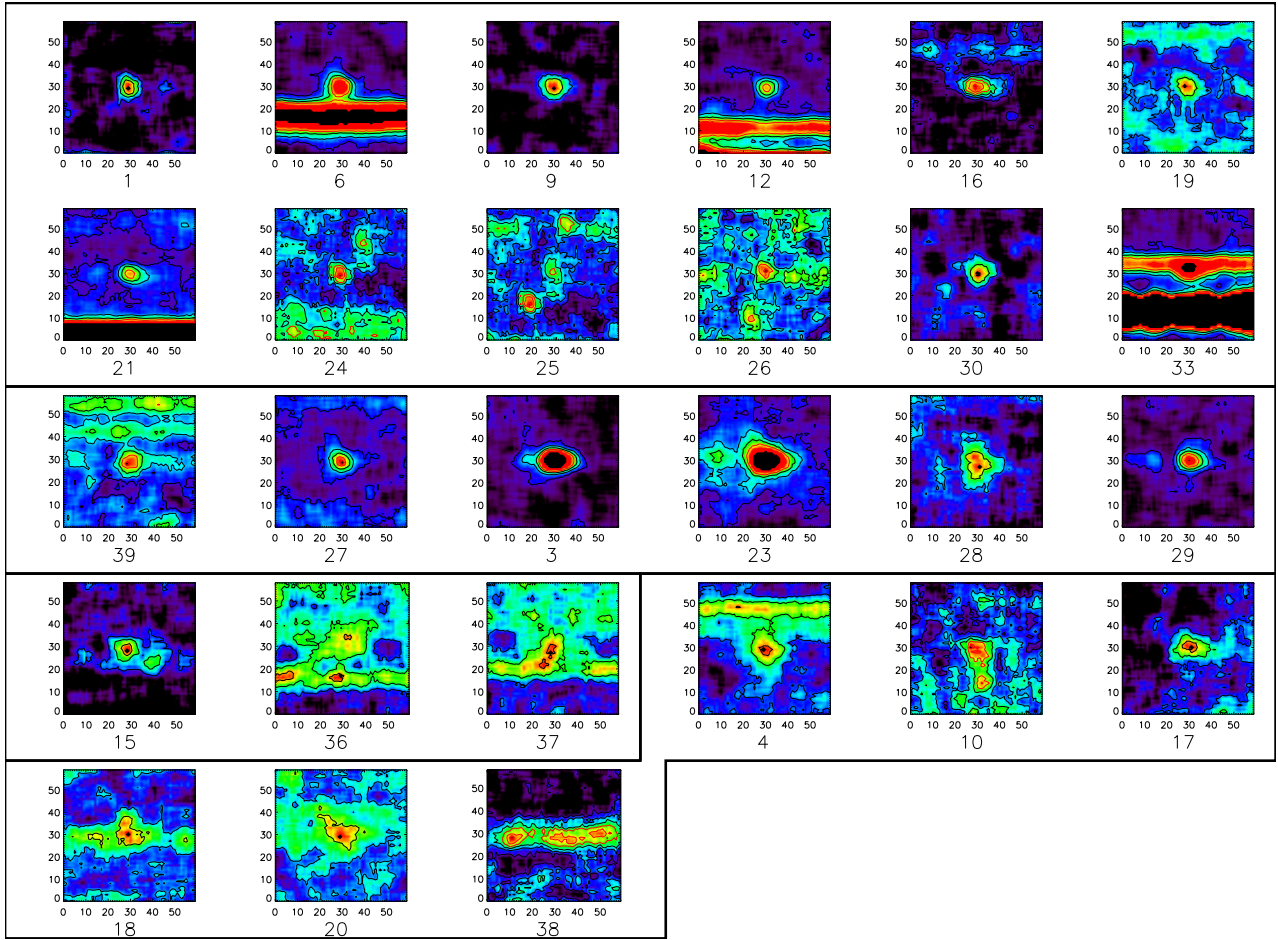}
\caption{\small short spectra
for the single line emitters. The coordinates are in pixel units ($0.252\arcsec \times 0.67 \AA $).
The sections of the spectra shown here are 15.12" or 116 proper kpc wide in the spatial direction and about
2266 kms$^{-1}$ long in the spectral direction (i.e., horizontally). The spectra have been heavily smoothed with a 7x7 pixel boxcar filter.
The areas within the light grey contours  (turqoise in the color version) have a flux density greater than approximately $1.5\times 10^{-20}$ erg cm$^{-2}$ s$^{-1}$ $\AA $. The numbers refer to the column entry
'ID' in table \ref{bigtable}.  The spectra are grouped together such that the first 12 of them
(top box) appear to have a single central peak; the next six (IDs 39, 27, 3, 23, 28, and 29, second box from top) show a clearly asymmetric red peak with a much weaker blue
counter-peak; the following three (third box to the left) either have a stronger blue than red peak (ID 15) or emission features blueward of an absorption line (36, 37); the remaining six are
unclassifiable, sometimes amorphous objects.\label{allpanel}}
\end{figure}

Fig. \ref{allpanel} shows the
two dimensional spectra of all remaining candidate  Ly$\alpha$ emission line regions. Numbers correspond to the 'ID' entry (column (2))
in table \ref{bigtable}. The 'stamps' are $60\times60$ pixels wide, with an individual pixel size of $0.252\arcsec\times 0.64 \AA $. Thus each stamp corresponds to 15.12 \arcsec (116 physical kpc at
the central redshift, z=3.2) in the spatial
direction (vertical) and 38.4\AA\ (2266 kms$^{-1}$ at the CCD center)  in the dispersion direction (horizontal; wavelength increases toward the right). The spectra have been heavily smoothed in both the spectral and spatial direction
(with a 7x7 boxcar filter)  for display purposes and to emphasize coherent regions of extended emission.
All spectra are displayed to within the same color stretch to demonstrate the variety in their appearance. Pixels within the  light grey (turquoise in the color version) areas correspond to a flux density  $> 1.5\times10^{-20}$erg cm$^{-2}$ s$^{-1}$ $\AA^{-1}$.

A closeup of the spectra, showing 7.56\arcsec\ by 25.6 \AA\ in a less highly smoothed ($3\times3$ pixels) version is seen in fig. \ref{indivpanel}.  Here the colorstretch was done individually for each image to emphasize the individual dynamic range. The QSO Ly$\alpha$ emission line region (ID 14) is also shown for reference but was not included in the actual analysis.
Next to the 2-d spectra are the object IDs, a background-subtracted one dimensional spectrum produced by collapsing the box along the slit direction
(vertical) in units of erg s$^{-1}$cm$^{-2}$\AA$^{-1}$, followed by a spatial surface brightness  profile along the slit direction that was obtained by collapsing the box in the dispersion direction 
and averaging the bottom and top sections to improve the signal to noise ratio.
The velocity and spatial zeropoints are the left edge of the box and the
center of the box, respectively, where the box is centered on the peak of a Gaussian fit to the spectral and spatial light  profile.

The 1-d spectra often look poor which is due to the faint signal
and the box size, folding in a lot of noise, and sometimes wings of other objects.
Optimal extraction has also been performed but with surprisingly
little improvement. 

To help judge whether a surface brightness profile is extended, the spatial profiles to the right also
show a model fit, where the surface brightness $S(y)$
is represented by a Gaussian core plus
power law wings (dotted profile). First the Gaussian with the fixed width of the point spread function (dashed profile) is fit to the innermost two pixels, to determine
the amplitude parameter A for the core of the emission. Then, beyond a distance $y_t$ along the slit from the center of the emission, a power law replaces the Gaussian. The transition
distance along the slit from the center of the emission, $y_t$, and the power law index are treated as free parameters:   
\begin{eqnarray}
S(y) & = A \exp [- y^2 / (2\sigma^2)],&  y < y_{t},\nonumber\\
     & = A \exp [- y_t^2 / (2\sigma^2)] \left(\frac{y}{y_t}\right)^{\alpha},&  y > y_{t}.\label{eq:model1}
\end{eqnarray}

In one instance the spatial  fit failed, because there were other objects nearby.

Most of the objects seem to have a relatively well defined emission peak, often surrounded by diffuse spatial emission, sometimes with
rather broad emission lines (fig. \ref{allpanel}). We have tried to losely classify these objects visually, according to whether they are consistent with
being point sources (i.e., have a Gaussian seeing profile with a FWHM $\sim 1.07\arcsec$, as derived from the QSO), labelled
as 'PS' in table \ref{bigtable}; whether they are dominated by amorphous extended emission, labelled as 'EXT', or a mix between
the two, centrally dominated ('CD') emission with yet an extended halo around them. The distinction is often subjective,
and used only to introduce some nomenclature to aid the discussion. In this scheme, objects with IDs 1,4,9,10,15,16,17,18,20,
28, 36, 37 and 39 are clearly extended. IDs 6,9,16, 19, 23, 29, and 30 could be classified as centrally dominated.
IDs 3,12,21,24,25,26,27, and 33 are in the somewhat better defined 'point source' class.

Several among the brighter, mostly PS and CD sources (IDs 3, 23, 27, 28,
29, 39) exhibit the classic asymmetric line
profiles known from previous studies of starbursting galaxies (e.g., Franx et al 1997; Mas-Hesse et al 2003), with a blue cutoff and a more extended red wing. A subset of those (3, 23, 28 and 29) appear to have a weaker
blue emission line in addition to the red dominant one, perhaps a sign of the double humped emission profile expected
from a static, externally illuminated slab (e.g. Neufeld 1990, Zheng \& Miralda-Escude 2002).

Intriguingly, at least one object (ID 15) shows the opposite situation, a stronger blue line opposed by a weaker red one, and there
are two other bizarrely shaped objects (36, 37) where the emission seems to occur blueward of the absorption the objects cause
in a nearby galaxy continuum. In those two cases the emission seems to drift redward toward larger distances from the absorber.

The object 38 shows a large emission region on top of a diffuse continuum, but because of its low S/N ratio
remains unclassifiable.

The properties of the detected sources are listed in table \ref{bigtable}.
The columns give: (1) the number of the entry in the table; (2)  the identification number of each object, as used
throughout the paper (including the figures); (3) the redshift, assuming the emitter is Lyman $\alpha$ 1215.67 \AA ;
(4) the sky-subtracted flux $F$ in units of $10^{-18}$erg cm$^{-2}$ s$^{-1}$ measured in a   $2\arcsec \times 2 \arcsec \times 755$ km s$^{-1}$ aperture; 
(5) the ratio of that flux to the one measured in a larger ($2\arcsec\times 7.6\arcsec \times 1510 $ km s$^{-1}$) aperture;
(6) the maximum surface brightness along the slit in units of $10^{-18}$erg cm$^{-2}$ s$^{-1}$ arcsec$^{-2}$; (7) the FWHM velocity width of a single component Gaussian fit to the optimally 
extracted emission line as a crude measure of overall velocity width, even where the line shape was distinctly non-Gaussian; (8) the Gaussian amplitude
of the central emission region for the surface brightness model profile described in the text; (9) the turnover distance between Gaussian center and power law wings for that model;
(10) the power law index for that model; (11) objects are losely classified as PS (point source), EXT (extended) or CD (centrally dominated), and peculiarities
are noted.  

The table shows a few instances where the ratio between the fluxes
measured in the smaller and larger apertures was formally larger than
unity. This happens when horizontal streaks on the CCD and sky residuals
enter the larger window but not the small one, or when the background
subtraction windows where in different positions for the two apertures.

The errors quoted are standard deviations
propagated in the usual way from the original pixel photon fluctuations.
The $1\sigma$ noise of the sky-subtracted flux (entry (4)) is quoted for the $2\arcsec \times 2 \arcsec \times 755$ km s$^{-1}$ aperture,  the noise for the maximum surface brightness (entry (6))
is per pixel. The signal-to-noise ratio attained
is totally dominated by the sky background with minor contributions from
the detector noise and suppression of cosmic rays.

A histogram of the wavelength distribution, together with a plot of the relative sensitivity
of the observation, is given in fig. \ref{zhist}.

\subsection{Spatial Profiles}

The size of the emitters can be characterized by a contour at which the surface brightness has dropped to a particular level. We define somewhat arbitrarily a projected "size" along the slit  as the distance $y$ from the center of the object along the slit at which the surface brightness of the fit model 
(\ref{eq:model1}) has risen to
$S(y) = 1 \times 10^{-19}$ erg cm$^{-2}$s$^{-1}$$\sq\arcsec^{-1}$ (approximately the standard deviation in surface brightness in a $1\sq\arcsec$ aperture),  going inward.
This is close to the surface brightness level that corresponds to a $3\sigma$ detection
threshold in an opening given by the product of  slitwidth( $2\arcsec$) and FWHM
($1.07\arcsec$), i.e., it would be the approximate detection threshold for an unresolved
source. With "size" we mean, in the following,  half the total extent along the slit, analogous to the "radius" of an object. Note that this is generally an underestimate of the true radius. The two are only identical if the slit were centered on the emitter.

 The distribution of these "sizes" for our 27 objects is given in fig. 
\ref{arcsecradii}.

Most of the sizes occur near the spatial resolution limit (ca. 0.54\arcsec, Half WHM), but there is a considerable tail to much larger radii. Four large objects that clearly extend beyond our
fitting range are collected in the bin at 3.6\arcsec. This bin  comprises the objects 4, 15, 23, and 29. 
The median "radius" is 0.99\arcsec, or 7.7 physical kpc  for a Lyman $\alpha$ emitter at redshift 3.2.

\subsection{Number Densities and Fluxes}

Some of our 27 emitters are unambiguously identifiable (as HI Ly$\alpha$ 1216 \AA ) just 
based on the line profile shapes.  The same is true for the identification of  the additional four, bright double component objects as [OII] 3726,3729 \AA\ emitters. Nevertheless,  the low signal-to-noise ratio and the mostly invisible continuum do not permit us to apriori exclude the identification of the majority
of the sources with either of those two classes of objects,  $2.667 < z < 3.751$ Ly$\alpha$, 
or $ 0.196 < z < 0.550$ [OII]. The  16 reddest of the 27 sources are at least in
the right redshift range ($0 < z < 0.16$) to be also eligible for [OIII] 5007 \AA , as the weaker 
[OIII] transition and H$\beta$ usually would be  too faint to be seen.
Although these possibilities are not equally likely, as we shall argue below, it is instructive
to look at the implied number densities, luminosity functions, and fluxes for the three extreme
interpretations.

For the purposes of this paper we adopt a flat cosmology with $\Omega_m = 0.3$,  $\Omega_{\Lambda} = 0.7$, and $H_0 = 70$ km s$^{-1}$ Mpc$^{-1}$. 

For the following we have not applied  upward corrections for slit-losses to the luminosities, which occur when an object is larger than the slit.
These corrections can be important, but 
are generally uncertain for a population of objects with an apparently large range of sizes 
and profile shapes. For a detailed discussion  of
slit losses we refer the reader to the Appendix. We emphasize here that the luminosities for objects with characteristics encountered in our sample may be underestimated by up to  factors 2-5.  

The redshift range of the spectrum, assuming HI Ly$\alpha$, is $\Delta z = [2.667,3.751] = 1.085$. For [OII] and [OIII] these values are $\Delta z = [0.196,0.550] = 0.354$, and $\Delta z = [0.,0.154] = 0.154$,  respectively.

The solid angle subtended by the slit is 0.252 \sq\arcmin. 

The number of objects per unit redshift per square arcminute for Ly$\alpha$ is given by

\begin{eqnarray}
\frac{\partial^2 {\cal N}}{\partial z \partial \Omega} = 3.66 \times {\cal N}_{Ly\alpha} = 98.7,
\end{eqnarray}
where ${\cal N}_{Ly\alpha}$ = 27 is the total number of objects.
For [OII] this value is

\begin{eqnarray}
\frac{\partial^2 {\cal N}}{\partial z \partial \Omega} = 11.2 \times {\cal N}_{[OII]} = 302.7,
\end{eqnarray}
for 27+4=31 putative [OII] emitters.

For [OIII],
\begin{eqnarray}
\frac{\partial^2 {\cal N}}{\partial z \partial \Omega} = 25.77 \times {\cal N}_{[OIII]} = 412.3,
\end{eqnarray}
with ${\cal N}_{[OIII]} = 16$ objects in the right wavelength range.

For the case of  Ly$\alpha$ 
the (cumulative) distribution $\partial^2 {\cal N}_{> F_{Ly\alpha}}/{\partial z \partial \Omega}$ of those objects exceeding a  line flux $F_{Ly\alpha}$ [erg s$^{-1}$ cm$^{-2}$] is given by fig. \ref{eff}.

The comoving survey volume is 
given by 
\begin{eqnarray}
dV_c = \frac{c}{H_0}(1+z)^2 D_A^2(z) d\Omega \frac{dz}{E(z)},
\end{eqnarray}
with the angular diameter distance $D_A(z)$
\begin{eqnarray}
D_A(z) = \frac{c}{H_0 (1+z)}\int_0^z  \frac{dz'}{E(z')}, 
\end{eqnarray}
and
\begin{eqnarray}
E(z) = \sqrt{\Omega_m (1+z)^3 + \Omega_{\Lambda}}.
\end{eqnarray}

Then the comoving volumes represented by the
two-dimensional spectrum are $V_c$ = 885 Mpc$^3$ h$^{-3}_{70}$ for Ly$\alpha$,  
$V_c$ = 57.5 Mpc$^3$ h$^{-3}_{70}$ for [OII], and  $V_c$ = 1.82 Mpc$^3$ h$^{-3}_{70}$ for [OIII].

The total comoving number density of objects detected, $d{\cal N}/dV_c$, is then
0.030, 0.53, and 14.9 Mpc$^{-3}$ h$^{3}_{70}$, for HI, [OII], and [OIII].

The line luminosity is given by 
\begin{eqnarray}
L = 4\pi D_L^2(z) F,
\end{eqnarray}
with the line flux $F$ and the luminosity distance $D_L = (1+z)^2 D_A$, and $D_A$ as given above.

The histogram of luminosities for HI Ly$\alpha$ is given in fig. \ref{luminoshist}.
Figs. \ref{luminos}, \ref{luminosoii}, and \ref{luminosoiii} show the  cumulative comoving density of objects versus luminosity, again under the assumption that the objects are entirely either HI Ly$\alpha$, [OII], or [OIII] emitters.
All  luminosity functions are calculated from the fluxes measured 
in the larger, $2\arcsec\times 7.6\arcsec \times 1510 $ km s$^{-1}$
aperture, which is related to the fluxes from the smaller $2\arcsec\times 2\arcsec\ $ one  (table \ref{bigtable}, column (4)) by the factors given in column (5).

\subsection{Rate of Incidence per Unit Redshift}

Each population of emitters produces a total 'footprint' in the plane of the sky, that
can be used to calculate the rate of incidence along a line of sight, e.g., in a QSO spectrum, once the number density and cross section on the sky are known.

The contribution of the emitter population to the rate of incidence per unit redshift
$dN/dz$, is given by

\begin{eqnarray}
\frac{dN}{dz} = \sum_i\frac{{\sigma_{i}}}{V_i}  \frac{dl}{dz},
\end{eqnarray}
where 
the sum is over emitters with index i. $V_i$ is the comoving volume in which object i
could have been detected, $\sigma_i$ the comoving spatial cross-section of emitter i, and the comoving distance per unit redshift at  redshift $z_i$ is
\begin{eqnarray}
\frac{dl({z_i})}{dz} = \frac{c}{H_0 E(z_i)}.
\end{eqnarray}

We use the distribution of sizes (fig. \ref{arcsecradii}) to compute the comoving spatial
cross sections $\sigma_i$ and plot the resulting cumulative $dN/dz$ as a function of
object size in  fig. \ref{dndzdist}, where each graph is derived
as if all objects where either entirely HI Ly$\alpha$, or [OII]3727, or [OIII]5007 emitters.
The short vertical lines denote the resolution limit, telling us that
objects with nominal sizes smaller than that may not be contributing 
as much to the cross-section on the sky and thus to the $dN/dz$.
This correction is relatively small for all cases.

\section{The Identity of the Emitters}
\label{whataretheysec}

Most of our objects are technically single line objects, too close to the detection
threshold to study the precise line shape  or detect other weaker lines that should also be 
present (as in the case of [OIII]). Moreover, for most of them 
neither  the spectrum nor a deep Keck LRIS V band image (see below) show significant continua. 

Judging from the emission line profiles alone, we estimate conservatively that at least about six objects are likely to be 
high redshift Lyman $\alpha$ because of their pronounced asymmetric emission profile.
A further three objects seem to coincide closely with QSO Ly$\alpha$ forest absorption systems (see 
below), which makes them relatively secure HI identifications.

Thus, from the spectroscopic evidence discussed so far it is not possible to exclude that the majority of our objects are low redshift contaminants from [OII] or [OIII]/Balmer series emitters. We shall now present
a number of arguments that will help to judge the plausibility of these alternatives.

\subsubsection{Are the emitters dominated by [OII] at low redshift ?}

In addition to the 4 objects already eliminated based on their bright [OII] doublets and continua
(see section 3)
there is  a similarly small number among the 27 remaining objects the line profiles of which seem at least to be consistent with [OII].
We can only barely resolve the [OII] 3736,3738 doublet in our spectra, but we estimate that we
see up to four objects with potential multiple emission peaks (probably including noise spikes)
at the right wavelength separation, that are at least consistent with  [OII] (though none of 
them has to be). Given our rather poor spatial resolution of 4-8 kpc at 
these redshifts we should probably  expect any additional [OII] emitters 
to reside amongst our spatially unresolved  sources.

If our emitters were [OII], the luminosities would range from  $3.7 \times
10^{38}$erg s$^{-1}$  to $2\times  10^{40}$ erg s$^{-1}$ and the redshifts range from
$z=0.196$ to $z=0.550$.  Using a standard calibration for the relation
between star formation rate and [OII] luminosity,
\begin{eqnarray}
\mathrm{SFR(M}_{\odot}\mathrm{yr}^{-1}) =1.4\times 10^{-41} L_{[OII]} (\mathrm{erg\ s}^{-1}) \label{eq:loii}
\end{eqnarray} 
(Kennicutt 1998), 
these luminosities correspond
to star formation rates between  $5\times10^{-3}$ and 0.3 M$_{\odot}$ yr$^{-1}$.
The total inferred [OII]  luminosity density of
$1.0\times10^{41}$ erg s$^{-1}$  Mpc$^{-3}$  would correspond  to
a star formation rate density of $ 1.42 $M$_{\odot}$yr$^{-1}$Mpc$^{-3}$. 

The space density  correponds to  $0.53 (N/31)$ 
Mpc$^{-3}$.  
We can estimate the number of expected [OII] detections from
the field galaxy luminosity function of Trentham, Sampson \& Banerji (2005).
We are able to detect emitters with line fluxes $> 10^{-18}$erg cm$^{-2}$s$^{-1}$. Most [OII] emitters (Hogg et al 1998) do not exceed
a rest equivalent width of 50 \AA . If we use this value to convert
our line flux detection threshold into continuum magnitudes, and adopt the redshift 0.364 that divides
the volume where we can detect [OII] into a lower and higher redshift half,
we find $M=-12.8$ as the faintest continuum magnitude where we would
be able to detect the corresponding [OII] emission.
The total space density of local field dwarf galaxies down 
to an absolute magnitude of $M_{R}=-13$ is about $ 8\times10^{-2}$Mpc$^{-3}$
(Trentham, Sampson \& Banerji 2005).  Based on that estimate,
about five of our emitters are indeed likely to be due to [OII] emission
in low redshift dwarf galaxies. This number fits in well 
with  the four [OII] emission line galaxies which we have already identified, outside 
of the 27 faint emitters. Just from Poissonian arguments (again assuming that we are not looking at a cluster), the probability  to have 2 more additional [OII]
emitters in our volume is about 22 \%, the probability to have more than 4 more is only about 7 \%.

The rate of incidence of low redshift damped
Ly$\alpha$ systems  is another (albeit somewhat uncertain) indication that our emitters are not predominantly
[OII].  If we assume that all star-forming galaxies are embedded in DLA (are a subset of DLAs), and that the radius of optically thick gas is very likely to be larger than the one of [OII] emission, the product of number density
and cross-section of [OII] emitters cannot be larger than  that of DLAs.
Our inferred total dN/dz  (fig. \ref{dndzdist}) for [OII] at 0.93 is more than
14 times larger than that of its contemporary damped Ly$\alpha$ systems, estimated from the Sloan Digital Sky Survey,  $dN/dz(\overline{z}=0.37)$ = 0.066 (Rao, Turnshek \& Nestor 2006), and so
even if DLAS were not larger than [OII] emission regions, and not more numerous, only about 7\%
of our emitters (i.e. two objects in total, or none in the remaining
sample of 27) should  be [OII] to not violate the $dN/dz$ constraint from DLAS.

\subsubsection{Are the emitters dominated by [OIII] at even lower redshift ?}

As for [OIII] 5007\AA , there is a total of 16 emitters in the redder 769 \AA\ long part of the spectrum where  [OIII]
could be detected. The remaining
12 emitters occupy the bluer 550 \AA , where the wavelength is below the rest frame
wavelength of [OIII] and obviously cannot be [OIII].  The ratio between the numbers of objects per wavelength in the [OIII] region versus those in the non-[OIII] region
is then $0.95\pm 0.23$, i.e., there is no significant enhancement of the line density, making a dominant
contribution from [OIII] emitters unlikely (a similar line of reasoning can be employed against
the emitters being [OIII] 4959 \AA\ etc).

The corresponding luminosities would range from  $2.0 \times
10^{36}$ erg s$^{-1}$   to $3.2 \times  10^{38}$ erg s$^{-1}$ at redshifts between
$z=0$ and $z=0.154$.  The space density
corresponds to $9 (N/16)$ Mpc$^{-3}$, about 40 times  higher than the local space density of dwarf galaxies down 
to an absolute magnitude of $M_{R}=-9$ (0.23 Mpc$^{-3}$; Trentham et al 2005). As we have found already one galaxy in the right redshift range
outside of our emitter sample
(though one that had only H$\beta$ and no actual [OIII] emission, see section \ref{obssec}), the Poissonian probability to have one or more additional
ones hidden in our sample is less than 7\%, the probability to have
two or more less than 1\%.
There is no obvious foreground cluster in our field and it seems 
thus very unlikely that even one of these emitters is due to [OIII]
emission from an HII region in low redshift dwarf galaxies.  
Note, however,  that curiously the lower end of the
inferred luminosities at the smallest distances corresponds to 
that of bright planetary nebulae (PNe; 4 objects). 
Gerhard et al. (2005, 2007) have searched for PNe in the core of the Coma
cluster  with a multiple slitlet technique to similar limiting fluxes.
They found  35 PNe candidates in a similarly-sized volume  
but centered on the core of the Coma 
cluster, where the overdensity of PNe should be very large. 
It appears thus unlikely that in a random field we should have found [OIII] emission from 
bright PNe.  
The rate of incidence of DLAs (at mean redshift 0.08) constrains the fraction of 
[OIII] emitters, too, limiting it to be less than 15\% of our emitters. 
\medskip

We conclude that we are likely to have found (and eliminated already) most of the [OII] contaminants in our emitter sample,
as predicted by the space density of the local galaxy population, and the rate of incidence
of low redshift damped Ly$\alpha$ systems.
It is even less likely that there are [OIII] 5007 \AA\ contaminants in our sample.

Therefore, from here on we shall treat the remaining emitters as  HI Ly$\alpha$ and discuss
the implications, keeping in mind that some of the objects may still be misidentified.

\section{Lyman $\alpha$ Emitters at Redshifts $ 2.666 < z < 3.751$ }
\label{lyasec}
If our objects are Ly$\alpha$ emitters the observed fluxes correspond to
luminosities between $7.9 \times 10^{40}$ erg s$^{-1}$ to $1.6\times10^{42}$
 erg s$^{-1}$.

If caused by star formation,
the range of luminosities  corresponds to star 
formation rates  of  $7\times10^{-2}$  to 1.5 M$_{\odot}$ yr$^{-1}$ ,
where we have used the standard relation
\begin{eqnarray} \mathrm{SFR(M}_{\odot}\mathrm{yr}^{-1}) =  9.1 \times 10^{-43} L_{Ly\alpha} (\mathrm{erg\ s}^{-1})\label{eq:llya}
\end{eqnarray} 
for the Ly$\alpha$ luminosity as a function of star formation rate
(based on Kennicutt (1998) and case B assumptions for the conversion of H$\alpha$
and Ly$\alpha$ ; Brocklehurst 1971).
Note again that the actual
values could be  larger by a factor of a few due to slit losses. 
The total Ly$\alpha$ luminosity density of 
$1.4\times10^{40}$ erg s$^{-1}$Mpc$^{-3}$ corresponds to a star formation 
rate density of $1.2\times 10^{-2}$M$_{\odot}$yr$^{-1}$Mpc$^{-3}$, about 36\% of the 
value (uncorrected for dust) inferred for B-band dropouts in the Hubble Ultra Deep Field by Bouwens et al 2007.  

The inferred space density is $3.0 (N/27) 10^{-2}$ Mpc$^{-3}$, a factor
three smaller than the total space density  of local dwarf galaxies, but   
an order of magnitude  larger than the space density of previously known Ly$\alpha$ 
emitters  at this redshift (but with our study going down to much lower flux limits). 
If our detected emission line objects
are  primarily due to Ly$\alpha$ this correponds to a
significant steepening of the luminosity function of  Ly$\alpha$
emitters at luminosities below $\sim 10^{42}$ erg s$^{-1}$. 

Is such a numerous population  of  Ly$\alpha$ emitters plausible? 
The inferred space density is  similar to the space density of 
B droup-outs in the HUDF  at slightly larger redshifts (Bouwens et al. 2007)
(see also below, and fig. \ref{lumfunc}), and in fact, less by factors $10 - 30$
than the number density inferred by Stark et al (2007) for $z\sim 8-10$ objects. 
Intriguingly, the abovementioned survey for planetary nebulae by
Gerhard et al (2005) also found 20 "background objects" that
could not be identified otherwise, in a similar volume. 
Other dedicated surveys for Ly$\alpha$ emitters at $z\sim 2.5 - 3.5$
(e.g., Hu, Cowie \& McMahon 1998; Kudritzki et al 2000; Steidel et al 2000; 
Stiavelli et al 2001; Fujita et al 2003; van Breukelen et al 2005, Gronwall et al 2007, Ouchi et al 2007) appear consistent with our survey (there is little overlap in the range of fluxes reached). Our objects
have about 20 times the volume density of, for example,  the objects found by the shallower
Gronwall et al survey (their detection threshold is $1.5 \times  10^{-17}$ erg cm$^{-2}$ s$^{-1}$). 

A comparison of our cumulative luminosity function (solid line, with $1\sigma$ errors indicated by the dotted lines) with the best fit, z=2.9  luminosity function from the IFU survey by van Breukelen (dash-triple-dotted line), the  z=3.1 (narrow band filter) luminosity function of Ouchi et al 2007, and the predictions by LeDelliou et al (2006, dash-dotted line) is shown in fig \ref{ouchi}. Note again that our luminosity function does not include corrections
for slit losses. At the bright end of our sample  (near $\sim 10^{42}$ erg s$^{-1}$) there is good
agreement with the two observed luminosity functions, and there is initial agreement in
the slope as well in the small region of overlap, but our sample becomes steeper going toward fainter magnitudes. The solid line appears to flatten again toward  luminosities below
$2\times10^{41}$, (or a flux $3\times10^{-18}$ in the usual units).
Objects at half that flux are still clearly detectable for sources with characteristics similar to ours.
This suggests that the flattening may be real, and the numbers may start
to decline.  One possibility is that we may already be seeing the bulk  of the currently star forming galaxies.

From a theoretical point of view the CDM picture of structure
formation predicts  a rather steeply rising mass function
at low masses, but note that even for a linear light-to-total halo-mass  
relation the luminosity function required to 
explain our inferred space density requires an  even steeper
faint end slope. 
Near $10^{42}$erg s$^{-1}$, our observed density of objects is
in agreement with the CDM based model population of  Ly$\alpha$
emitters from  LeDelliou et al. (2005, 2006; dash-dotted line in fig \ref{ouchi}), but then it steepens over the next decade in luminosity considerably, relative to the models based on a constant Ly$\alpha$ escape fraction, perhaps suggesting that this fraction may not be constant after all. The discrepancy approaches about a factor 5 at $2\times10^{41}$erg s$^{-1}$ and then decreases
again toward fainter magnitudes. Such a steepening of the
luminosity function  could perhaps  be explained if dust extinction
becomes increasingly less important for fainter emitters.   

Unfortunately, the stellar and total masses of the emitters are very uncertain. If
the  Ly$\alpha$ emission is due to star formation at the rates
estimated above, the accumulated stellar mass within $10^9$yr  is in 
the range  $7\times10^{7}$ M$_{\odot}$  to   $1.5 \times 10^{9}$M$_{\odot}$.  Another
estimate of the mass can be obtained by comparing  our inferred space 
density with that of dark matter halos  predicted by CDM
models. The space density inferred by our sample of objects  
corresponds to the cumulative space density of dark matter
halos  with total mass $>3\times 10^{10}$ M$_{\odot}$  and circular velocities 
$v_{\rm c}>50$kms$^{-1}$ (e.g. Mo \& White 2002, Wang et al. 2007). 

In the next chapter we shall examine the competing Ly$\alpha$ production mechanisms,
before returning to a discussion of the nature of the emitters, in the larger
scheme of galaxy populations.

\section{Astrophysical Origins of the Ly$\alpha$ Emission}
\label{origsec}

At the faint detection threshold attained here a number of different physical processes can produce Ly$\alpha$
emission at comparable fluxes, and it is not certain that we are necessarily seeing the result
of star formation. The faintest of these competing mechanisms is Ly$\alpha$ fluorescence, induced 
by the general UV background. However, our fluxes (see table \ref{bigtable}) typically exceed the predicted
surface brightness limit for individual objects (Gould \& Weinberg 1996) by an order of magnitude. 
A second source of Ly$\alpha$ photons arises from the presence of a QSO in our field. 
The QSO locally enhances the UV flux and can in principle boost the surface brightness of HI to much higher levels where it can be readily detected. A third effect expected to rear its head at our sensitivity threshold is cooling radiation; 
gas falls into a galactic potential well and sheds part of its potential energy in the
form of Ly$\alpha$ line radiation. These processes are observationally distinct from star formation in that the
latter is the only one that actually produces a significant (stellar) 
continuum as well, which can serve as a discriminant among the various sources of Ly$\alpha$.
This question will be addressed briefly in the next section, followed by an investigation of the
role of the QSO's local radiation field, and of cooling radiation.

\subsection{Stellar Continuum Emission from the Line Emitter Sample}

To check for continuum emission  we could avail ourselves of V band images
taken with the LRIS instrument
on the Keck I telescope, with a total exposure time of 5610 s. The combined image was flux-calibrated with the photometric data
from Hall et al (1996). The slit coordinate system was mapped onto the two-dimensional image and the V band fluxes were measured in  appropriately positioned apertures of size $2\arcsec\times2\arcsec$.  These apertures are expected to have a typical spatial uncertainty on the order of half a slit width perpendicular to the slit, as the spectrum allows
us only to derive the coordinate {\em along} the slit. 

The $1\sigma$ detection threshold in this aperture is $3\times10^{27}$h$_{70}^{-2}$ erg s$^{-1}$Hz$^{-1}$. 
Detected V band luminosities (crosses with error bars) together with
3-$\sigma$ upper limits for undetected objects (arrows) are shown as a function of the Ly$\alpha$ luminosity in fig. \ref{vversuslya}.
Objects 1,2,3 and 39 fell off the edges of the V band image and
where not constrained.  However, object 39 has a detectable continuum in the spectrum itself, and shows a clear Lyman $\alpha$ forest
decrement. Its data-point in the plot gives  the 1500\AA\  continuum flux measured directly from the spectrum. 

As for the other objects, visual inspection shows that very few objects selected by the presence of Ly$\alpha$ emission in the spectrum show up in the
V band image.  At the $3\sigma$ flux level, only two objects have automatically detectable V-band counterparts, namely 9, and 33,  both
of which could be low redshift continuum sources or high redshift line emitters
experiencing chance coincidences with
lower redshift continuum sources.  Interestingly, these are the same two
objects picked out by eye as having clear continuum counterparts. 
In the spectrum itself, several objects coincide with apparent continuum traces (all very faint), many of which, with the exception of the above mentioned object 39, are consistent with bad rows
or charge transfer problems, or accidental spatial coincidence with unrelated continuum sources. 

It is instructive (and sobering) to consider where  in the continuum - line luminosity  diagram (fig.\ref{vversuslya}) star forming galaxies should reside, were we able to detect them
both in continuum and line emission. Adopting again eqn. \ref{eq:llya}
for the Ly$\alpha$ luminosity as a function of star formation rate, and 
\begin{eqnarray}
L_{UV} (\mathrm{erg\ s}^{-1}\mathrm{Hz}^{-1}) = 8\times 10^{27}\  \mathrm{SFR(M}_{\odot}\mathrm{yr}^{-1}) \label{eq:luv}
\end{eqnarray} 
for the UV continuum luminosity (for a Salpeter IMF, and solar metallicity;  Madau, Pozzetti \& Dickinson (1998)),
we equate the star formation rates in these relations to obtain 
the dashed line in the bottom RHS corner of fig. \ref{vversuslya}. It delineates
the positions of galaxies where both UV continuum flux and Ly$\alpha$ line flux are entirely
due to star formation, and is given by

\begin{eqnarray}
\log(L_{UV}) = -14.14 + \log(L_{Ly\alpha}).\label{eq:linecont}
\end{eqnarray}
The Ly$\alpha$ restframe equivalent width formally implied in equation (\ref{eq:linecont})  is 68 \AA , a
value high for color-selected galaxies (Shapley et al 2003) but not exceptional
even for much brighter Ly$\alpha$ emitters (e.g., Gronwall et al 2007).
There is also an upper diagonal dashed line, showing the locus for a rest equivalent width of 20 \AA .
The dotted vertical line to the left gives the Ly$\alpha$ flux for a star formation
rate of 1/10 M$_{\odot}$yr$^{-1}$, the RHS one for   4/10 M$_{\odot}$yr$^{-1}$.

Unfortunately, most of our objects are predicted to be  too faint in the continuum to be able to
test whether star formation is the origin of the Ly$\alpha$, so the continuum flux measurement
or, equivalently, the equivalent widths are not helpful here. The non-detection of the
continuum is of course fully consistent with star-formation induced Ly$\alpha$ emission.

The position of object 39 so far (to the left) of the SF locus suggests
that the Ly$\alpha$ emission is heavily suppressed, e.g., by dust,
as seems to be the case for massively starforming galaxies (e.g., Shapley et al 2003).

The situation is summarized in fig. \ref{lumfunc}, where we compare the {\em cumulative} UV continuum luminosity functions of Steidel et al (1999; asterisks) and 
the Hubble Ultra Deep Field $z\sim 4$ B-band dropouts (Bouwens et al 2007; dashed line) with the
line emitters in our survey (solid line). For all but object 39 (where we
have an actual measurement) we have
assigned "continuum magnitudes" based on eqn. (\ref{eq:linecont}).
Note that this is just a scheme to show the predicted
continuum if both UV continuum and Ly$\alpha$ line radiation would be entirely due
to star formation, ignoring any extinction effects.

Given our small survey volume, only about one object in our entire survey  should be bright enough to have shown up in a
ground-based, broad-band-color survey, i.e., as a "Lyman Break" galaxy (Steidel  et al 1999), and this is what we find (namely number 39).
Object \# 39 brings
up the number of galaxies brighter than -20.3 AB magnitudes to unity, virtually identical to 
the prediction from the integrated continuum luminosity functions for a volume of our size. Our volume is too small to have a much brighter galaxy in it.
The total number density of our emitters is comparable to the number density of the Bouwens et al study
at magnitudes brighter than $M_{AB} = -16.5$. The "luminosity function" for the line emitters appears
to be steepening between -20  and  -18, a behavior
already seen above in our comparison with the Ly$\alpha$ emitters. It  could
be indicative of dust extinction decreasing towards fainter magnitudes. A correction for dust would reduce the slope of the line emitter
luminosity function,  
bringing it into better agreement with the Bouwens et  al curve. 
We caution, however, that our objects cannot be not strictly identical
to the class of B-band dropouts as half of them are at lower  redshift.

\subsection{Ly$\alpha$ fluorescence induced by the QSO ?}

We turn next to the possibility that the Ly$\alpha$ radiation arises from patches of optically thick hydrogen gas, induced to Ly$\alpha$ fluorescence by the ionizing radiation from  
the QSO in our field.
The basic idea is that the partial conversion of the QSO UV radiation field into Ly$\alpha$ photons at the surface of optically thick hydrogen bodies raises the emission from clouds or galaxies in the
QSO vicinity above the detection level.  This effect has been studied by a number of authors (Fynbo et al. 1999, Francis \& Bland-Hawthorn 2004, Francis \& McDonnell 2006,
Adelberger et al 2006, Cantalupo et al 2007).  The spatial extent of the zone of influence is obviously inversely proportional
to the square root of the intensity of the fluorescent emission. Following Cantalupo et al 2005 we can express the enhanced
Ly$\alpha$ flux in terms of a boost factor, i.e., the ratio of the surface brightness enhanced by the QSO ($S$) to the one caused
by the general UV background ($S_{bg}$),
\begin{eqnarray}
\frac{S}{S_{bg}} = \left(0.74 + 0.5 b^{0.89}\right),
\end{eqnarray}
where
\begin{eqnarray}
b=15.2 \times \frac{ L_{LL}}{10^{30}{\mathrm erg} {\mathrm s}^{-1} {\mathrm Hz}^{-1}}  \frac{0.7}{\alpha} \left(\frac{r}{{\mathrm phys. Mpc}}\right)^{-2}.
\end{eqnarray}
Both the luminosity of the QSO at the Lyman limit, $L_{LL}$,  as well as its precise systemic redshift are critically important 
ingredients in this calculation. Estimating the latter from the position of the OI $\lambda 1302$ \AA\  emission line 
we determine the QSO systemic redshift as $z_{em}$ = 3.32209. 
The luminosity per unit wavelength is given by
\begin{eqnarray}
L(\lambda/(1+z_{em})) = 4\pi (1+z_{em}) D_L^2(z) f(\lambda),
\end{eqnarray}

With $f(1040\AA )$ = $2.2\times 10^{-17}$erg cm$^{-2}$ s$^{-1}$\AA$^{-1}$ measured directly from our fluxed QSO spectrum
we arrive at a luminosity per unit wavelength
$L(1050\AA )$ = $9.405 \times 10^{42}$erg s$^{-1}$\AA$^{-1}$. 
To measure the number of HI ionizing photons we still need to determine the luminosity at the Lyman limit and the power law dependence for wavelengths below the ionization threshold.
According to the study by Scott et al (2004), the power law index for a QSO with
$\log\lambda L(1050\AA )\approx 9.8 \times 10^{45}$ erg s$^{-1}$ where
\begin{eqnarray}
L(\nu)= L(\nu_0)\left(\frac{\nu}{\nu_0}\right)^{\alpha}
\end{eqnarray}
is (statistically) consistent with $\alpha\sim -1.5$ ( in good agreement with $\alpha\sim -1.57$ for the radio-quiet sample
from Telfer et al 2002).

Extrapolating the luminosity from 1050 \AA\ to the Lyman limit with 

\begin{eqnarray}
L(\lambda)= \frac{c}{\lambda_0^2}L(\nu_0) \left(\frac{\lambda_0}{\lambda}\right)^{\alpha+2}
\end{eqnarray}
we get

\begin{eqnarray}
L(\nu)=  2.798\times 10^{30} \left(\frac{\nu}{\nu_{LL}}\right)^{-1.5} {\mathrm \ \ erg\ s}^{-1}{\mathrm Hz}^{-1}.
\end{eqnarray}

Inserting these results in the above relation for the boost factor gives
\begin{eqnarray}
r = \frac{3.018}{\left(S/S_{bg} - 0.75\right)^{0.5618}} {\ \ {\mathrm phys. Mpc}}.
\end{eqnarray}

For the relatively faint QSO DMS2139.0-0405 to boost the surface flux from the background value $S_{bg} = 3.67 \times 10^{-20}$ erg cm$^{-2}$s$^{-1}$$\sq\arcsec^{-1}$
to a typical  surface brightness of $S \geq 10^{-18}$ erg cm$^{-2}$s$^{-1}$$\sq\arcsec^{-1}$  as observed would require the object to be within 
only 0.479 proper Mpc  or 2.07 comoving Mpc. This distance corresponds radially to 254.8 pixels spatial pixels along the slit
(about 1/4 of the length of the field), but only 4.3 (!) pixels in the dispersion direction. Fig. \ref{ellipse} shows the highly excentric elliptical contour within which to expect the enhancement to  $10^{-18}$ erg cm$^{-2}$s$^{-1}$$\sq\arcsec^{-1}$.
Only one object, ID 16, falls within the ellipse, and with its maximum surface brightness of $\sim 10^{-18}$ and absence of a continuum
is consistent with fluorescing in the ionizing field of the QSO. The overwhelming majority of our sources, however, appear to be 
oblivious to the QSO's proximity.

\subsection{Signs of Radiative Transfer, and Cooling Radiation}

The spectral line shapes and sizes of our emitters suggest that the Lyman $\alpha$ photons may have been processed by radiative transfer through an optically thick HI medium.
The trapping by and protracted escape of line radiation from such a medium should lead to random walk in the spatial and frequency domain. The result 
may be observable as emission broadened in frequency space and extended in the spatial direction beyond the extent of the actual source of Ly$\alpha$ photons (e.g., Adams 1972; Neufeld 1990; Zheng \& Miralda-Escud\'e 2002;  Dijkstra et al 2006a; Tasitsiomi 2006).
The data appear to show  some evidence for these mechanisms at work. The large velocity widths (see table \ref{bigtable}) and radial extent (median projected radius along the slit 7.7 kpc proper, and considerably larger in individual cases) that we have observed are thus suggestive of  the signatures of radiative
transfer.
The FWHM velocity widths measured from optimally extracted spectra of the individual emission line regions are plotted in fig. \ref{alphavsvfwhm} versus the power law index of the
surface brightness model (equation (\ref{eq:model1})). In some cases these widths are underestimates because only a single peak was fitted, as opposed to a double humped or more complex structure. 
In that plot, the area to the left of the spectral resolution, about 286 km s$^{-1}$ FWHM, is visible as a zone of avoidance, and between a third and half of the measured velocity widths clearly exceed the resolution. Sources with large velocity widths seem to prefer smaller power law indices,
i.e., spatial surface brightness profile that drop less rapidly with radius.

\subsubsection{Spatial Surface Brightness Profiles and Fluxes}

To learn more about the topography of the Ly$\alpha$ source and the origin of the radiation we can attempt to compare our average measurements of the sizes, peak surface
brightness,  and total 
fluxes to the models by Dijkstra et al (2006).
With the number of free parameters and the simplifications in these models and the observational complication of the long-slit technique it is difficult to make a quantitative comparison,
but we can at least check whether the observables agree at an order of magnitude level.
As far as we can tell given the limited spatial resolution, our typical surface brightness profile requires that the
sources are at least somewhat centrally concentrated, similar to the  model 4 of Dijkstra et al (see the surface brightness profile in their fig. 5). Our median observed "radius" (= half the extent along the slit) at
the $10^{-19}$erg cm$^{-2}$ s$^{-1}$ $\sq\arcsec^{-1}$ surface brightness contour is about $1.0\arcsec$, the median power law slope $\alpha_{med}=-2.0$.  
Making appropriate corrections for the slit losses and distortions of the surface
brightness profile we find good agreement with the surface brightness profile
shape of Dijkstra et al if we scale down their total flux to $1.1\times10^{-17}$erg cm$^{-2}$ s$^{-1}$.
 The mass dependence of the flux for their model 4 at z=3.2 is $4.16 \times 10^{-18} (M_{tot}/10^{11} M_{\odot})^{5/3} $erg cm$^{-2}$ s$^{-1}$. Our corrected median flux $1.1\times10^{-17}$erg cm$^{-2}$ s$^{-1}$ would then correspond to 
a cooling halo with total mass $\sim 1.8\times10^{11}  M_{\odot}$.

Thus, the median surface brightness profiles and total fluxes observed appear broadly consistent with the Ly$\alpha$ arising predominantly as cooling radiation. 
The typical halo mass required to produce the luminosity function of our emitters, is, however,  
uncomfortably large for cooling radiation to be the dominant source of Ly$\alpha$ for a majority
of our objects.
Dijkstra et al. (2006)  estimate the expected cooling radiation
assuming that the gas in DM halos cools on a free-fall time scale and 
cooling is predominantly by  Ly$\alpha$ emission. The Ly$\alpha$
emission is then a strong function of the virial velocity of the halo,  $L_{Ly\alpha} 
\sim 1.6 \times 10^{39} (v_{c}/35$ kms$^{-1}$)$^5$ erg s$^{-1}$.   Note that  the free  
fall time scale is  shorter than the time that
corresponds to the redshift interval  $2.667<z< =3.751$, and  a newly
collapsed DM halo would only emit for about 40\% of the redshift range 
where we can observe it.  Even if we assume that all DM
halos present at the lower end of the redshift interval
have  collapsed and started cooling in our redshift interval, all
halos  with $v_{\rm c}>35$ kms$^{-1}$ would be necessary to account for the
observed space density of emitters.  The  Ly$\alpha$  luminosity for the typical object
would generally be more than a factor ten lower than we observe even if we neglect
the slit losses. This does not preclude that the flux of a few of our emitters in more massive halos is
dominated by Ly$\alpha$ cooling radiation  but it is very unlikely that
this is a large number. Ly$\alpha$ radiation powered by star
formation appears to be the energetically most favorable explanation for the majority of our emitters.

\subsubsection{Evidence for Radiative Transfer Mechanisms from Spectral Line Profiles}

Irrespective of the origin of Ly$\alpha$ photons, radiative transfer of line photons from
a central source within an optical thick halo should have observational
signatures characteristic of the kinematics of the gas.

 Several of our objects (ID \# 3, 12, 21, 23, 28, 29, and 39; see figs. \ref{allpanel} and \ref{indivpanel})  exhibit strong, spatially concentrated emission peaks (even though their emission often extends further out) with 
asymmetric  line profiles showing a steep drop in the blue and an extended red shoulder; such profiles have been seen previously in low- and high redshift star-forming galaxies
and are generally  considered to be consistent with radiative transfer in the expanding supershells of galactic winds (e.g., Lequeux et al 1995; Mas-Hesse et al 2003)   
At various stages of their evolution the line profiles may resemble single emission line peaks, PCygni profiles, or double component profiles with a dominant red component (Tenorio-Tagle et al 1999; Ahn et al 2003, Ahn 2004). 

Several of these asymmetric emitters (3, 23, 28, and 29) show a weaker blue peak opposing the red one, which may be evidence for a wind shell or more generally
radiative transfer through an expanding optically thick medium (Zheng \& Miralda-Escud\'e 2002; Dijkstra et al 2006a,b; Tasitsiomi 2006).

\subsubsection{Individual Candidates for Emitters dominated  by Cooling Radiation}

Most other  emission profiles look amorphous and defy classification because of the low signal-to-noise level,
but there is a small group  of emitters (IDs 15, 36 and 37; fig.  \ref{smudgegroup}) fortuitously projected near the QSO trace, that shows a number of intriguing
properties different from those of the other sources. 
The three continuum traces visible in that figure are the QSO (with four strong Ly$\alpha$ forest absorption lines of rest equivalent widths 2.0 \AA\ (A), 1.3 \AA\ (B), 1.2 \AA\ (C), and 0.9 \AA\ (D); an unrelated, featureless, presumably low redshift continuum object just above the QSO trace; and further up a faint high redshift object from which
two faint emission smudges (36 and 37) seem to protrude. 
Even though their appearance seems unusual, the identification of all three smudges with 
HI Ly$\alpha$ is relatively secure because of their close alignment in redshift with QSO absorption systems:
the dip in the emission region of object 15 and the blue starting point of the emission region of object 37
both coincide to within less than 100 kms$^{-1}$ with the strong absorption line C in the QSO spectrum between them. The projected transverse (here: vertical)  distances from the QSO are 150 (object 37) and 90 (object 15) physical kpc.
The object 36, at a similar transverse distance from the QSO as 37, also coincides closely  in redshift with
another QSO absorption system (B).

Object 15, below the QSO trace,  appears to consist of  a strong blue emission component, separated by a dip in flux from a fuzzy redder emission bit, which also may be rotated slightly. Both, the velocity shift between the
blue emission line and the dip, and the FWHM of the blue peak each amount to approximately 200 kms$^{-1}$. 
It is difficult to be sure of what we are seeing here (perhaps two merging protogalactic clumps), but the
signature of strong blue peak, central dip, and weak red peak is not unlike the one expected for cooling radiation
from gas falling into a galactic halo. If this is what we are are seeing, then,
according to the simulations by Dijkstra et al (their fig. 8) such relatively small values for blueshift and FWHM
of the blue peak may indicate a cooling halo with relatively small infall velocities and HI optical depths. 

Object 37 (cf. figs.  \ref{allpanel} and \ref{indivpanel})
shows a bar-like emission region projecting out at about a $60\deg$ angle on one side from the blue edge of a strong absorption line in the nearby, faint background galaxy,
as if a "door" had been opened anti-clockwise in the continuum of that galaxy. The transverse extent of the
emission is at least about 3.3\arcsec or 26 physical kpc. The spectral width of the tilted emission bar is about 320 kms$^{-1}$
(i.e., it is possibly unresolved), and it projects out from the continuum object, starting about 490kms$^{-1}$ blueward of the centroid of the absorption line (rest EW=3.9 \AA ) in the faint continuum object, shifting to the red with increasing distance from the continuum
object by between 280 and 470 km s$^{-1}$ (the uncertainty arises from the difficulty of 
estimating the spatial extent of the emission region).

Object 36 is another broad smudge of emission,  losely (the S/N is poor) lining up in redshift with QSO absorption system B (a weak absorption feature appears in the continuum object, about 110 kms$^{-1}$  blueward of the
QSO absorber). Again, going outward from the continuum object the emission can be traced spatially to a similar extent as object 37, but in this case extends over a larger wavelength range, becoming redder by up to 1500 kms$^{-1}$. 
The feature is clearly resolved in velocity, with a width of about 1000 kms$^{-1}$.

All three objects show emission blueward of the absorption centroids of either the QSO absorption lines or the
absorption in the continuum emitter. 
In addition, the color  gradient from red to blue when approaching  the absorption systems could be
understood in terms of infalling halo gas, that accelerates and cools when approaching smaller radii, 
as described by Dijkstra et al. However, it is not clear that the absorption systems really represent the
centroids of the halos and do not rather arise in the outskirts. A scenario with out- rather than inflows
and a different topology cannot be excluded, at least not for objects 36 and 37.

There are two more QSO absorbers in that group
that span a total (from A to D) of  52.75 h$^{-1}$ Mpc (or 20 physical Mpc), if the redshift difference is due to the Hubble flow. The fact, that the three most unusual
emitters, together with a cluster of strong QSO absorption lines occur in a spatially relatively narrowly but apparently highly elongated region (even the line of sight distance between B and C is
7 physical Mpc (or 18.43 h$^{-1}$ comoving Mpc) suggests that we may be looking along a large scale filament or sheet, with the QSO absorbers representing the outskirts
of the three galaxies whose emission regions we see.

\section{Correspondence between the Line Emitters and Optically Thick Ly$\alpha$ Forest Absorption Systems}
\label{dlassec}

The existence of two independently identified classes of optical thick objects in the universe, Ly$\alpha$ emitters
and Lymit Limit absorption systems, enables us to establish a correspondence between them and constrain
their properties. 

If we assume that the emitting and absorbing regions are identical in size (in reality,
the absorption cross section may be an upper limit to the emission cross section, for optically
thick gas), we can equate the rate of incidence per unit redshift, $dN_{LL}/dz$, of Lyman limit absorbers above a certain HI column density,
and the product of the spatial comoving density of Ly$\alpha$ emitters, their emission cross-section, and the redshift path, 
\begin{eqnarray}
\frac{dN_{LL}}{dz} = \sum_i\frac{\sigma_{i}}{V_i}  \frac{dl}{dz},
\end{eqnarray}

where 
\begin{eqnarray}
\frac{dl({z_i})}{dz} = \frac{c}{H_0 E(z_i)}.
\end{eqnarray}

The comoving density of objects is again $dN/dV_c = {\cal N}_{obj}/(885 h^{-3}_{70} Mpc$). 
From fig. \ref{dndzdistcor}, our  total observed rate of incidence of Ly$\alpha$ emitters would be $dN/dz$ = 0.30, if all 27 objects were Ly$\alpha$ emitters. 

They same cautions as mentioned above about slit losses apply to the estimate
of the radius from the size along the slit. If the emitters had a spherical, sharp-edged outline in the plane of the sky, then by adopting the total extent along the slit
for the diameter of the object we would underestimate the latter by a factor $\pi/4$.
The finite sizes of the objects also affect the detectability on the slit,
as an extended object may be detected even if its center falls outside the slit.
This increases the effective comoving volume and decreases the space density
for the objects. We use a simple model to compute the comoving 
volume, where an object of a given radius
can be detected if part of it fills the slit.
Our correction can only be indicative of 
the true corrections. The little we know about the emitters at present does not
warrant a more detailed approach. 

The resulting $dN/dz$, corrected for these effects,  is shown in fig. \ref{dndzdistcor} as the dotted line. 
The correction emphasizes the relative contributions to $dN/dz$ from objects smaller than the slitwidth and
reduces the relative contribution from larger objects.  The total correction (larger cross-section and smaller comoving volume) reduces the overall $dN/dz$ by about 22\%.
to $dN/dz = 0.23$. About half of the contribution to $dN/dz$ arises from the four most extended objects.

Interestingly, the observed $dN/dz$ for our emitters and the one for DLAS
with neutral hydrogen column densities $N_{HI} > 2\times10^{20}$cm$^{-2}$ (
 $dN/dz_{DLA}$ = 0.26; Peroux et al 2005; Storrie-Lombardi \& Wolfe 2000)
{\em are comparable}. 

In other words,  the combination of large sizes {\em and} the 
high space density of the emitters together imply that the total
rate of incidence is sufficient to explain the majority  of damped Ly$\alpha$ systems. 
It thus appears that we may have
finally  detected  the star formation associated with most of the rate of incidence of damped Ly$\alpha$
absorption systems in the redshift range $2.667 < z < 3.751$. 
The low star  formation rates  of  0.07  to 1.5 M$_{\odot}$ yr$^{-1}$ 
would explain the low success rate of direct searches for 
the host galaxies of DLAS.  If the interpretation of the emission as 
being due to the hosts of DLAS is correct, then we have  for the first 
time established the typical size and space density of DLA  host
galaxies. Within the CDM model of  structure formation we can then also infer their
masses  ($3\times 10^{10}$ M$_{\odot}$ total, $5\times 10^{9}$ M$_{\odot}$ in baryons) 
and virial velocity scale ($\sim 50$ km s$^{-1}$). 
The typical values for size, mass and virial velocity 
agree well with the  predictions of the model of 
Haehnelt, Steinmetz \& Rauch (1998, 2000), who interpreted
 the observed kinematic properties 
of  the neutral gas in DLAS as probed by low-ionization species
within the context of CDM models. 
The inferred star formation rates 
and the inferred masses are consistent with the
low observed metallicities of DLAS which are generally  overpredicted 
in models assuming larger star formation rates.  Note that
if all the observed emission is due to Ly$\alpha$, 
the total  star formation rate density corresponds already to
$36 (f_{\rm slit}/1.0)$ percent of the total non-dust-corrected star 
formation rate inferred from drop-out studies at these redshifts (e.g. Bouwen et
al. 2006; $f_{\rm slit}$ is the factor by which the observed flux needs to be multiplied
to correct for slit-losses).
Unfortunately, we have no direct observational handle to decide
whether the star formation itself is  extended. If all the emission is
Ly$\alpha$,  our total star formation rate density is, however, higher 
by at least an order of magnitude than the upper  limits
obtained by Wolfe and  Chen who searched for  {\em extended} continuum
emission from DLAS in the HUDF. 
At the same time, our star formation density is close to (about 60\% of)  the 
value needed to explain the heating of DLAS (Wolfe, Gawiser \& Prochaska (2003)).  As suggested by Wolfe et al
these discrepancies can be reconciled if
the star
formation in question is confined to a {\em compact} region at the centre of
DLA hosts, rather than arising in large stellar disks. 
The Ly$\alpha$ emission  in our objects often
appears extended, but this does not necessarily mean
that the sources of the ionizing photons responsible for producing 
the Ly$\alpha$ photons are similarly large.
The extended  nature of the Ly$\alpha$ emission would be  
due to resonant line scattering, with
Ly$\alpha$ photons random walking their way out to radii that have never
seen a star. 
Alternatively, some of the large sizes could be  due to unresolved
merging protogalactic clumps, a situation that is 
common in a CDM scenario and may explain the observed kinematics of DLAS (Haehnelt et al. 1998).
 Finally, the identification of the emitters with DLAS,
which are known to be essentially dust-free (e.g. Murphy \& Liske 2004),
and the realization that Ly$\alpha$ from emitters brighter by an order of magnitude 
have Ly$\alpha$ emission reduced by a factor 3, presumably by dust (e.g., Gronwall et al 2007),
would also explain at least in part the steep rise of the number of emitters over a decade
in surface brightness as a drop in dust content.

\section{The Hogan-Weymann Effect}
\label{fluores}

Sofar we have not addressed the  effect of Ly$\alpha$ fluorescence, induced by the
general UV background (Hogan \& Weymann 1987). The original expectation was that the number of objects
lit up by the UV background corresponding to optically thick LLS would yield about the same number of objects (30, with a radius of 2.5\arcsec\ at column density
$10^{19}$ cm $^{-2}$) as we have found,
albeit at considerably lower surface brightness. A possible explanation is 
that many LLS may have ongoing local, low level star formation.
This would be consistent with  LLS having somewhat higher metallicities (Steidel 1990) than the general Ly$\alpha$ forest (Simcoe et al 2004). 
Given our surface brightness
profiles it is possible that the underlying fluorescence is simply swamped by star formation Ly$\alpha$. 

There are several conceivable approaches to searching for low light level emitters in the field. Originally we had considered the possibility of a blind
search (e.g. Bunker et al 1998), which, however would have reached only a sensitivity a factor two above
the recently revised (lower!) estimates for the anticipated signal, based on the opacity
of the Ly$\alpha$ forest (Bolton et al 2005). One possibility to increase the sensitivity, suggested by the number of objects
already detected is to search in their immediate neighborhood for an extended
signal of diffuse emission surrounding the brighter star forming regions that is
in agreement with the expected surface brightness (Gould \& Weinberg 1996)
\begin{eqnarray}
9\times 10^{-20}\left(\frac{\eta}{0.5}\right) \left(\frac{J}{4.3\times10^{-22}}\right) {\mathrm erg \ cm}^{-2} s^{-1}\sq\arcsec^{-1},\label{eq:jay}
\end{eqnarray}
at $<z>=3.2$ for a QSO dominated, Haardt \& Madau 1996 UV spectrum with slope 1.73; $\eta$ is the 
fraction of the energy of the impinging UV photons converted into Ly$\alpha$,
and $J$ is the UV background. The value for $J$ in equation (\ref{eq:jay})  is scaled
to the recent measurement  of the photoionization rate at $z\sim 3$ in the Ly$\alpha$ forest
(Bolton et al 2005). An ionizing spectrum in which galaxies and QSOs each produce half
of the ionizing flux would result in a surface brightness lower by a 
factor 2 than the fiducial value in eqn. (\ref{eq:jay}).

Our observation is deep enough to probe the surface brightness for general fluorescence if we combine the signal from all sources to improve the signal-to-noise ratio. Boxes of spatial width $2\arcsec$ and spectral length 1500 km s$^{-1}$ (to include  
the likely extent of the double humped emission profiles predicted; e.g., Cantalupo et al 2005)  have been  placed on either side of our emitters, at varying distances
along the slit direction.
The boxes were sky-subtracted once more locally, using windows  below and
above (along the slit) the box used for the signal, but further out from the
emitter than the signal boxes. Because of the strong presence of 
weak continuum objects in the 2-D spectrum only a subset of the emitters
(usually 12-14) were in a sufficiently clean area of the field to be
useful for this analysis.
The mean and median surface brightnesses and their statistical errors were
extracted. The results are presented in fig. \ref{hog_wey}. The open squares show 
the medians, and the crosses with error bars give the total weighted mean surface brightness 
of the boxes used, in units of $10^{-19}$ erg cm$^{-2}$ s$^{-1}$\sq\arcsec$^{-1}$, as a function of angular distance in arcsecs along the slit. 
We caution that the error bars shown are merely statistical noise errors
and do not include the  fluctuations of the
sky level reflected in the difficulty of finding a "clean" patch of sky to place the background subtraction windows on.
The combined profile is not very meaningful in the innermost 2\arcsec\ because
of the wide variation in amplitudes, but if there were a universal fluorescent
glow we would expect the outskirts of the objects to take the appearance
of  annuli of uniform surface brightness. In practice, such a plateau
should be washed out by the distribution in sizes of the emitters.

There is a hint of a flattening  between 3 and 4\arcsec, but the signal continues to dive beyond 5\arcsec . However,  at a level of $\sim 2\times 10^{-19}$erg cm$^{-2}$ s$^{-1}$\sq\arcsec$^{-1}$, the surface brightness is rather higher by a factor of $2 - 4$ than expected 
for fluorescent emission with the favored range of the UV background intensity. Close inspection of the frame shows that this signal appears due
to some genuinely very extended individual objects, i.e., these
are not artifacts of the seeing. This is consistent with the large extent
of the $1.5\times10^{-19}$erg cm$^{-2}$ s$^{-1}$\sq\arcsec$^{-1}$ surface brightness contour, corresponding to a flux density $1.5\times 10^{-20}$erg cm$^{-2}$ s$^{-1} \AA\ $ seen in fig. \ref{allpanel}.

It is intriguing that the corresponding physical radius is 30 kpc, about four times
larger than we had estimated based on the individually modelled
surface brightness profiles above. If this were identical to the radius out to which  all objects have optically thick HI, our sample of 27 emitters would project a dN/dz $\sim 1.4$, and   would 
correspond to all LLS with column densities larger than about $3\times 10^{18}$ cm$^{-2}$, where the fraction of Ly$\alpha$ photons per ionizing photon is just flattening off to attain the maximum conversion rate (Gould \& Weinberg 1996). We will defer a more detailed analysis to future study, and conclude
that the large lateral median extent of our emitters is fully consistent with
them being surrounded by optically thick, Lyman limit zones that radiate
to within a factor two at the level predicted by Gould \& Weinberg (1996), updated
by latest estimates for the photoionization rate. 

\section{Conclusions}
\label{concsec}

Our longslit search for Ly$\alpha$ fluorescence from the intergalactic medium, taking
advantage of moderate seeing periods with FORS2 at the VLT,
has yielded a sample of 27 faint emitters with line fluxes of a few $\times 10^{-18}$ erg s$^{-1}$cm$^{-2}$ over a redshift range $2.66 < z< 3.75$. 

At least a third of the sample shows emission line profiles or an association
with absorption systems in the nearby QSO, strongly suggesting identification with Ly$\alpha$.
Spectroscopic features and the absence of detected continua down to $3-\sigma$ flux limits of $\sim 1.5 \times 10^{-19}$ erg s$^{-1}$cm$^{-2}$ make a direct identification of the other emitters (as HI Ly$\alpha$, [OII] doublet, or [OIII]/HI Balmer emission lines) difficult, but comparison with known galaxy populations
and other statistical arguments indicate that the  majority of emitters is likely to be
Ly$\alpha$ at mean redshift 3.2. 

If this identification is correct, the emitters present a steeply rising luminosity function
with a total number density more than 20 times larger than the comoving density of Lyman break galaxies
($M_R < 25.5$) at comparable redshifts. About half of the profiles are extended, possibly owing to
radiative transfer of Ly$\alpha$ photons from a central source, and there are candidates for both
outflows and infall features. We have investigated several mechanisms for the Ly$\alpha$ production
and find star formation to be the energetically most viable process, with a few objects being
candidates for cooling radiation. 

The inferred low star formation rates, large line emission cross-sections, high number density, and a fitting
total cross-section per unit redshift on the sky seem to provide an excellent match to the low luminosities, low metallicities, low dust content, and rate of incidence of damped Ly$\alpha$ systems,
the main reservoir of neutral gas at high redshift. 
This suggests that our objects are the long-sought counterparts of DLAS in emission. The properties
of the objects paint the DLAS host galaxies as a  population of low mass proto-galactic clumps
as suggested by some of us (Haehnelt et al 1998, 2000; Rauch et al 1997b), disfavoring a model
dominated by large disk galaxies (Prochaska \& Wolfe 1997).
Recent, apparently contradictory limits on spatially extended star-formation 
from DLAS (Wolfe \&  Chen 2007) and on the heating of DLAS (Wolfe, Gawiser \& Prochaska 2003) are consistent with the amount of star-formation we measure if it is confined to a small
unresolved region, irrespective of the fact that both the absorption and the Ly$\alpha$ emission
cross-sections appear much larger, the latter because of the random walk of photons to the edge
of optically thick gas. 
The physical origin and the nature of the extended optically thick gas is uncertain, but it is intriguing that quite a few objects show Ly$\alpha$ emission line profiles consistent with galactic outflows. Thus the absorption cross section of DLAS could be enhanced by  winds, 
a possibility raised by Nulsen, Barcons \& Fabian (1998) and discussed in the context of Lyman break galaxies by Schaye (2001).

Finally, adding up the surface brightness profiles of all
objects in the outer 2-6\arcsec\ we detect radiation at the  $> 2\times 10^{-19}$ erg cm$^{-2}$ s$^{-1}$\sq\arcsec$^{-1}$ level out to 4\arcsec . 
The light level is higher by a factor $2 - 4$ than the Ly$\alpha$ fluorescence signal expected 
for a UV background intensity consistent with current estimates for
the photoionization rate of the intergalactic medium.
The large sizes
can be explained if radiative transfer of Ly$\alpha$ lights up the outskirts of our objects out to a radius where the conversion
from UV to Ly$\alpha$ starts becoming inefficient. A radius of 4\arcsec\ combined
with the observed comoving density of our sample can explain the rate of
incidence of DLAS and LLS with column densities as low as $3\times10^{18}$cm$^{-2}$,
consistent with the possibility  that many  LLS  arise in the outskirts of DLAS which
in turn are to be identified with the faint emitters. With our interpretation
the gas in DLAS is the reservoir from which typical $L_*$ galaxies must have
formed in a CDM based universe.

Some of our conclusions here are speculative. Further study is highly desirable,
but in any case it should be obvious that long spectroscopic exposures are a 
promising way of discovering low-mass galaxies. Performing single field, blank sky searches for
Ly$\alpha$ emitters for longish amounts of observing time (but not longer than regularly
used by  radio astronomers and now even in space based UV, optical and X-ray astronomy)
can bring a whole new range of astrophysical phenomena within the range of existing ground based, optical telescopes, and obviously, could provide one of the
most exciting science projects for a future generation of ultra-large telescopes.
\medskip

\acknowledgments

We would like to thank the following people for helpful advice and discussions during the 
course of this project: James Bolton, Scott Burles, Bob Carswell, Hsiao-Wen Chen, Sandro D'Odorico, Johan Fynbo, Rob Kennicutt,  Juna Kollmeier, 
Rob Simcoe, and Neil Trentham. 
We thank the Director General of ESO, Catherine Cesarsky, for keeping this project afloat, and we
are grateful to 
the observatory staff at ESO and Gemini for performing the observations, and to the time assignment committees of both observatories.
MR thanks the IoA for hospitality in spring and summer 2007.
GDB and MR were supported by the National Science Foundation through grant AST 05-06845. This work was partially supported by the EC RTN network "The Physics of the
Intergalactic Medium".

\clearpage

\appendix

\section{Slit Losses and Selection Effects}
The determination of total fluxes and surface brightnesses from long slit spectra
is generally not possible for individual objects, the problem  being the
unkown spatial shape, and overall extent, and the random position  of the slit center
relative to the position of the underlying object in the plane of the sky. The spatial variation of the surface brightness
along the slit is bound to generally be different from the intrinsic surface brightness dependence on radius.
However, performing simulations with a given radial surface brightness distribution, a size, and radial symmetry,  one can
obtain an impression of the {\em average} slit losses and of the relation between the actual radial 
surface brightness dependence and the typical distribution of surface brightness as a function of position along the slit.
The results of Monte Carlo simulations of a 2" wide, long slit, randomly positioned over a set of disks
with a given size and surface brightness dependence, are shown  in figures \ref{haloslit1} and \ref{haloslit2}.
The model distribution assumes the surface brightness to have a Gaussian core with a width to match the measured seeing,
replaced by a power law at larger radii:
\begin{eqnarray}
S(r)=\propto \exp (- r^2 / (2\sigma^2), {\ \ \ \ \ \ \mathrm for\ \ \ } r < r_{turn},
\end{eqnarray}
and 
\begin{eqnarray}
S(r)= \propto r^{\alpha}, {\mathrm \ \ \ \ \ \ for\ \ \ } r > r_{turn}.
\end{eqnarray}
For simplicity, the transition  between the Gaussian and the power law at the turnover radius, $r_{turn}$, was chosen such that
the amplitude and slope of the two regimes were continuous at $r_{turn}$, corresponding to $r_{turn} = \sqrt{-\alpha \sigma}$.

Fig. \ref{haloslit1} shows a number of radial  model surface brightness distributions (smooth curves) and the predicted average observed distributions (binned curves)  along a 
2" wide long slit, derived from a Monte Carlo simulations with 100 realizations (= random slit positions) per model.  For the input distribution, the abscissa in all panels is the radius from the center of the emitter; for the
output predicted surface brightness it is the spatial coordinate along the slit  (both in units of 0.252" wide pixels). Note that for non-trivial functional forms these distributions should not agree unless the slit is infinitely narrow and runs precisely radially with respect to the underlying emitter.

The left column of panels shows the form of these distributions for three different external power law slopes $\alpha$ = -0.5 (top), -1.5 (middle), and -3.0 (bottom), representing profiles increasingly dominated by a central peak. The right column of panels shows the ratios between the same input radial surface brightness distributions and the predicted "along-the-slit" distributions.
The multiple curves in each panel represent emitters with different overall radial extent, going from R = 5 to 30 pixels in steps of 5. With a pixel scale
of 0.252", the largest emitter model considered would then be $30\times 0.252" \approx 7.5"$ in radius. The dotted vertical lines in the LHS panels
show the location of the transition radius between central Gaussian and external power law.  The RHS panels show the {\em ratios} between input radial and output slit distribution.
From the RHS panels it is clear that the observed surface brightness along the slit  is significantly distorted from the actual radial distribution.
In particular, the central peak and the outer edges are suppressed in the observed profile, in both cases because they do not subtend a large area and are
difficult to hit by a randomly positioned slit. Besides, the overall surface brightness is depressed typically by a factor 2 or more. 

Fig. \ref{haloslit2} shows the dependence of the {\em average total flux} received through the slit as a function of the radius of
the underlying emitter (again in pixels).  The spectrum contains about 43 \% of the flux for a source cut off radially at 5 pixels ($\sim 1.25"$), but only 20\% for a source
with an extent of 16 pixels ($\sim 4"$). Thus within the R=4" where there is detectable flux in some of the sources discovered here we would underestimate the total flux by
up to a factor 5.

When applied to a luminosity function
this upward correction in flux by a factor 2 - 5 will not affect the numbers of emitter in the faint bins (say below F=10$^{-18}$), because
the cumulative distribution is flat (incomplete) here anyway, but it will increase each individual luminosity in the brighter bins, leading
to a larger abundance of relatively bright objects, and will just shift the abundance of emitters  to a
flux range brighter by  a factor 2 - 5.








\clearpage

\begin{figure}[p]
\includegraphics*[scale=0.78,angle=0.]{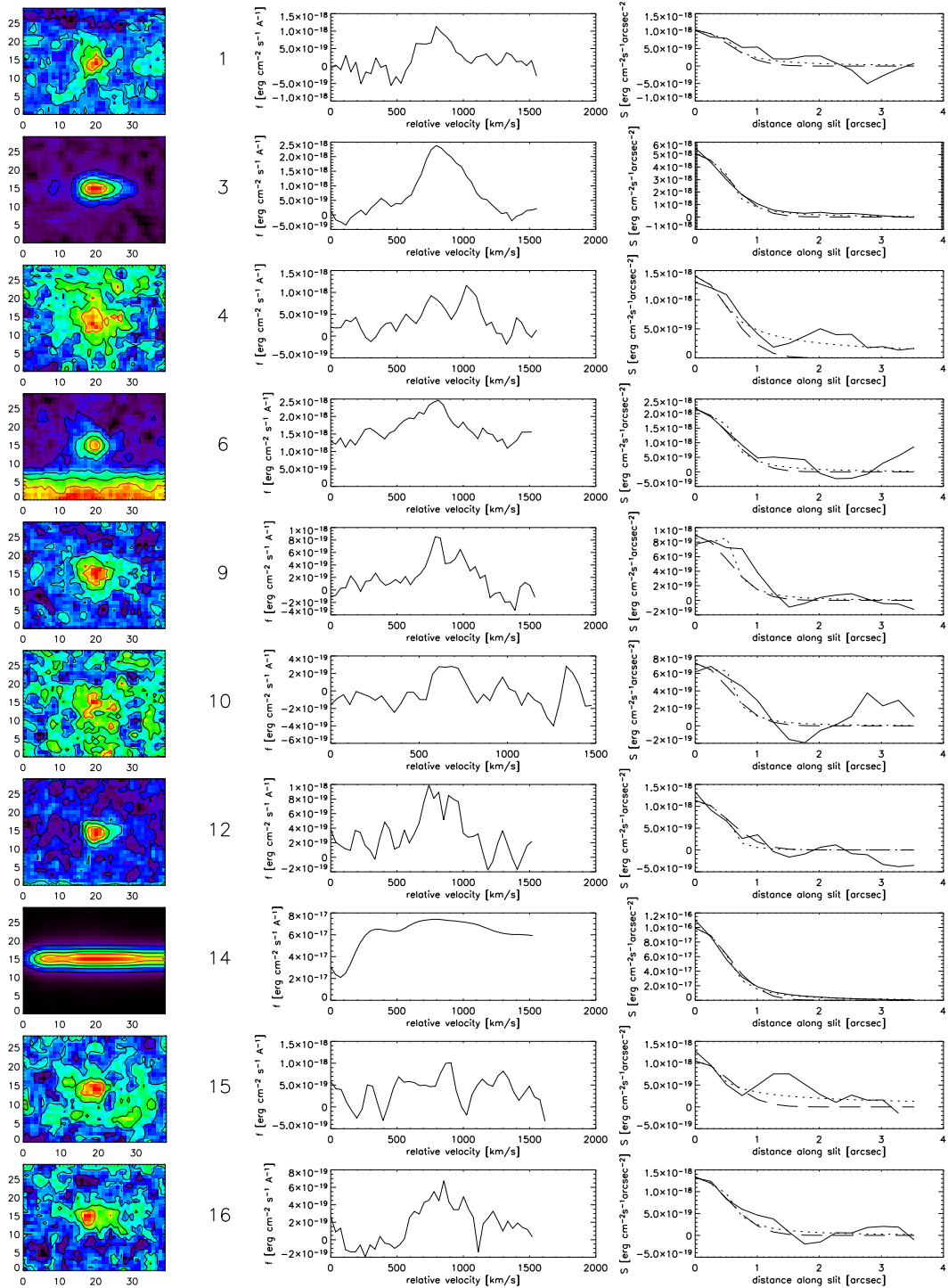}
\caption{\small two-dimensional spectra, extracted 1-d spectra (flux density in erg cm$^{-2}$ s$^{-1}$ $\AA^{-1}$), and spatial surface brightness cross-sections 
(in erg cm$^{-2}$ s$^{-1}$ $\sq\arcsec^{-1}$) along the slit.
The spectra are about 25.6 \AA\ long in the horizontal direction and 7.56" vertically, and are smoothed with a 3x3 pixel filter. The solid, dashed, and dotted 
lines show the actual data, a Gaussian PSF normalized to the surface brightness in the central 2 pixels, and a fitted model consisting of that central Gaussian and a power law continuation further out, respectively.
\label{indivpanel}}
\end{figure}
\addtocounter{figure}{-1}
\begin{figure}[p]
\includegraphics*[scale=0.78,angle=0.]{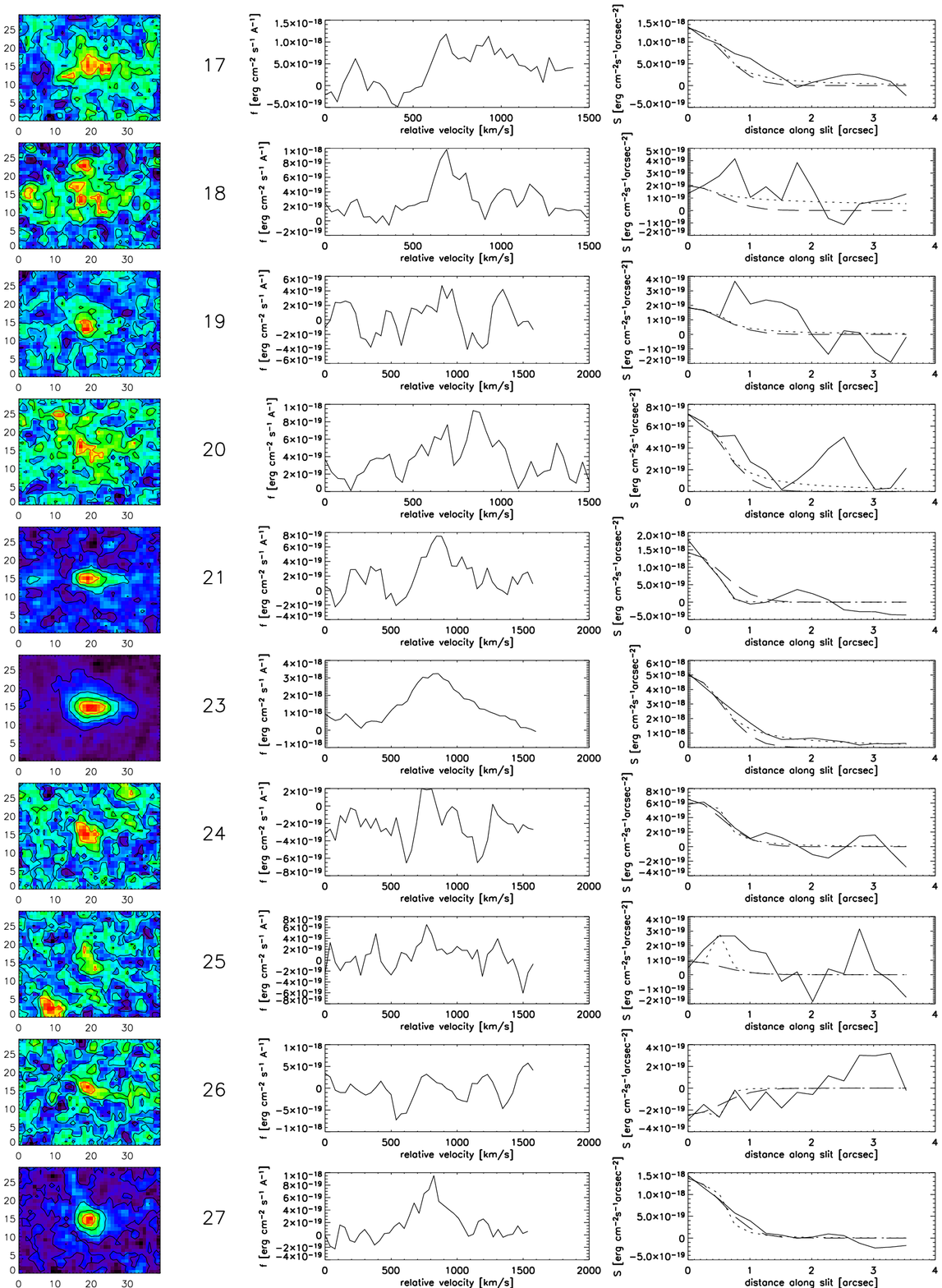}
\caption{\small continued. }
\end{figure}
\addtocounter{figure}{-1}
\begin{figure}[p]
\includegraphics*[scale=0.78,angle=0.]{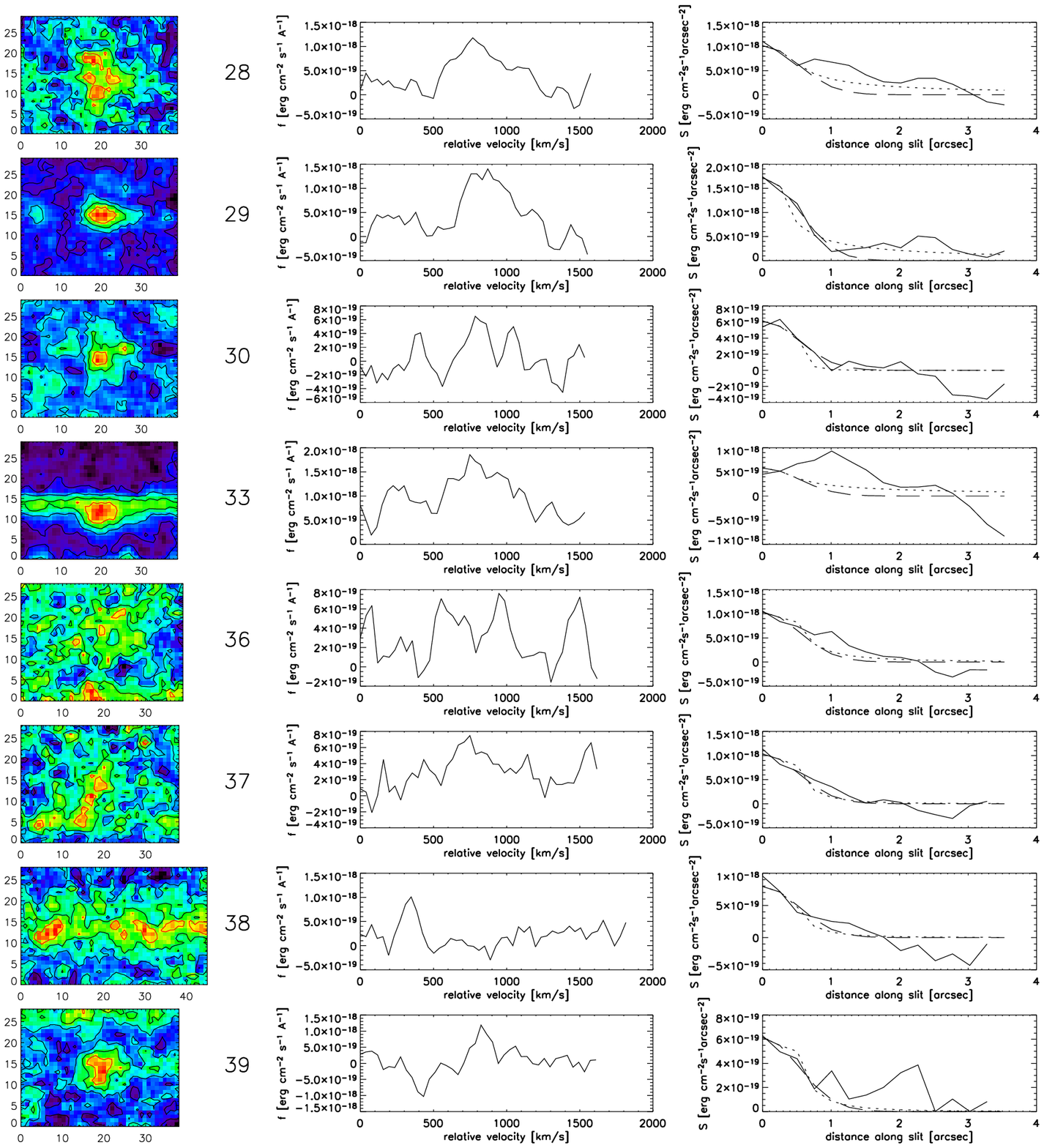}
\caption{\small continued. }
\end{figure}
\clearpage

\begin{figure}[p]
\includegraphics*[scale=0.70,angle=0.]{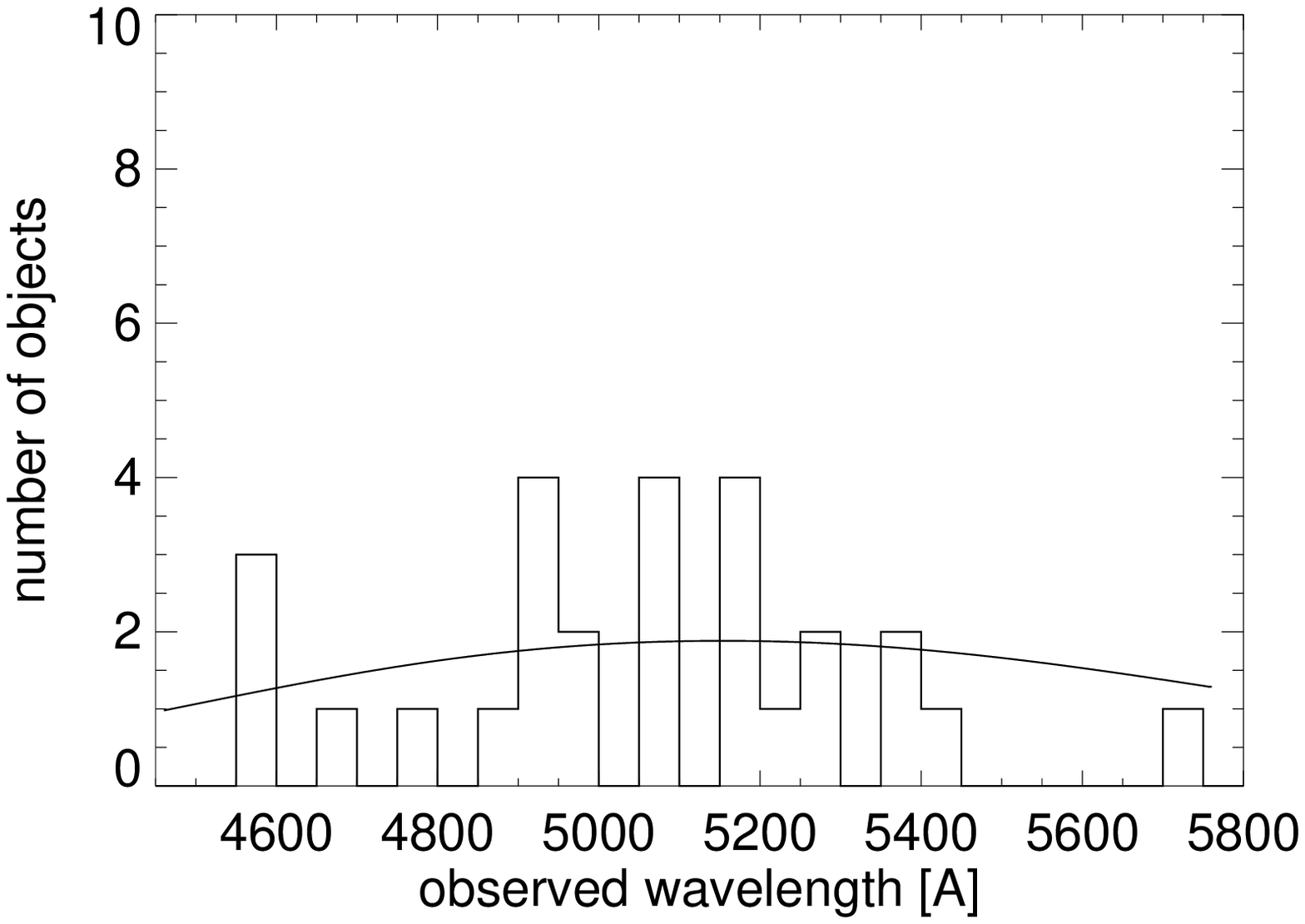}
\caption{\small Distribution of the emitters in observed wavelength. The solid line is the sensitivity of the
instrument, in arbitrary units. The actual detection threshold probably drops faster toward the edges because of illumination, dithering losses,  and detector artifacts. \label{zhist}}
\end{figure}

\begin{figure}[p]
\includegraphics*[scale=0.9,angle=0.]{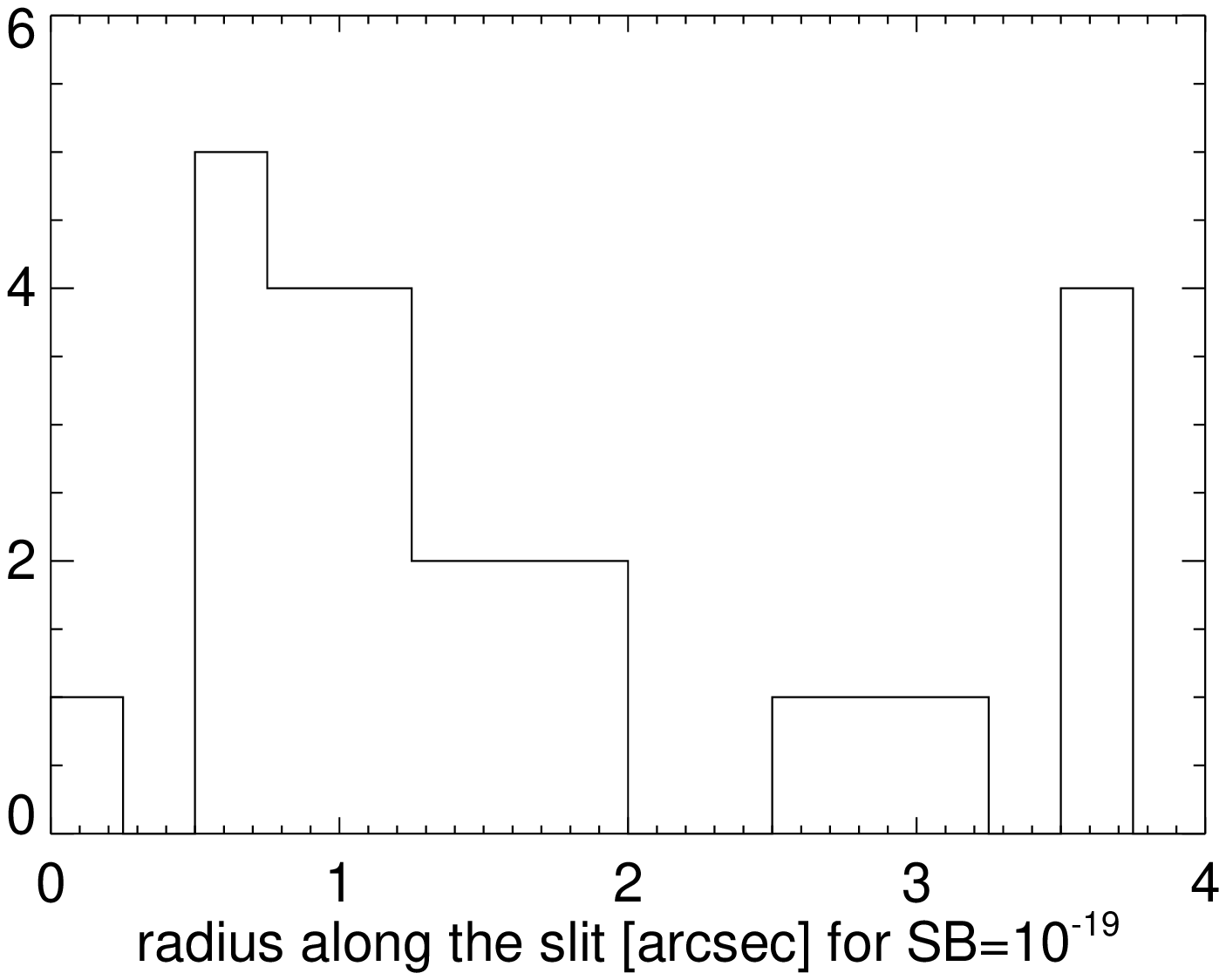}
\caption{\small distribution of the projected radius along the slit, i.e., the distance of the $10^{-19}$ surface brightness contour (based on the 
model profile) from the center of the emitter. The four cases where the contour extends beyond our fitting range are collected in the  3.6\arcsec bin, but the contour
may reach considerably larger distances than that.  
\label{arcsecradii}}
\end{figure}

\begin{figure}[p]
\includegraphics*[scale=0.9,angle=0.]{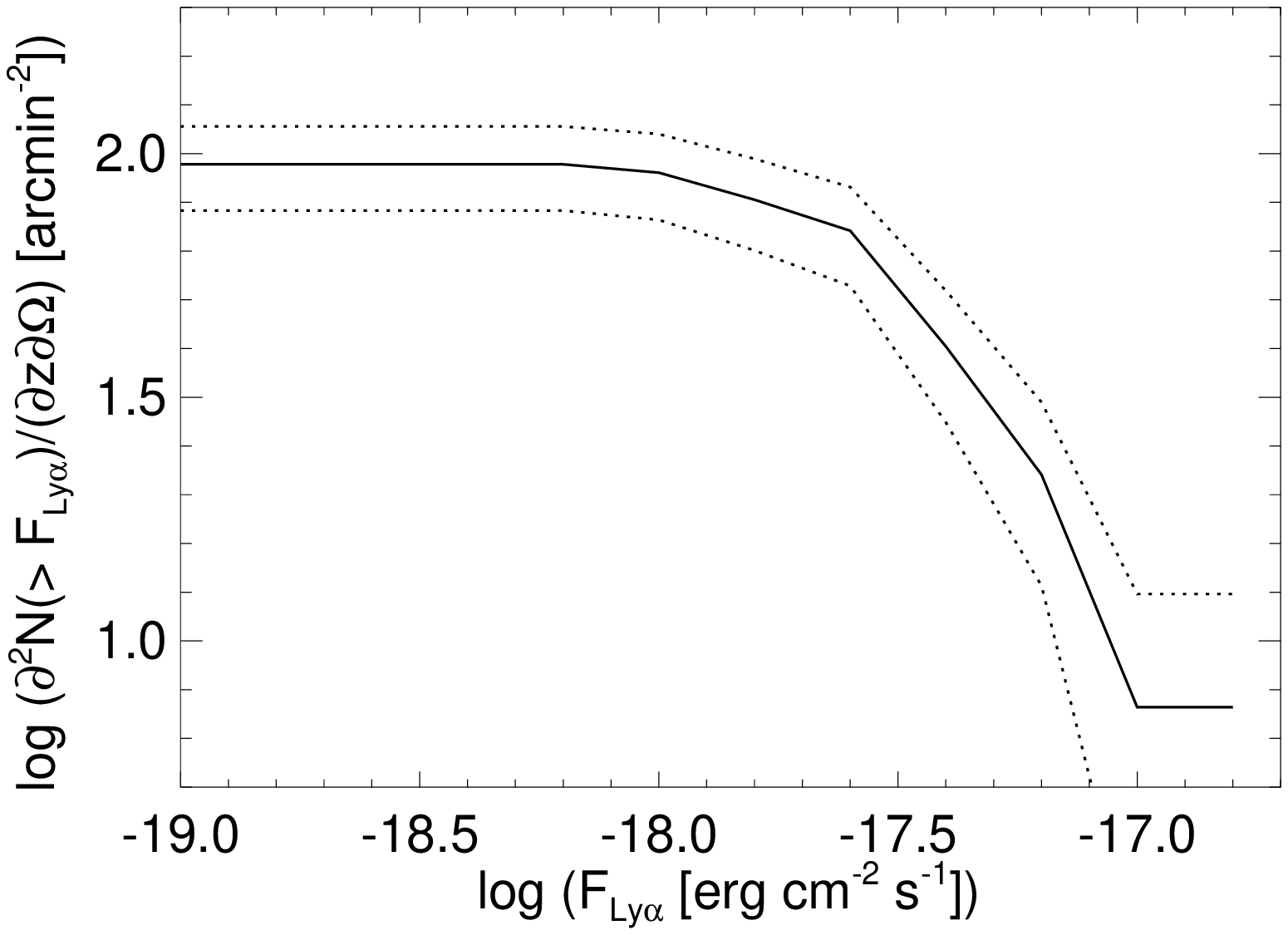}
\caption{\small Number of emitters per unit redshift and square arcminute
with a line flux exceeding $F_{Ly\alpha}$. \label{eff}}
\end{figure}

\begin{figure}[p]
\includegraphics*[scale=0.9,angle=0.]{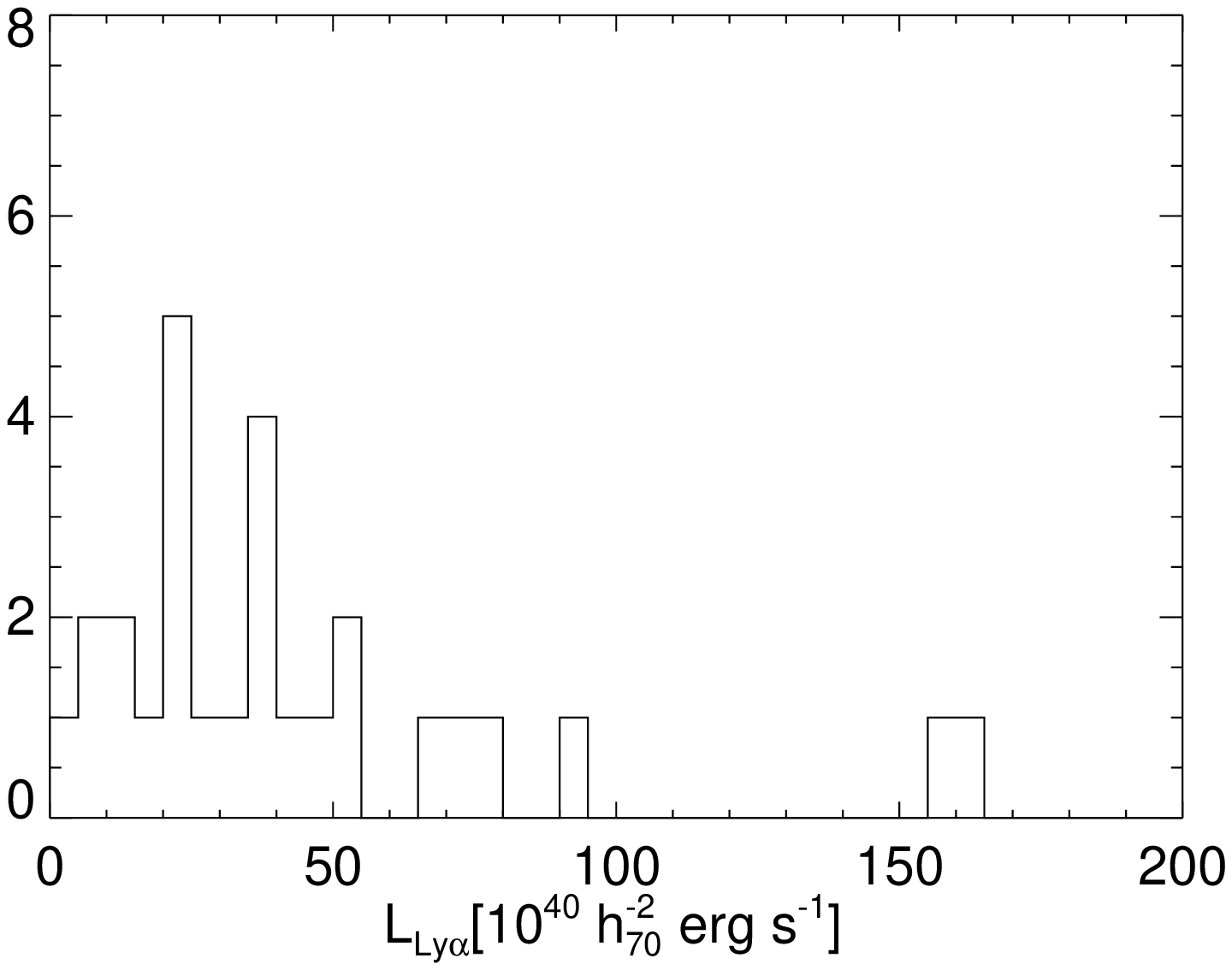}
\caption{\small  Frequency distribution of  luminosities
$L_{Ly\alpha}$. \label{luminoshist}}
\end{figure}

\begin{figure}[p]
\includegraphics*[scale=0.9,angle=0.]{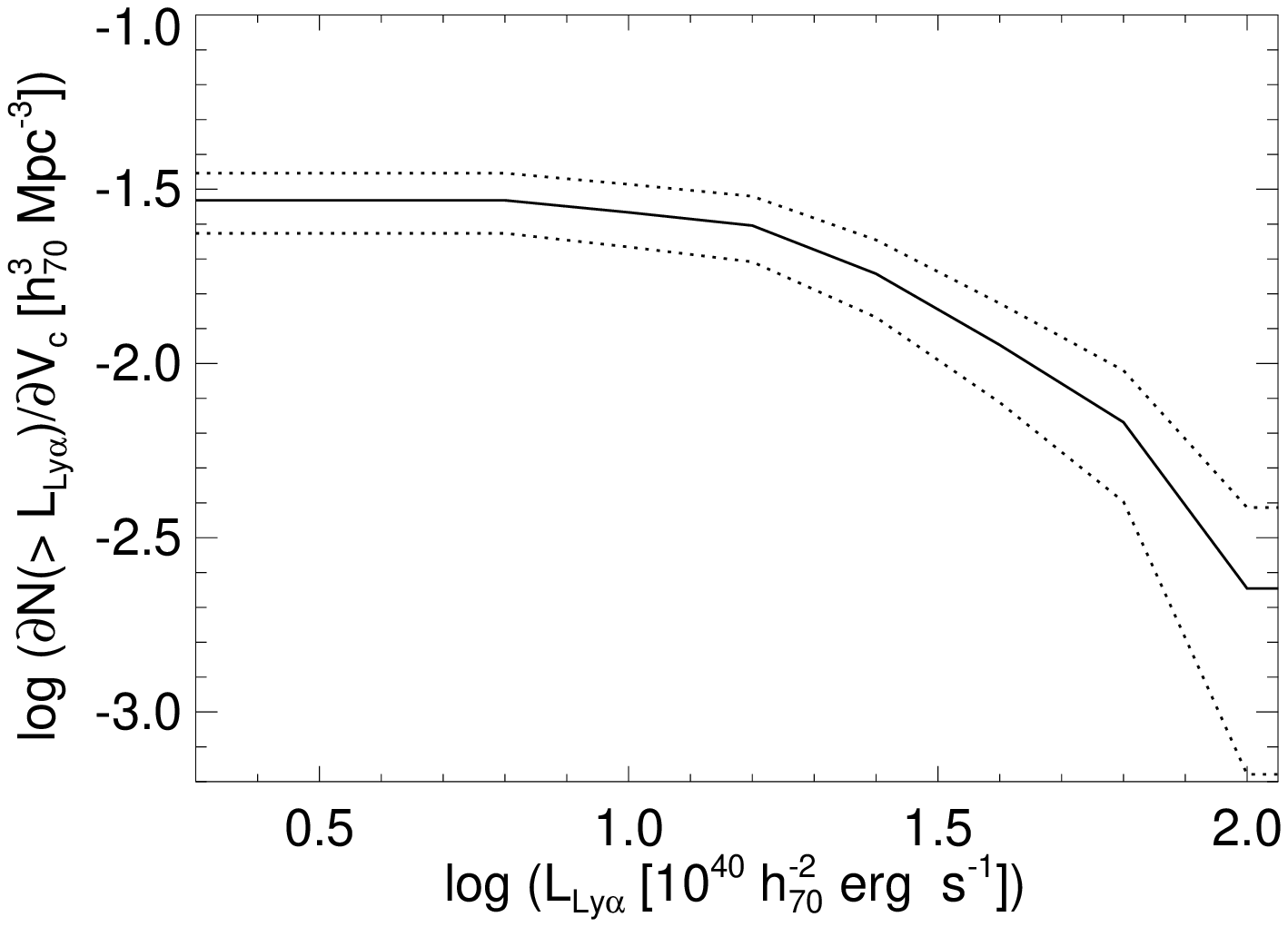}
\caption{\small  Comoving density of emitters with a luminosity
exceeding $L_{Ly\alpha}$. \label{luminos}}
\end{figure}

\begin{figure}[p]
\includegraphics*[scale=0.9,angle=0.]{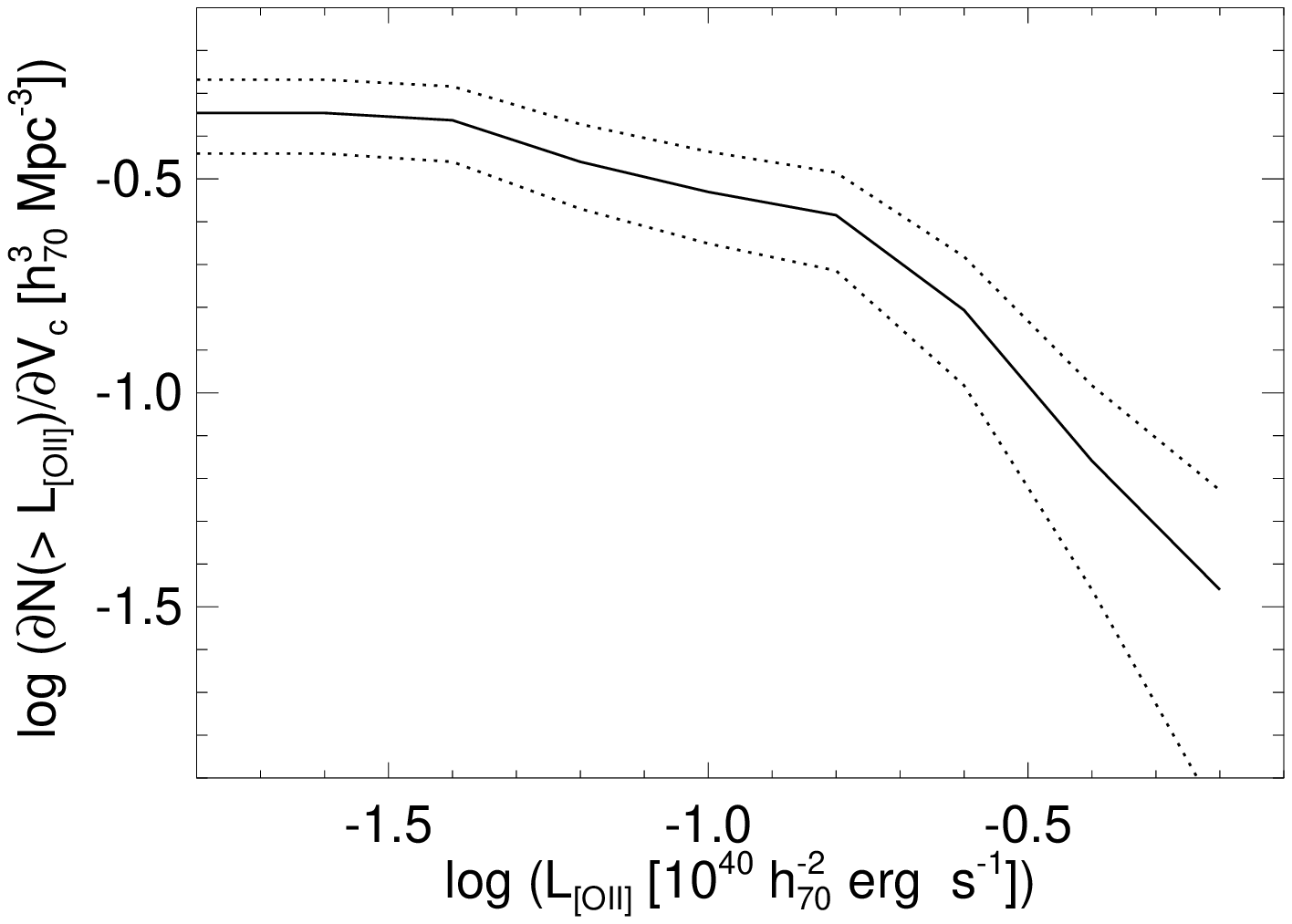}
\caption{\small  Comoving density of emitters under the assumption that they are
[OII] 3728 \AA , with a luminosity
exceeding $L_{[OII]}$. \label{luminosoii}}
\end{figure}

\begin{figure}[p]
\includegraphics*[scale=0.9,angle=0.]{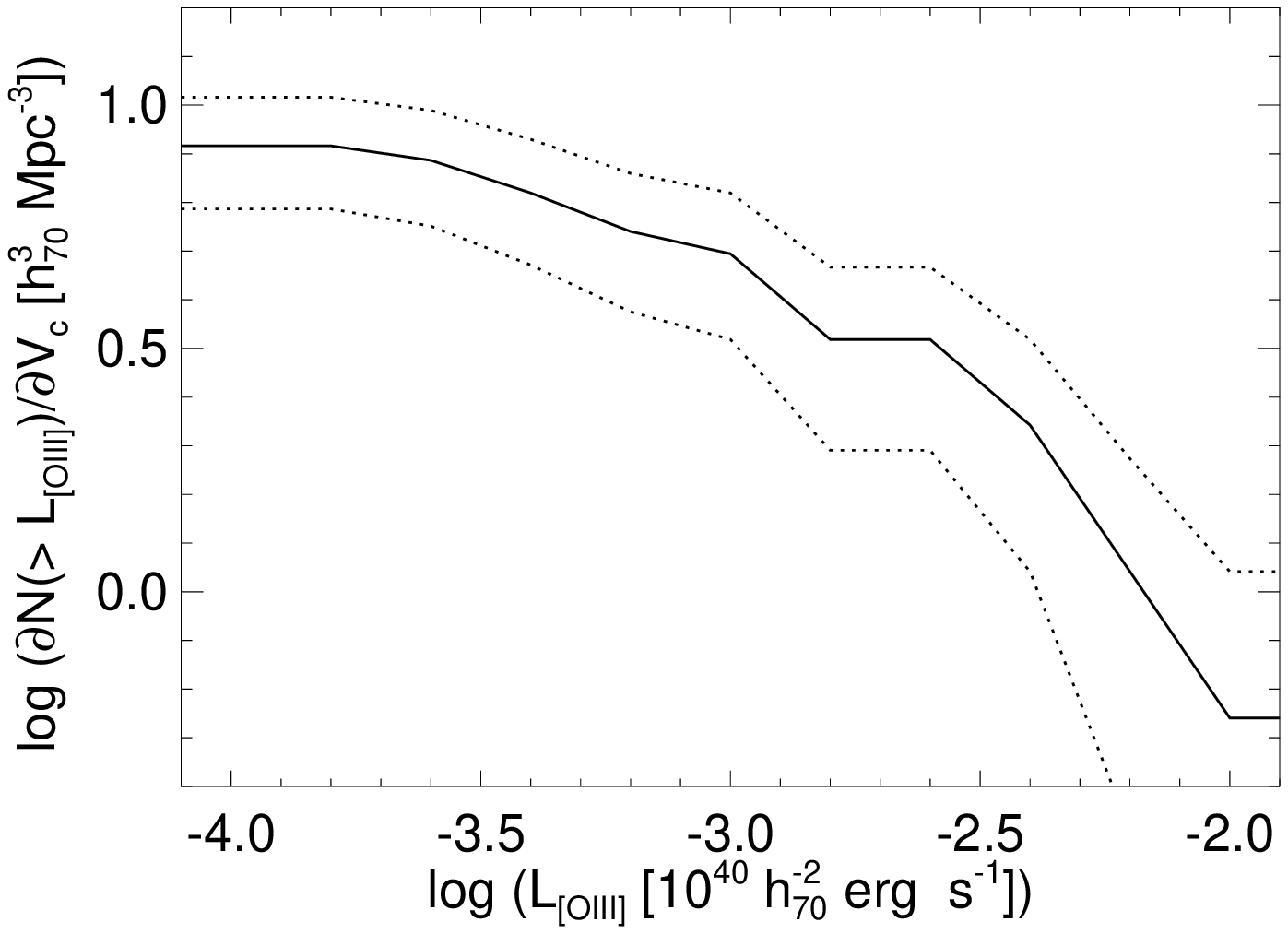}
\caption{\small  Comoving density of emitters under the assumption that they are
[OIII] 5007 \AA , with a luminosity
exceeding $L_{[OIII]}$. \label{luminosoiii}}
\end{figure}

\begin{figure}[p]
\includegraphics*[scale=0.9,angle=0.]{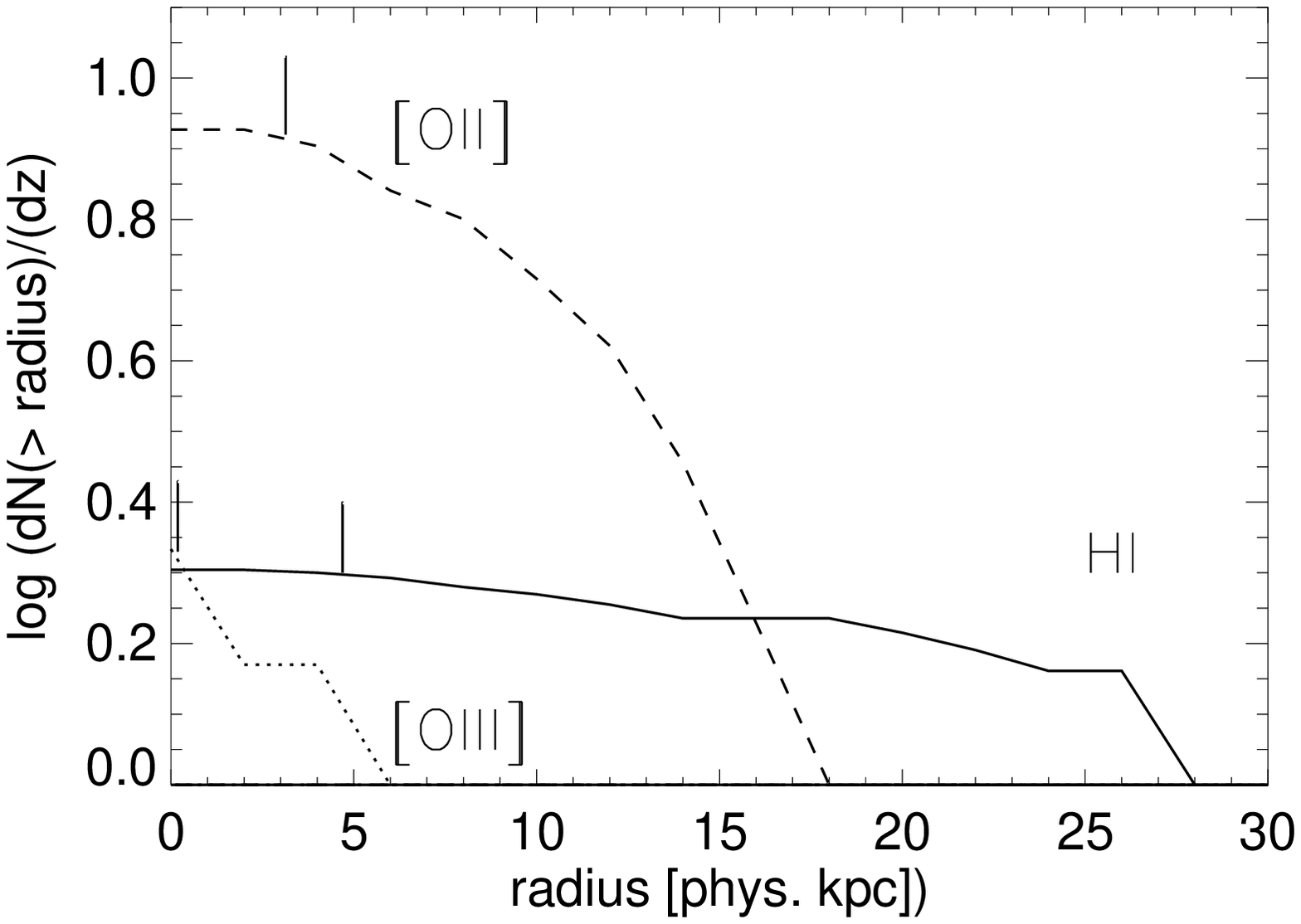}
\caption{\small Contribution of objects of different sizes to the rate of incidence per unit redshift, $dN/dz$, for HI (solid line), [OII] (dashed line), and [OIII] (dotted line). The short vertical lines riding on top of the curves indicate the spatial resolution limit along the slit. \label{dndzdist}}
\end{figure}

\begin{figure}[p]
\includegraphics*[scale=0.9,angle=0.]{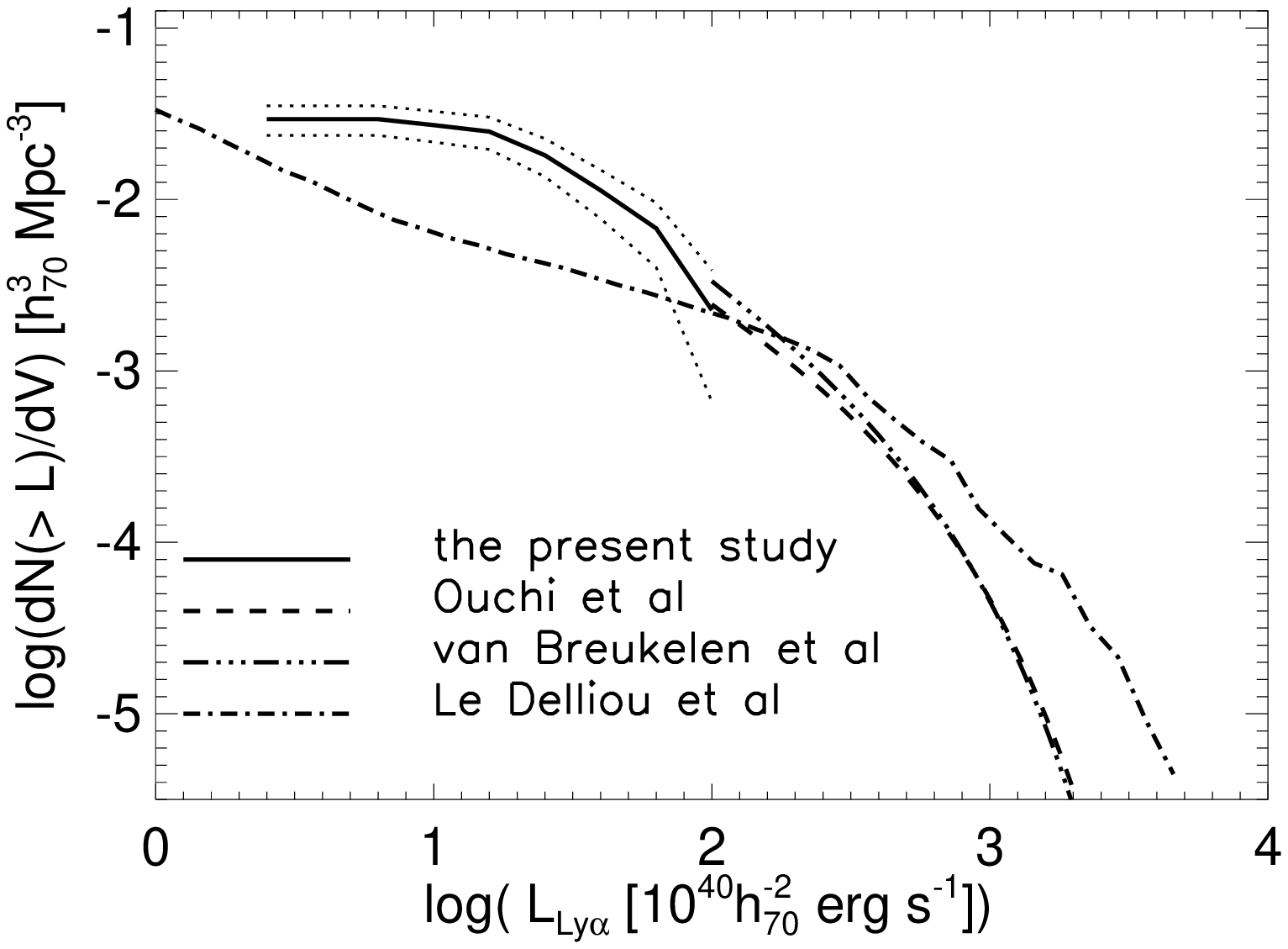}
\caption{\small Observed cumulative luminosity function at z=3.1 from Ouchi et al (2007; dashed line), z=2.9 from van Breukelen et al (2005; dash-triple dotted line),  predictions from LeDelliou et al (2005, 2006; dash-dotted line) and our sample (solid line, with the dotted lines representing the $1-\sigma$ error contours). There is almost continuity in amplitude
and slope in the overlap region with the  brighter observational data.
The van Breukelen et al function with its adopted $\alpha =-1.6$ slope if continued beyond its measured range is never more than 0.2 dex below our curve. 
The theoretical curve from  LeDelliou et al for a constant escape
fraction is shallower than
all the observed distributions at all luminosities, but the gap between
it and  our function steepens toward fainter luminosities. Our observed  distribution  starts to flatten near $3\times10^{41}$erg s$^{-1}$, well above the detection threshold, so the turnover may
be intrinsic. \label{ouchi}}
\end{figure}


\begin{figure}[p]
\includegraphics*[scale=0.9,angle=0.]{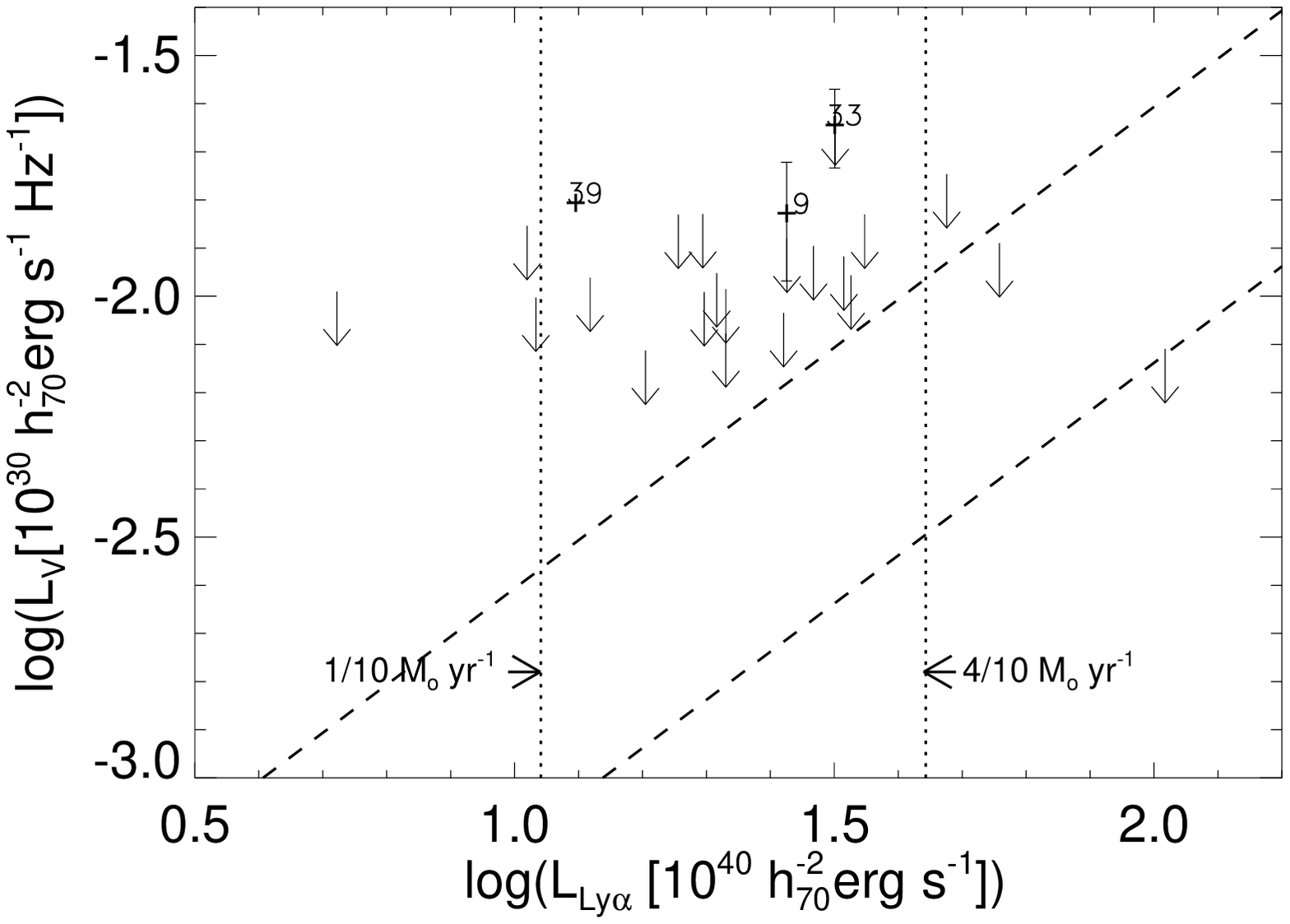}
\caption{\small V-band luminosity density versus Ly$\alpha$ line luminosities, under the assumption that all sources are high redshift
Ly$\alpha$. The arrows are upper limits, i.e., detected line emitters without V band counterparts
at a  $3-\sigma$ significance level (in a $2\arcsec\times2\arcsec$ aperture). Of the three positive detections 39, 9, and 33, the latter is somewhat
off center and may  be a low z source or a chance coincidence of a high redshift emitter with an lower redshift
continuum source.   
Three objects, 1, 3 and 39 are not covered by the V band image. The spectrum of 39, however, shows a continuum consistent with the expected Ly$\alpha$ forest
decrement. We have given here its 1500\AA\ rest frame luminosity measured from the spectrum instead of the V band luminosity. The lower, dashed, diagonal  line is the expected locus of Ly$\alpha$ emitters assuming that both,  UV luminosity and Ly$\alpha$ flux, were produced by
star formation only (see text). This line corresponds to a rest frame equivalent width of 68 \AA\ and would intercept the y-axis at -4.14.
The higher dashed line with the same slope delineates EW=20\AA . 
The vertical dotted lines in Ly$\alpha$ emitters indicate star formation
rates of 1/10 (left) and 4/10 (right) $M_{\odot}$ yr$^{-1}$.
\label{vversuslya}}
\end{figure}

\begin{figure}[p]
\includegraphics*[scale=0.9,angle=0.]{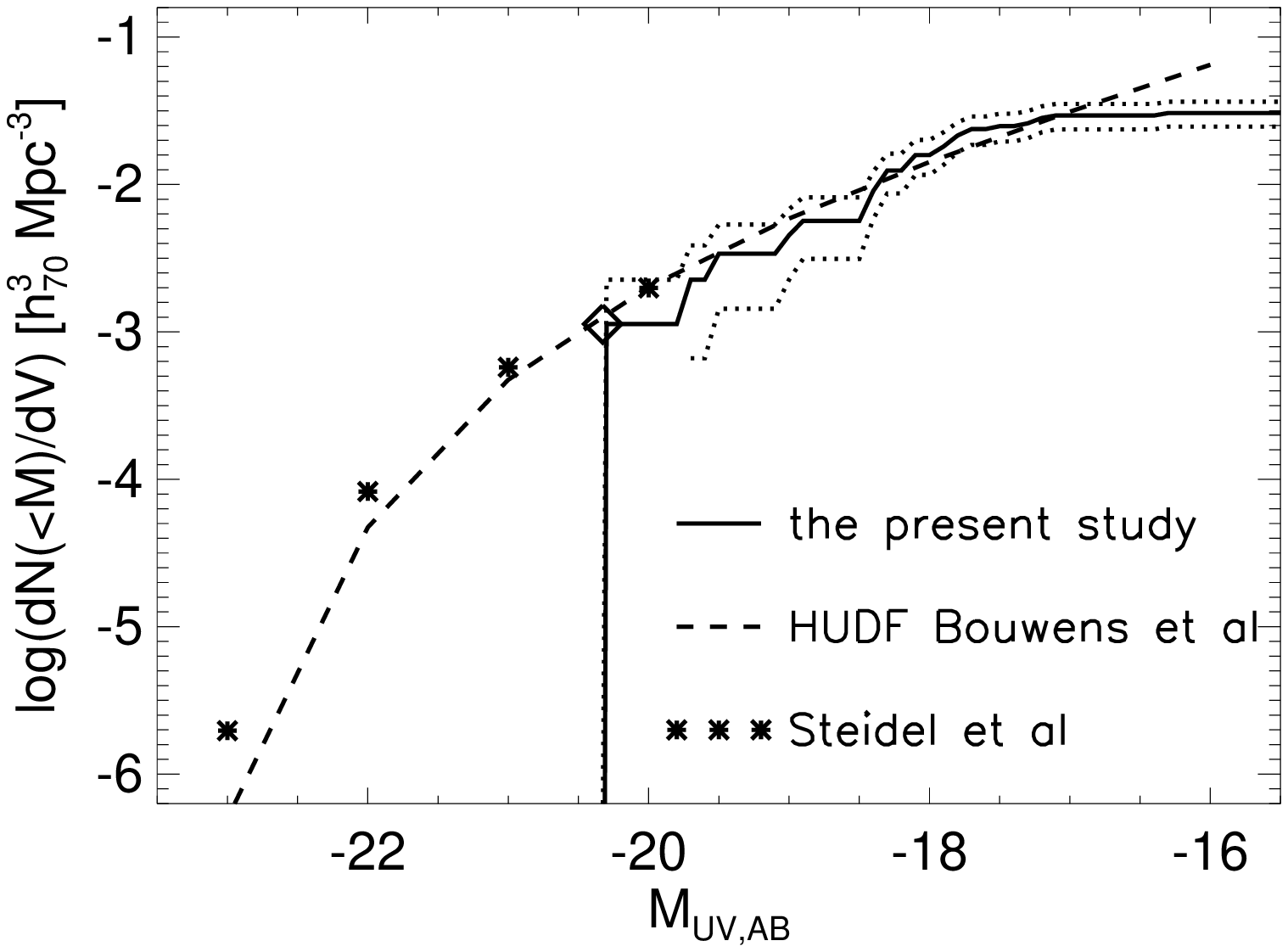}
\caption{\small 
{\em Cumulative} UV continuum luminosity functions of Steidel et al (1999; asterisks),
the Hubble Ultra Deep Field (Bouwens et al 2007; dashed line), and the
cumulative distribution of our survey (solid line; dotted lines are $\pm 1\sigma$ errors). The diamond symbol shows object \# 39. 
The emitters are entered with a {\em continuum magnitude predicted by their
Ly$\alpha$ line flux} as we have only upper limits on the continuum - see text. The number density of our emitters closely corresponds to the number density of  Bouwens et al (2007).
The absence of objects brighter than -21 is consistent
with our small survey volume.
\label{lumfunc}}
\end{figure}




\begin{figure}[p]
\includegraphics*[scale=0.6,angle=0.]{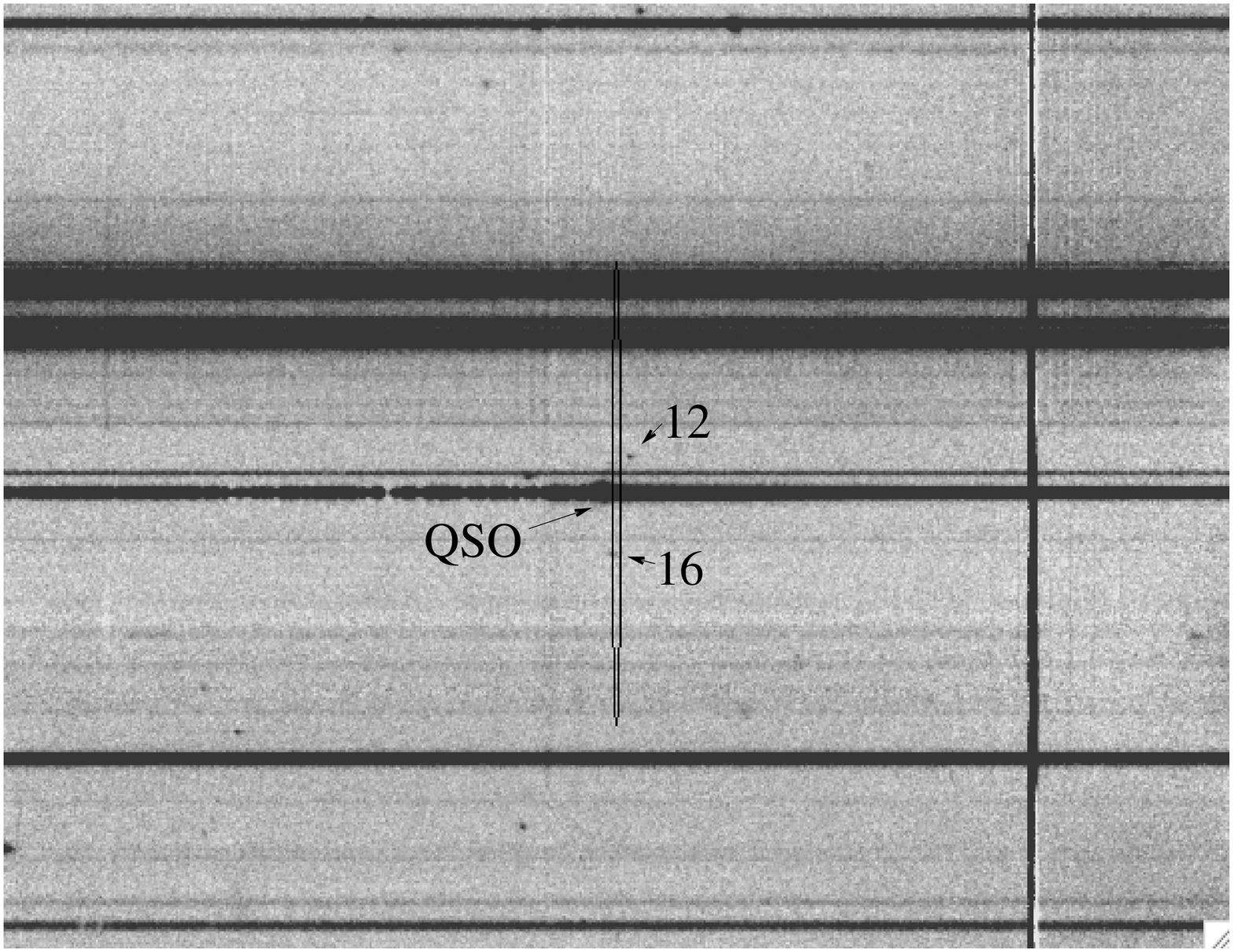}
\caption{\small The elliptical 'zone of influence' near the QSO where the
ionizing radiation would be sufficient to cause fluorescence at the $10^{-18}$erg cm$^{-2}$ s$^{-1}$ arcsec$^{-2}$ surface brightness level.\label{ellipse}}
\end{figure}

\begin{figure}[p]
\includegraphics*[scale=0.9,angle=0.]{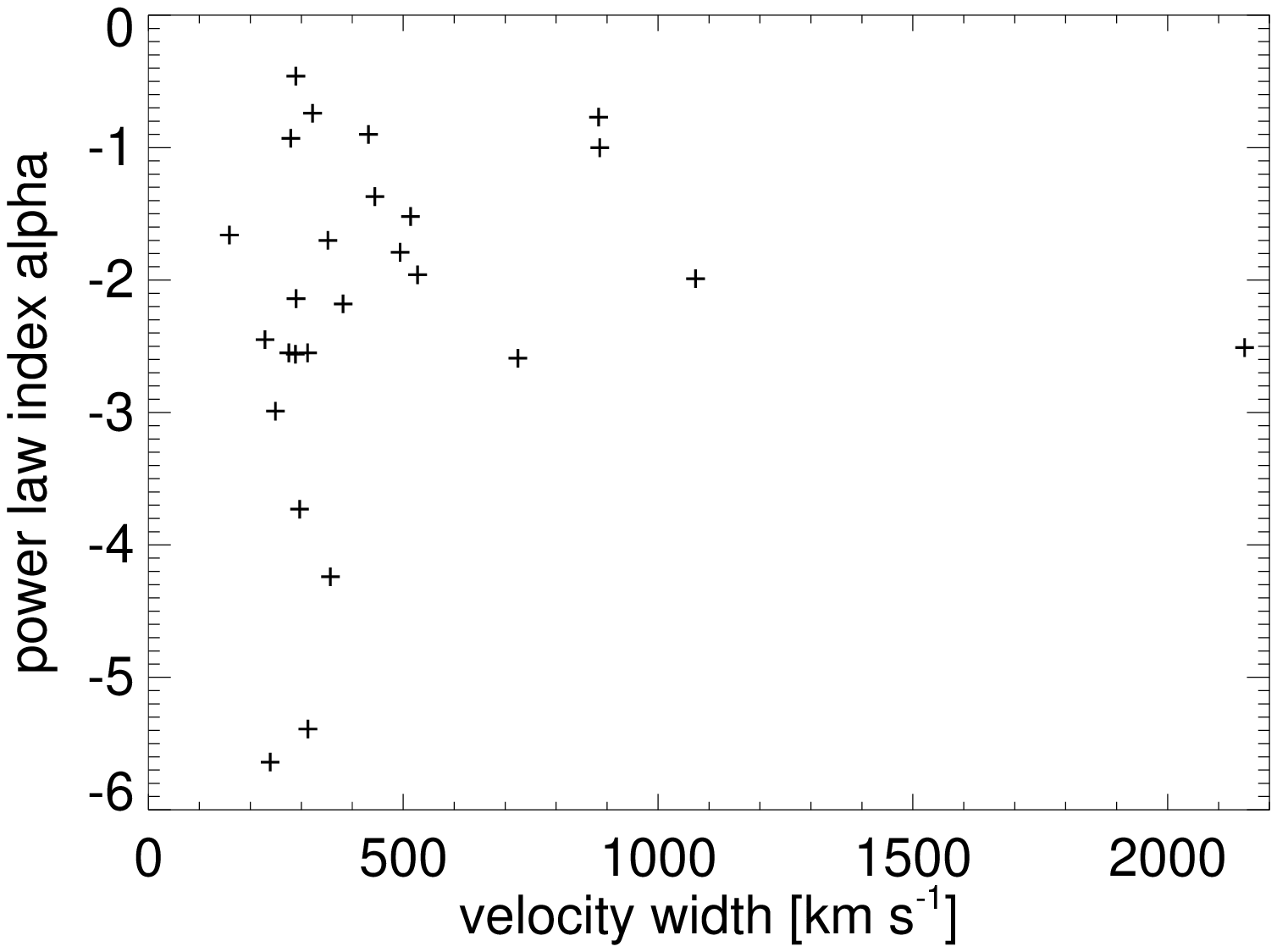}
\caption{\small Power law slope for the Gaussian + power law fit to the 
spatial surface brightness profile, versus Gaussian velocity width of the emission line. There
is no significant correlation between the two, but the objects with the largest velocity widths
also seem to have  small power law indices, i.e., the slowest radial decline in surface brightness .
\label{alphavsvfwhm}}
\end{figure}

\begin{figure}[p]
\includegraphics*[scale=0.6,angle=0.]{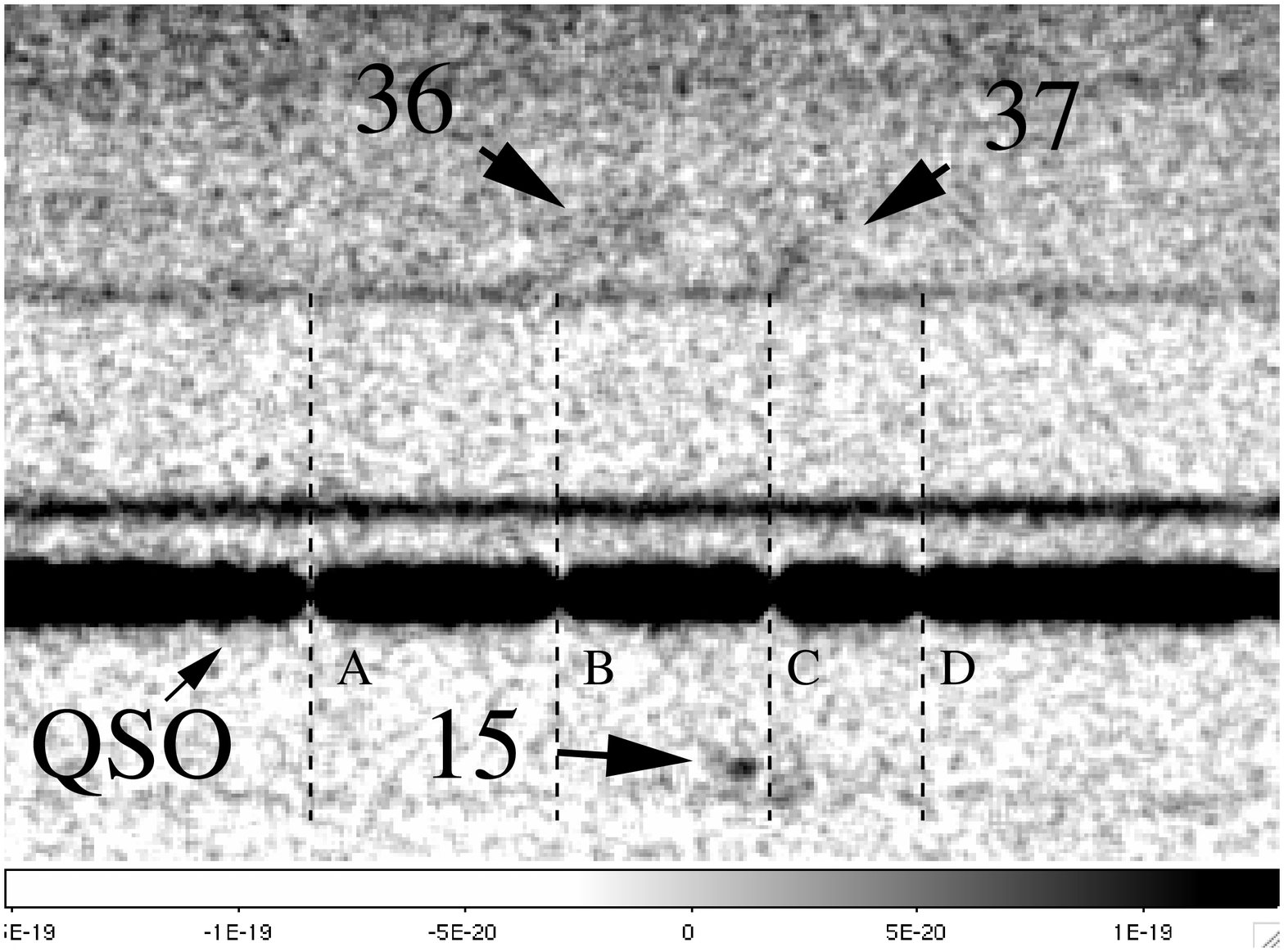}
\caption{\small The immediate surroundings of the smudges \# 36, 37, and 15. The labels A,B,C, and D refer to the strong absorption lines in the 
QSO spectrum, partly lining up with the emitters (see text). The featureless continuum object just above the QSO is a low redshift object, the fainter
continuum further up appears to show absorption by the emitting smudges 36 and 37.\label{smudgegroup}}
\end{figure}

\begin{figure}[p]
\includegraphics*[scale=0.9,angle=0.]{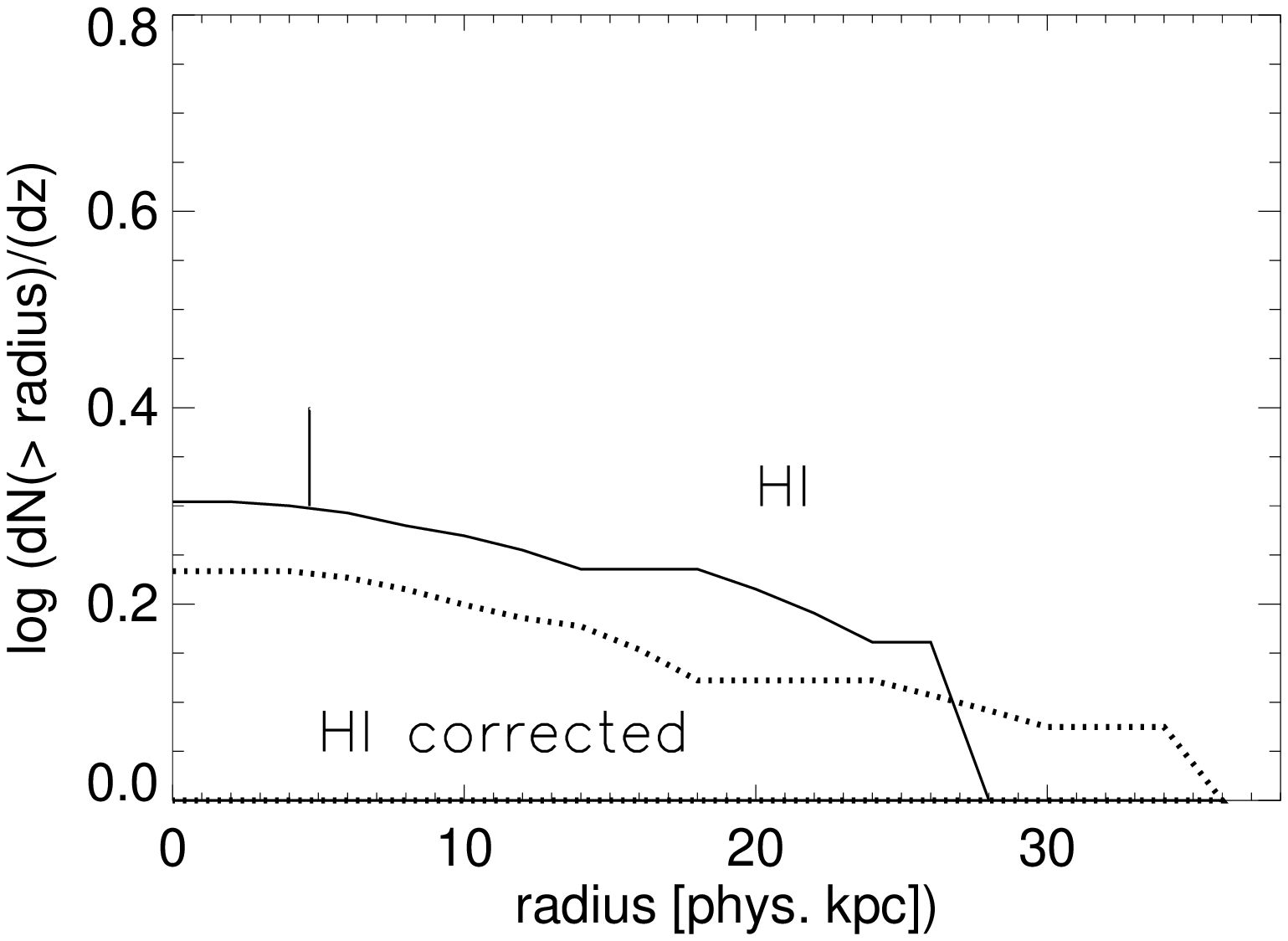}
\caption{\small Contribution of objects of different sizes to the rate of incidence per unit redshift, $dN/dz$, for HI with (dotted line) and without (solid line)
correction for the extended sizes and our underestimating  the radius.
The short vertical line riding on top of the uncorrected curve indicates the spatial resolution limit along the slit. \label{dndzdistcor}}
\end{figure}

\begin{figure}[p]
\includegraphics*[scale=0.95,angle=0.]{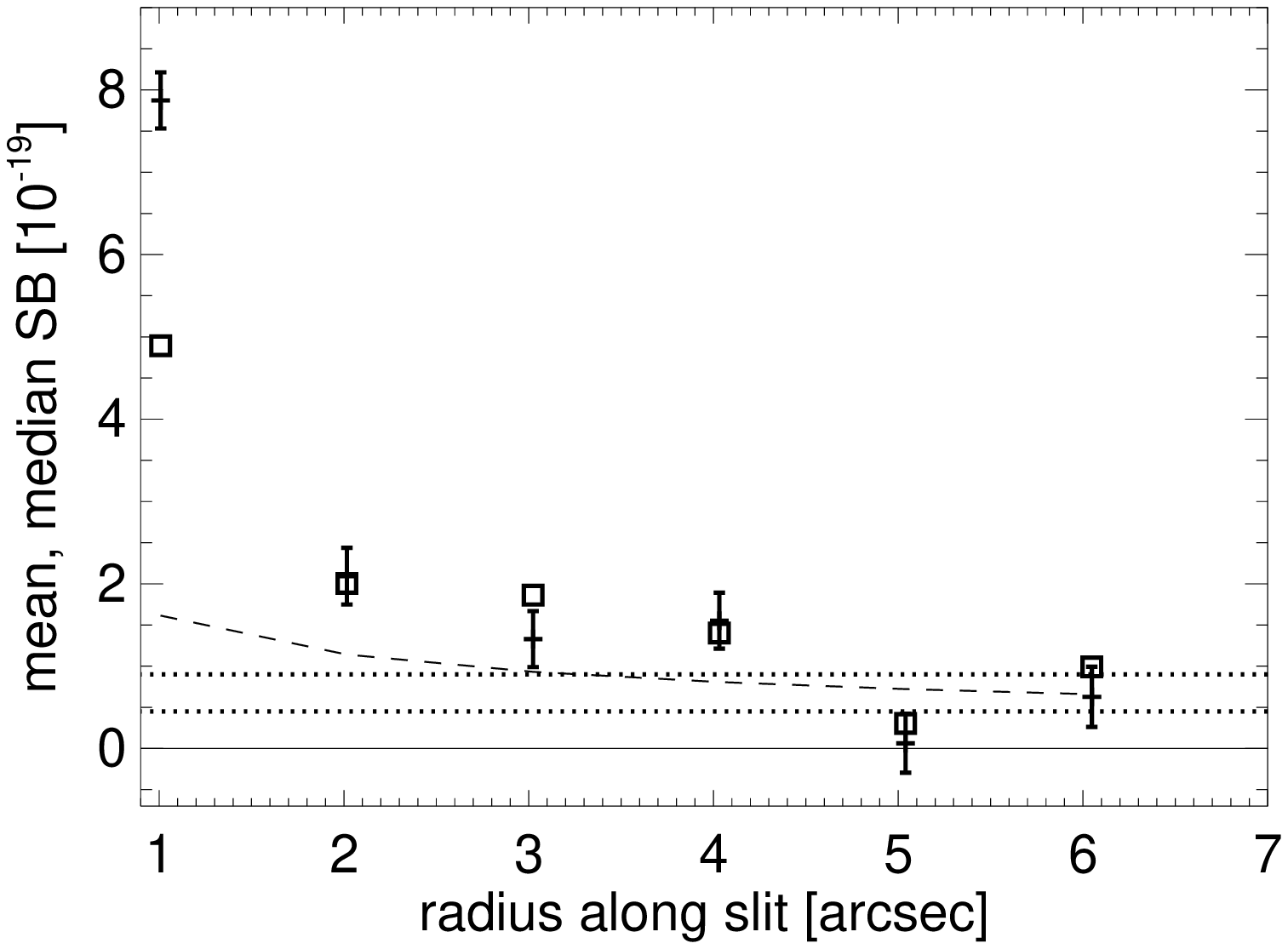}
\caption{\small Mean (points with error bars) and median (open squares) surface 
brightness measurements in units of $10^{-19}$ erg cm$^{-2}$s$^{-1}$\sq\arcsec$^{-1}$ for the combined surface brightness profiles,
as a function of angular distance in arcsecs from the center of emission along the slit. The dotted lines give the range of the expected 
surface brightness based on the Bolton et al (2005) photoionization rate
of the $z\sim 3$ IGM. The upper dotted line is for a QSO type UV spectrum,
the bottom line for a spectrum where 50\% of the flux is contributed by 
galaxies. The dashed line is the $4\sigma$ surface brightness detection
threshold for individual objects.
 \label{hog_wey}}
\end{figure}

\begin{figure}[p]
\includegraphics*[scale=0.75,angle=0.]{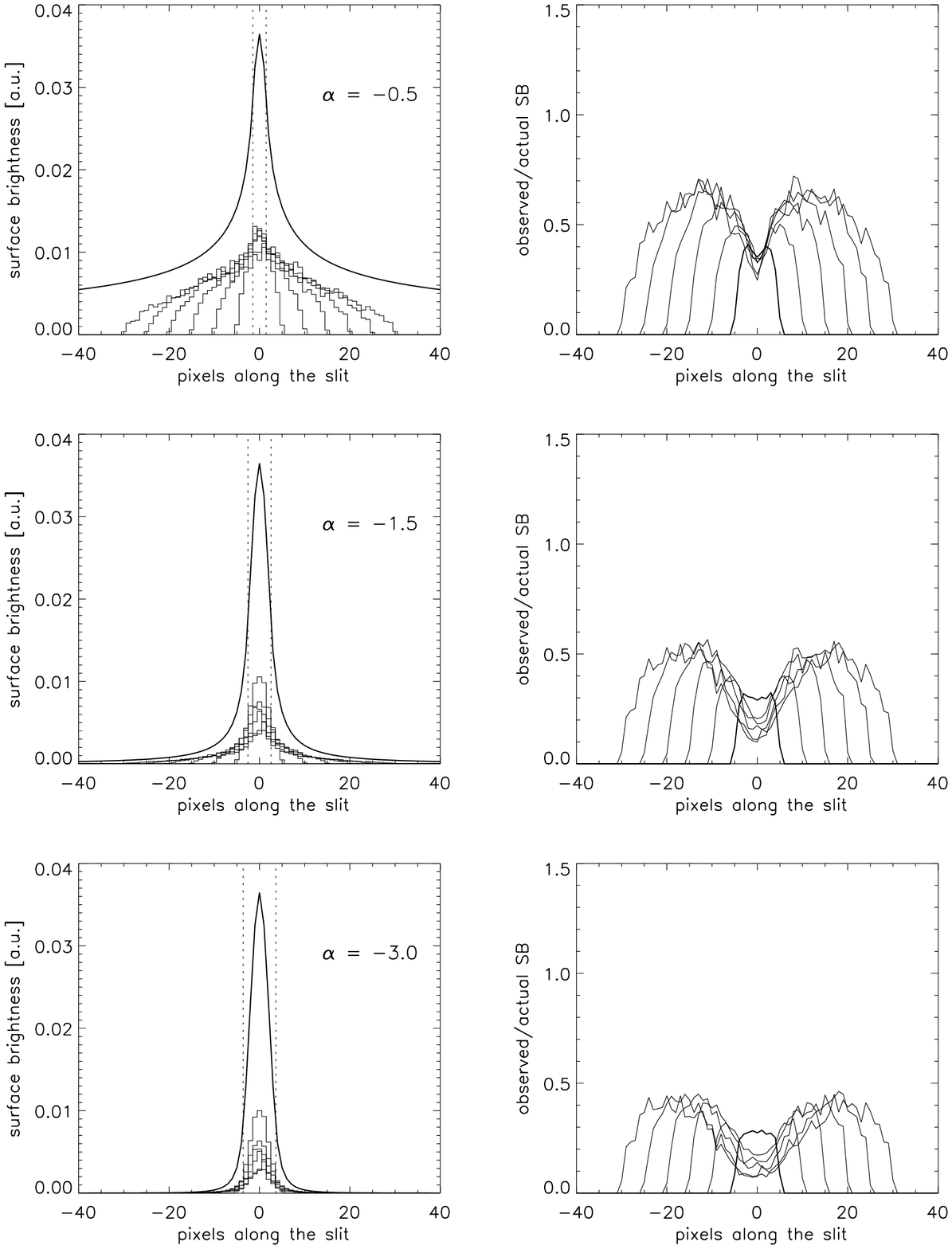}
\caption{\small Simulation of slit losses. Left hand side column: radial surface brightness profile of the input model (smooth curve) and 'observed' average profiles (binned curves) along the slit after passage through a finite slit (2\arcsec or about 8 pixels wide).
RHS column: ratios
between the input and observed profile as a function of distance along the slit. The three rows of panels in either column show models with external slope -0.5 (top row),-1.5 (middle row), and -3.0   (bottom row). The different output profiles within each panel arise from emitters with different overall radial extent. Dashed vertical lines show the position of turnover between Gaussian
core and power law wings.\label{haloslit1}}
\end{figure}

\begin{figure}[p]
\includegraphics*[scale=0.8,angle=0.]{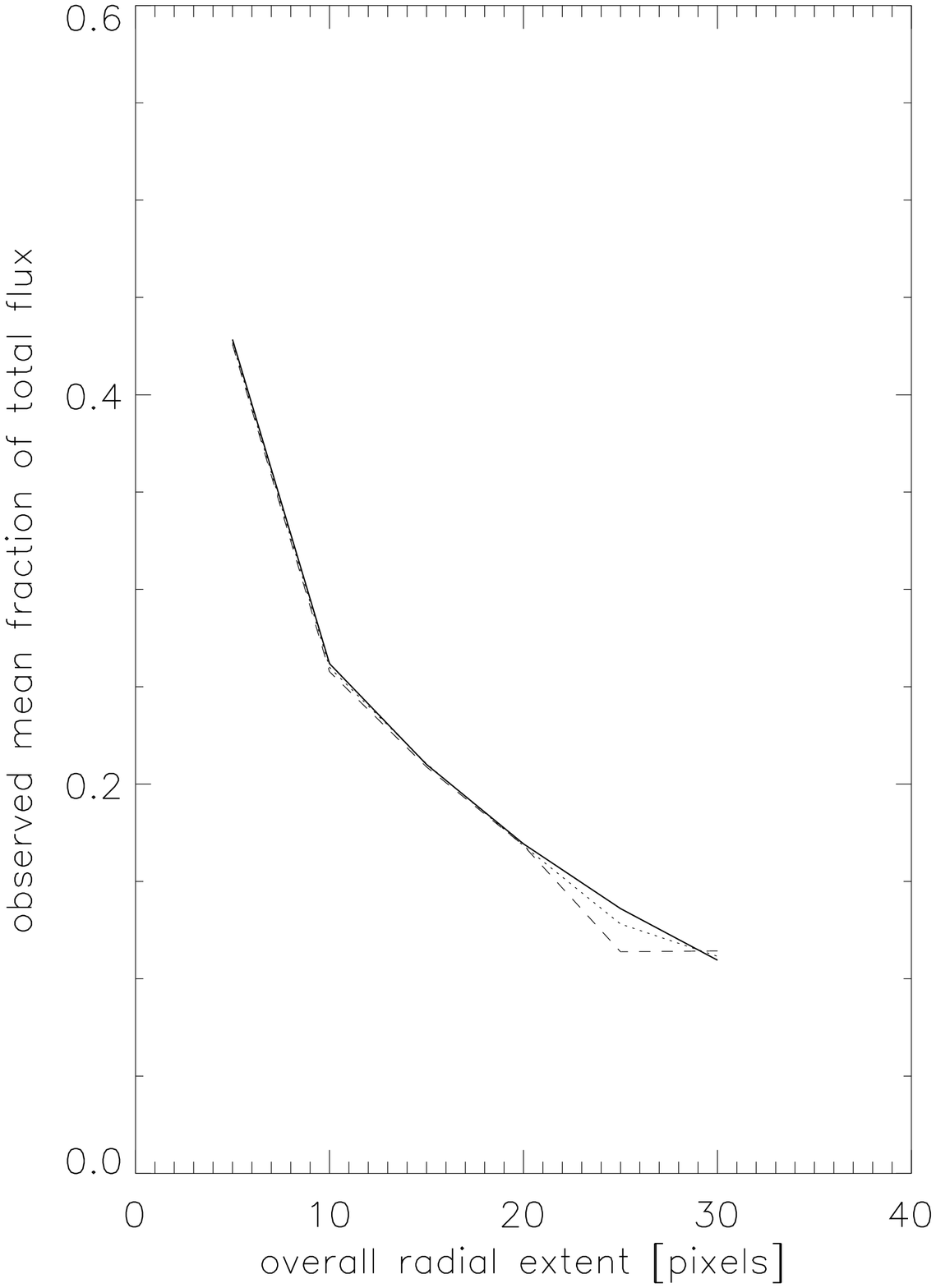}
\caption{\small Average fraction of total flux passing through the 2"(8 pixel)  wide slit, as a function of total radial extent of the emitter.\label{haloslit2}}
\end{figure}

\clearpage
\begin{deluxetable}{ccrrrrcccrcl}
\tabletypesize{\scriptsize}
\rotate
\tablecaption{Foreground Emission Line Objects \label{oiitable}}
\tablewidth{0pt}
\tablehead{
\colhead{(1)} & \colhead{(2)} & \colhead{(3)} & \colhead{(4)} & \colhead{(5)} &
\colhead{(6)} \\

\colhead{\#} & \colhead{ID} & \colhead{z} & \colhead{$F$ [$10^{-18}$]\tablenotemark{a}}  &
\colhead{S$_{max}$ [$10^{-18}$]}  & \colhead{Source of Identification}
}
\startdata
1&  7a& 0.3928  &   12.45  $\pm$  0.29  &  5.06  $\pm$ 0.22 &  [OII] doublet  \\
2&  7b& 0.4336  &   27.53  $\pm$  0.33  &  8.99  $\pm$ 0.23 &  [OII] doublet  \\
3&   8& 0.0458  &    2.60  $\pm$  0.28  &  1.10  $\pm$  0.21 &  H$_{\beta}$ and Mgb triplet  \\
4&  22& 0.1980  &   18.49  $\pm$  0.31  &  8.30  $\pm$  0.23 &  Balmer lines, [OII] doublet \\
5&  31& 0.4019  &   26.55  $\pm$  0.32  &  9.24  $\pm$  0.23 &  [OII] doublet  \\
\enddata
\tablenotetext{a}{total flux for both lines where doublet}
\end{deluxetable}

\clearpage
\begin{deluxetable}{ccrrrrcccrcl}
\tabletypesize{\scriptsize}
\rotate
\tablecaption{Properties of  Single-Line Emitters \label{bigtable}}
\tablewidth{0pt}
\tablehead{
\colhead{(1)} & \colhead{(2)} & \colhead{(3)} & \colhead{(4)} & \colhead{(5)} &
\colhead{(6)} & \colhead{(7)} & \colhead{(8)} &
\colhead{(9)} & \colhead{(10)} &
\colhead{(11)}\\

\colhead{\#} & \colhead{ID} & \colhead{z} & \colhead{$F$ [$10^{-18}$]} & \colhead{flux ratio} &
\colhead{S$_{max}$ [$10^{-18}$]}  & \colhead{v$_{FWHM}$} &
\colhead{ A [$10^{-18}$]} & \colhead{y$_{t}$} &
\colhead{$\alpha$} & \colhead{Comments}
}
\startdata
   1 &   1 &3.1801 & 2.32 $\pm$ 0.57&  0.53 $\pm$ 0.13&  0.88 $\pm$  0.4&   352.2 & 1.04 &  3.0 & -2.12 & somewhat EXT\tablenotemark{a}, CD\tablenotemark{b}    \\
   2 &   3 &3.1916 &12.77 $\pm$ 0.34&  0.73 $\pm$ 0.02&  4.81 $\pm$  0.24&   382.1 & 4.81 &  3.9 & -2.07 & PS\tablenotemark{c}; red-dominated em. w. faint blue peak           \\
   3 &   4 &3.1797 & 3.38 $\pm$ 0.34&  0.47 $\pm$ 0.05&  1.14 $\pm$  0.24&   431.8 & 1.28 &  2.9 & -0.50 & EXT             \\
   4 &   6 &3.3362 & 6.04 $\pm$ 0.35&  0.67 $\pm$ 0.04&  2.09 $\pm$  0.24&   289.8 & 2.26 &  3.9 & -2.12 & CD   \\
   5 &   9 &3.2378 & 2.89 $\pm$ 0.35&  0.79 $\pm$ 0.10&  0.81 $\pm$  0.24&   288.3 & 0.92 &  2.2 & -3.78 & CD, coincidence with unrelated (?) continuum object\\
   6 &  10 &3.4833 & 1.91 $\pm$ 0.38&  0.56 $\pm$ 0.11&  0.67 $\pm$  0.26&   312.4 & 0.81 &  6.5 & -3.87 & EXT, "plug-shaped" em.  \\
   7 &  12 &3.3300 & 3.36 $\pm$ 0.35&  1.29 $\pm$ 0.13&  1.30 $\pm$  0.24&   296.7 & 1.15 &  6.5 & -3.75 & PS\\
   8 &  14 &3.3221       &                &                 &                 &          &      &      &       & QSO centered on broad em. line\\
   9 &  15 &2.7682 & 2.67 $\pm$ 0.41&  0.38 $\pm$ 0.06&  1.02 $\pm$  0.29&   883.4 & 1.04 &  2.4 & -0.91 & EXT "ring" with blue-dominated double comp.\\
  10 &  16 &3.3189 & 3.06 $\pm$ 0.37&  0.59 $\pm$ 0.07&  1.08 $\pm$  0.28&   528.3 & 1.48 &  4.1 & -2.20 & CD; broad em. line\\
  11 &  17 &3.6954 & 4.12 $\pm$ 0.45&  0.62 $\pm$ 0.07&  1.40 $\pm$  0.32&   493.9 & 1.48 &  2.7 & -1.68 & EXT; broad em. line\\
  12 &  18 &3.4373 & 1.14 $\pm$ 0.37&  0.50 $\pm$ 0.16&  0.48 $\pm$  0.27&   289.3 & 0.24 &  2.0 & -0.54 & EXT, amorphous  \\
  13 &  19 &3.0595 & 1.53 $\pm$ 0.44&  3.27 $\pm$ 0.94&  0.47 $\pm$  0.30&   158.8 & 0.13 &  3.1 & -2.20 & CD, somewhat EXT   \\
  14 &  20 &3.4023 & 1.79 $\pm$ 0.37&  0.37 $\pm$ 0.08&  0.68 $\pm$  0.26&   514.3 & 0.69 &  2.1 & -0.99 & EXT, amorphous\\
  15 &  21 &3.0809 & 2.77 $\pm$ 0.36&  0.72 $\pm$ 0.09&  1.29 $\pm$  0.25&   228.1 & 1.64 &  8.5 & -5.37 & PS; narrow line   \\
  16 &  23 &2.9075 &15.32 $\pm$ 0.37&  0.65 $\pm$ 0.02&  5.30 $\pm$  0.26&   444.4 & 5.45 &  2.5 & -1.46 & CD, ring; weak blue, strong red double comp.  \\
  17 &  24 &3.0593 & 1.48 $\pm$ 0.35&  0.69 $\pm$ 0.16&  0.58 $\pm$  0.25&   228.8 & 0.53 &  2.5 & -1.78 & PS, faint\\
  18 &  25 &3.0655 & 0.70 $\pm$ 0.36&  0.68 $\pm$ 0.35&  0.28 $\pm$  0.25&   357.1 & 0.11 &  4.2 & -4.05 & PS, faint\\
  19 &  26 &3.0692 &-0.04 $\pm$ 0.37& -0.22 $\pm$ 2.02&  0.19 $\pm$  0.25&   238.9 & fit & failed & --- & PS, faint\\
  20 &  27 &3.2617 & 3.51 $\pm$ 0.34&  0.78 $\pm$ 0.08&  1.23 $\pm$  0.24&   249.2 & 1.32 &  4.7 & -2.49 & PS; narrow line\\
  21 &  28 &3.0732 & 2.90 $\pm$ 0.36&  0.45 $\pm$ 0.06&  1.07 $\pm$  0.25&   885.7 & 1.09 &  2.2 & -1.09 & CD, somewhat EXT; weak blue, strong red double comp.\\
  22 &  29 &3.1819 & 3.96 $\pm$ 0.34&  0.48 $\pm$ 0.04&  1.36 $\pm$  0.24&   279.5 & 1.57 &  3.3 & -0.34 & CD; weak blue, strong narrow red double comp.\\
  23 &  30 &3.2715 & 3.05 $\pm$ 0.35&  1.56 $\pm$ 0.18&  1.27 $\pm$  0.25&   313.1 & 0.88 &  4.4 & -2.38 & CD\\
  24 &  33 &3.2646 & 3.50 $\pm$ 0.34&  1.00 $\pm$ 0.10&  1.88 $\pm$  0.24&   322.2 & 1.51 &  4.0 & -2.16 & PS; coincidence with unrelated lower z continuum obj.\\
  25 &  36 &2.7483 & 3.46 $\pm$ 0.52&  0.81 $\pm$ 0.12&  0.90 $\pm$  0.31&  1073.7 & 1.08 &  2.7 & -1.62 & very EXT, amorphous smudge, nearby QSO and gal. abs.  \\
  26 &  37 &2.7713 & 3.27 $\pm$ 0.57&  1.14 $\pm$ 0.20&  0.73 $\pm$  0.29&   725.0 & 1.02 &  4.9 & -2.76 & EXT, assym.;  em. ('trapdoor') blueward of abs.  \\
  27 &  38 &3.0322 & 3.43 $\pm$ 0.70&  2.69 $\pm$ 0.55&  1.44 $\pm$  0.51&  2151.1 & 0.79 &  5.2 & -2.70 & very broad em. on top of fuzzy continuum \\
  28 &  39 &2.8285 & 2.28 $\pm$ 0.43&  0.41 $\pm$ 0.08&  0.75 $\pm$  0.33&   275.6 & 0.86 &  2.2 & -1.07 & CD, narrow PCygni em. line w. continuum\\
\enddata
\tablenotetext{a}{EXT = 'extended'}
\tablenotetext{b}{CD = 'centrally dominated'}
\tablenotetext{c}{PS = 'point source'}
\end{deluxetable}


\clearpage



\begin{references}

\reference{} Adams, T. F. 1972, \apj, 174, 439

\reference{} Adelberger, K. L., Steidel, C. C., Kollmeier, J, A., Reddy, N. A., 2006, \apj,637, 74

\reference{} Ahn, S.-H., Lee, H.-W., Lee, H. M., 2003, \mnras, 340, 863

\reference{} Ahn, S.-H., 2004, \apj, 601, L25 
	
\reference{} Beckwith, S. V. W., Stiavelli, M., Koekemoer, A. M., Caldwell, J. A. R., Ferguson, H. C.,
Hook, R., Lucas, R. A. Bergeron, L. E., Corbin, M., Jogee, S., Panagia, N., Robberto, M., Royle, P., Somerville, R. S., Sosey, M., 2006, \aj, 132, 1729

\reference{} Binette, L., Wang, J. C. L., Zuo, L., Magris, C. G., 1993 ,\aj,105,797

\reference{} Bolton, J. S., Haehnelt, M. G., Viel, M., Springel, V.,2005,\mnras,357, 1178

\reference{} Bouwens, R. J., Illingworth, G. D., Franx, M., Ford, H., 2007, \apj, submitted, arXiv:0707.2080

\reference{} Bower, R. G., Morris, S. L., Bacon, R., Wilman, R. J., Sullivan, M., Chapman, S., Davies, R. L., de Zeeuw, P. T., Emsellem, E., 2004,\mnras, 351, 63

\reference{} Brocklehurst, M., 1971,\mnras, 153, 471

\reference{} Bunker, A. J., Marleau, F. R., Graham, J. R., 1998, \aj, 116, 2086

\reference{} Bunker, A. J., Marleau, F. R., Graham, J. R., 1999,
in: proceedings of the "Galaxies in the Young Universe II" workshop, Ringberg Castle 2-6 August 1999,
Lecture Notes in Physics, eds. H. Hippelein \& K. Meisenheimer


\reference{} Bunker, A. J., Warren, S. J., Clements, D. L., Williger, G. M., Hewett, P. C.,
1999, \mnras, 309, 875


\reference{} Bunker, A., Smith, J., Spinrad, H., Stern, D., Warren, S.,2003, \apss, 284, 357

	
\reference{} 	Bunker, A. J., Stanway, E. R., Ellis, R. S., McMahon, R. G., 2004, \mnras, 355, 374





\reference{} Cantalupo, S., Lilly, S. J., Porciani, C., 2005, \apj, 628, 61

\reference{} Cantalupo, S., Lilly, S. J., Porciani, C., 2007,\apj,657,135


\reference{} Christensen, L., Wisotzki, L., Roth, M. M., Sanchez, S. F., Kelz, A., Jahnke, K.,  2007, \aap, 468, 587

\reference{} Chen, H.-W., Kennicutt, R.C., Rauch, M., 2005,\apj, 620, 703 
	
\reference{} Chun, M. R., Gharanfoli, S., Kulkarni, V. P., Takamiya, M., 2006, \aj, 131, 686


\reference{} Cowie, L. L., Hu, E. M., 1998, \aj,115, 1319

\reference{} Dey, A., Bian, C., Soifer, B.T., Brand, K., Brown, M.J.I., Chaffee, F.H., LeFloc'h. E., Hill, G., Houck, J.R., Jannuzi, B.T., Rieke, M., Weedman, D., Brodwin, M., \reference{} Eisenhardt, P., 2005, \apj, 626, 654

\reference{} Dijkstra M., Haiman, Z., Spaans, M. , 2006, \apj, 649, 14

\reference{} Dijkstra M., Haiman, Z., Spaans, M. , 2006, \apj, 649, 37

\reference{} Erb, D. K., Steidel, C. C., Shapley, A. E., Pettini, M., Reddy, N. A., Adelberger, K. L.,
2006, \apj, 128


	
\reference{} 	Fardal, M. A., Katz, N., Gardner, J. P., Hernquist, L., Weinberg, D. H., DavŽ, R., 2001, \apj, 562, 605

\reference{} Francis P. J., et al., 2001, \apj, 554, 1001


\reference{} Francis, P. J., Bland-Hawthorn, J., 2004, \mnras, 353, 301

\reference{} Francis, P. J., McDonnell, S., 2006,\mnras, 370, 1372


\reference{} Franx, M., Illingworth, G. D.,  Kelson, D. D., van Dokkum, P. G.; Tran, K.-V.
1997, \apjl,486, 75 

\reference{} Fujita, S. S., Ajiki, M., Shioya, Y., Nagao, T., Murayama, T., Taniguchi, Y., Okamura, S., Ouchi, M., Shimasaku, K., Doi, M.,  and 18 other authors, 2003,\aj, 125, 13


\reference{} Furlanetto, S. R., Schaye, J., Springel, V., Hernquist, L., 2005, \apj, 622, 7	
	

\reference{} Fynbo, J. U., M\o ller, P., Warren, S. J., 1999, \mnras, 305, 849

\reference{} Fynbo, J. P. U., Ledoux, C., Mšller, P., Thomsen, B., Burud, I., 2003 \aap, 407,147

\reference{}  Gerhard, O., Arnaboldi, M., Freeman, K. C., Okamura, S., Kashikawa, N., Yasuda, N.,2005, \apj, 621, 93

\reference{}  Gerhard, O., Arnaboldi, M., Freeman, K. C., Okamura, S., Kashikawa, N., Yasuda, N., 2007, \aap,468, 815
	
\reference{} Gould, A., Weinberg, D. H., 1996, \apj, 468, 462

\reference{} Gronwall, C., Ciardullo, R., Hickey, T., Gawiser, E., Feldmeier, J. J., van Dokkum, P. G., Urry, C. M., Herrera, D., Lehmer, B. D., Infante, L., and 6 coauthors, 2007, \apj, in press, arXiv:0705.3917

\reference{} Haardt, F., Madau, P., 1996, \apj, 461, 20

\reference{} Haehnelt, M. G., Steinmetz, M., Rauch, M.,  1998,\apj, 495, 647

\reference{} Haehnelt, M. G., Steinmetz, M., Rauch, M., 2000, \apj, 534, 594

 \reference{} Haiman, Z., Rees, M. J., 2001, \apj, 556, 87	

\reference{} Haiman, Z., Spaans, M., Quataert, E., 2000, \apj, 537, 5

\reference{} Hall, P. B., Osmer, P. S. Green, R. F.; Porter, A. C.; Warren, S. J., 1996, \apj, 462, 614

\reference{}  Hennawi, J. private communication, 2007
	
\reference{}  Hogan, C. J., Weymann, R. J., 1987, \mnras, 225, 1

\reference{} Hogg, D. W., Cohen, J. G., Blandford, R., Pahre, M. A., 1998, \apj, 504, 622

\reference{}  Hu, E. M., Cowie, L. L., McMahon, R. G., 1998,\apj, 502, 99


\reference {} Johansson, P. H.; Efstathiou, G.,  2006, \mnras, 371, 1519

\reference{} Kauffmann, G., 1996, \mnras, 281, 475

\reference{} Keel, W. C., Cohen, S. H., Windhorst, R. A.,  Waddington, I., 1999, \aj, 118, 2547

\reference{}   Kelson, D. D.,2003, \pasp, 115, 688

\reference{}  Kennicutt, R.C., 1998, \apj,498, 541
    
\reference{}  	Kudritzki, R.-P., Mendez, R. H., Feldmeier, J. J., Ciardullo, R., Jacoby, G. H., Freeman, K. C., Arnaboldi, M., Capaccioli, M., Gerhard, O., Ford, H. C., 2000,\apj, 536, 19


\reference{} Kulkarni, V. P., Hill, J. M., Schneider, G. Weymann, R. J., Storrie-Lombardi, L. J., Rieke, M. J., Thompson, R. I., Jannuzi, B. T., 2000, \apj, 536, 36


\reference{}	Kulkarni V. P., Hill, J. M., Schneider, G., Weymann, R. J., Storrie-Lombardi, L. J., Rieke, M. J., Thompson, R. I., Jannuzi, B. T., 2001, \apj, 551, 37


\reference{} LeBrun v., Bergeron, J., Boiss\'e, P., Deharveng, J.M., 1997, \aap, 321, 733



\reference{} Le Delliou, M., Lacey, C., Baugh, C.M., Guiderdoni, B., Bacon, R., Courtois, H., Sousbie, T., and Morris, S.L., 2005, \mnras, 357, L11

\reference{} Le Delliou, M., Lacey, C.G., Baugh, C.M., and Morris, S.L., 2006, \mnras, 365, 712




\reference{} Ledoux C., Petitjean P., Bergeron, J., Wampler E. J., Srianand, R., 1998, \aap, 337, 51


\reference{} Lequeux, J., Kunth, J.M., Mas-Hesse, J.M., Sargent, W.L.W., 1995, \aap, 301, 18 


\reference{} Lowenthal, J., D. Hogan, C. J., Leach, R. W., Schmidt, G. D., Foltz, C. B., 1990,\apj,357,3



\reference{} Madau, P., Pozzetti, L., Dickinson, M., 1998,\apj, 498, 106

\reference{} Mas-Hesse, J.M., Kunth, D., Tenorio-Tagle, G., Leitherer, C., Terlevich, R.J., Terlevich, E., 2003, \apj, 598, 858

\reference{} Matsuda, Y., Yamada, T., Hayashino, T., Yamauchi, R., Nakamura, Y., 2006,\apj 640, 123

\reference{} M\o ller, P., Warren, S. J., Fall, S. M., Fynbo, J. U., Jakobsen, P.,
2002, \apj 574, 51

\reference{} Mo, H. J., White, S. D. M., 2002, \mnras, 336, 112

\reference{} Murphy, M.T., Liske, J., 2004,\mnras, 354, 31

\reference{} Nagamine K.,  Wolfe A.M.,  Hernquist L. , Springel, V., 2007, \apj, 660, 945   

\reference{} Neufeld, D.A., 1990, \apj, 350, 216
    
\reference{} Nilsson, K. K., Fynbo, J. P. U., M\o ller, P., Sommer-Larsen, J., Ledoux, C.,2006,\aap,452,23

\reference{} Nulsen, P.E.J., Barcons, X., Fabian, A.C., 1998, \mnras, 301, 168

\reference{} 	Ouchi, M., Shimasaku, K., Akiyama, M., Simpson, C., Saito, T., Ueda, Y., Furusawa, H., Sekiguchi, K., Yamada, T., Kodama, T., and 6 coauthors, submitted to ApJ, arXiv:0707.3161

\reference{} Peroux, C., McMahon, R. G., Storrie-Lombardi, L. J., Irwin, M. J., 2003, \mnras. 346, 1103
 
\reference{} Peroux C., Dessauges-Zavadsky M., D'Odorico S., Kim, T.-S.,, McMahon, R. G., 2005,\mnras,363, 479

\reference{} Prochaska, J. X., Wolfe, A. M., 1997, \apj, 487, 73

\reference{} Rao S.M., Turnshek D.A., Nestor D.B., 2006, \apj, 636, 610

\reference{} Rauch, M., Haehnelt, M. G., Steinmetz, M. 1997a, \apj, 481, 601	

\reference{} Rauch, M., Miralda-Escude J., Sargent, W. L. W., Barlow, T. A.; Weinberg, D. H., Hernquist, L., Katz, N., Cen, R., Ostriker J. P., 1997b, \apj, 489, 7

\reference{} Schaye, J., 2001, \apj, 559, 1

\reference{} Scott, J. E., Kriss, G. A., Brotherton, M., Green, R. F., Hutchings, J., Shull, J. M., Zheng, W., 2004, \apj, 615, 135

\reference{} Shapley, A. E., Steidel, C. C., Pettini, M., Adelberger, K. L., 2003, \apj, 588, 65

\reference{} Shapley, A. E., Steidel, C. C., Pettini, M., Adelberger, K. L., Erb, D. K., 2006, \apj, 651, 688

\reference{} Simcoe, R. A., Sargent, W. L. W., Rauch, M., 2004,\apj,606, 92

\reference{} 	Smith, D. J. B., Jarvis, M. J., 2007, \mnras,378, 49

\reference{} Stark, D. P., Ellis, R. S., Richard, J., Kneib, J.-P., Smith, G. P., Santos, M. R., 2007, \apj, 663, 10

\reference{} Steidel, C. C., 1990, \apjs, 74, 37

\reference{} 	Steidel, C. C., Hamilton, D., 1993, \aj, 105,2017

\reference{} 	Steidel, C. C., Adelberger, K. L., Giavalisco, M., Dickinson, M., Pettini, M., 1999,\apj, 519, 1


\reference{} 	Steidel, C. C., Adelberger, K. L., Shapley, A. E., Pettini, M., Dickinson, M., Giavalisco, M., 2000, \apj, 532, 170

\reference{} 	Stiavelli, M., Scarlata, C., Panagia, N., Treu, T., Bertin, G., Bertola, F., 2001,\apj, 561, 37

\reference{} Tasitsiomi, A., 2006, \apj, 645, 792

\reference{} Tapken, C., Appenzeller, I., Noll, S., Richling, S., Heidt, J., Meink\"ohn, E., Mehlert, D., 2007, \aap, 467, 63

\reference{} Telfer, R. C., Zheng, W., Kriss, G. A., Davidsen, A. F., 2002 \apj, 565, 773
	
\reference{} Tenorio-Tagle, G., Silich, S. A., Kunth, D., Terlevich, E., Terlevich, R., 1999, \mnras, 309, 332

\reference{} Trentham, N., Sampson, L., Banerji, M., 2005, \mnras, 357,783


\reference{} van Breukelen, C., Jarvis, M.J., Venemans, B.P., 2005, \mnras, 359, 895

\reference{} Wang, L., Li, C., Kauffmann, G., De Lucia, G., 2007, \mnras, 377, 1419

\reference{} Warren, S. J., M\o ller, P., Fall, S. M.; Jakobsen, P., 2001 \mnras, 326, 759

\reference{} Weatherley, S. J., Warren, S. J., M\o ller, P., Fall, S. M., Fynbo, J. U., Croom, S. M.,  2005, \mnras, 358, 985	

\reference{} Weidinger, M. M\o ller, P., Fynbo, J. P. U., 2004, Nature, 430, 999

\reference{} Wild, V., Hewett, P.C., Pettini, M., 2007, \mnras, 374, 292

\reference{} Weidinger, M., M\o ller, P., Fynbo, J. P. U., Thomsen, B., 2005, \aap, 436,825

\reference{} Wolfe, A. M., Lanzetta, K. M., Foltz, C. B., Chaffee, F. H., 1995, \apj,454,698

\reference{} Wolfe, A. M., Gawiser, E, Prochaska, J. X., 2003, \apj, 593, 235 


\reference{} Wolfe, A. M., Chen, H.-W., 2006, \apj, 652, 981


\reference{} Zheng, Z., Miralda-Escud\'e, J., 2002, \apj, 578, 33	

\end{references}
\end{document}